\documentclass[aps,pra,twoside,twocolumn,10pt,longbibliography,nofootinbib,floatfix]{revtex4-1}
\usepackage[colorlinks=true, citecolor=red, urlcolor=blue ]{hyperref}
\usepackage{times,epsfig,amssymb,amsfonts,amsmath,bm,subfigure,mathtools,amsthm,braket,soul,enumitem,color,geometry,comment}
\usepackage[normalem]{ulem}
\newcommand{\bigket}[1]{{\left|\left.#1\right\rangle\right.}}
\newcommand{\bigbra}[1]{{\left\langle\left.#1\right|\right.}}
\newcommand{\kstate}[1]{\bigket{\text{#1}}}
\newcommand{\bstate}[1]{\bigbra{\text{#1}}}
\newcommand{\bracket}[2]{\langle{#1}|{#2}\rangle}

\newcommand{\JK}[1]{{\color{black}#1}}
\newcommand{\note}[1]{{\noindent\textbf{Note #1.}}}
\newcommand{\propose}[1]{{\noindent\textbf{$\blacksquare$ Proposition #1.}}}
\newcommand{\corollary}[1]{{\noindent\textbf{$\diamond$ Corollary #1.}}}

\begin{document}

\title{Controlling gain with loss: Bounds on localizable entanglement in multi-qubit systems}
\author{Jithin G. Krishnan, Harikrishnan K. J. and Amit Kumar Pal}
\affiliation{Department of Physics, Indian Institute of Technology Palakkad, Palakkad 678 623, India}
\date{\today}

\begin{abstract}
We investigate the relation between the amount of entanglement localized on a chosen subsystem of a multi-qubit system via local measurements on the rest of the system, and the bipartite entanglement that is lost during this measurement process. We study a number of paradigmatic pure states, including the generalized GHZ, the generalized W, Dicke, and the generalized Dicke states. For the generalized GHZ and W states, we analytically derive bounds on localizable entanglement in terms of the entanglement present in the system prior to the measurement. Also, for the Dicke and the generalized Dicke states, we demonstrate that with increasing system size, localizable entanglement tends to be equal to the bipartite entanglement present in the system over a specific partition before measurement. We extend the investigation numerically in the case of arbitrary multi-qubit pure states. We also analytically determine the modification of these results, including the proposed bounds, in situations where these pure states are subjected to single-qubit phase-flip noise on all qubits. Additionally, we study one-dimensional paradigmatic quantum spin models, namely the transverse-field XY model and the XXZ model in an external field, and numerically demonstrate a cubic dependence of the localized entanglement on the lost entanglement. We show that this relation is robust even in the presence of disorder in the strength of the external field. 
\end{abstract}

\maketitle

\section{Introduction}
\label{sec:intro}

\begin{figure*}
    \centering
    \includegraphics[width=0.6\textwidth]{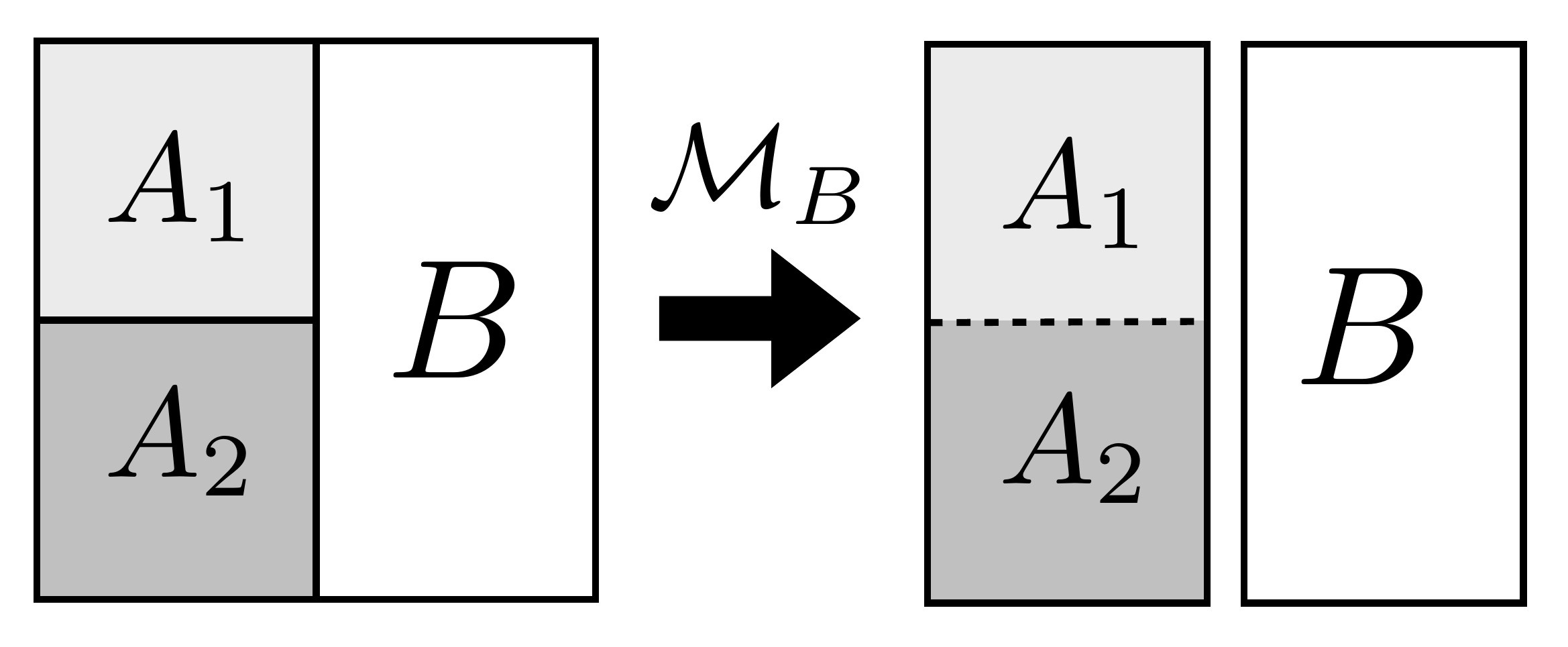}
    \caption{Consider a multi-qubit system divided into three subsystems, $A_1$, $A_2$, and $B$, with bipartite entanglement $E_{A_1A_2:B}$, $E_{A_1:A_2B}$, and $E_{A_2:A_1B}$ over different bipartitions. Measurement on all qubits in $B$, denoted by $\mathcal{M}_B$, decouples the subsystem $B$ from the qubits in $A\equiv A_1A_2$.}
    \label{fig:fig1}
\end{figure*}

In the last three decades, entanglement~\cite{horodecki2009,guhne2009} have been established as the key resource in quantum information processing tasks, including quantum teleportation~\cite{horodecki2009,bennett1993,bouwmeester1997}, super-dense coding~\cite{horodecki2009,bennett1992,mattle1996,sende2010}, and quantum cryptography~\cite{ekert1991,jennewein2000}. Concepts related to entanglement theory have also been used in areas that are seemingly different from quantum information theory, such as in probing gauge-gravity duality~\cite{hubeny2015,pastawski2015,almheiri2015,jahn2017}, in understanding time as an emergent phenomena from entanglement~\cite{page1983,gambini2009,moreva2014}, and even in studying systems like photosynthetic complexes~\cite{lambert2013} that are important from biological point of view. These have motivated enormous experimental advancements in creating and manipulating entangled states in the laboratory using various substrates, namely, photons~\cite{raimond2001,prevedel2009,barz2015}, trapped ions~\cite{leibfried2003,leibfried2005,brown2016}, cold atoms~\cite{mandel2003,bloch2005,bloch2008}, superconducting qubits~\cite{clarke2008,barends2014}, and nuclear magnetic resonance molecules~\cite{negrevergne2006}. Moreover, quantum many-body systems~\cite{Augusiak2012} have emerged as the natural choice for implementing quantum information processing tasks, and the necessity of studying the entanglement properties of these systems has also been realized~\cite{amico2008,dechiara2018}.

Entanglement over a subsystem $A$ of a composite quantum system in the state $\rho$ can be quantified by computing an appropriate entanglement measure $E$ over the reduced state $\rho_A=\text{Tr}_{B}\left[\rho\right]$ of the subsystem $A$, where $B$ is the rest of the system~\cite{horodecki2009,guhne2009}. While this approach has been successful in a wide variety of multiparty quantum states~\cite{horodecki2001a,horodecki2009,guhne2009}, there exist states like $N$-qubit Greenberger-Horne-Zeilinger (GHZ) states~\cite{greenberger1989}, graph states~\cite{hein2006}, and stabilizer states in quantum error correcting codes~\cite{raussendorf2003,fujii2015} for which the \emph{partial trace-based} avenue may lead to vanishing entanglement measure on the state $\rho_A$. In such situations, one may take a \emph{measurement-based} approach, where non-zero bipartite or multipartite entanglement can be \emph{localized} on the subsystem $A$ in the post-measured state of the system by performing measurements on $B$~\cite{divincenzo1998}. In the case of a multi-qubit system, this leads to the definition of the localizable entanglement~\cite{verstraete2004,verstraete2004a,popp2005}, defined as the maximum average entanglement localized over $A$ via local single-qubit projection measurements on all qubits in $B$, given by 
\begin{eqnarray}
\langle E_A\rangle=\max \sum_{k}p_k E(\tilde{\rho}^k_A).
\label{eq:loce_def}
\end{eqnarray}
Here, $k$ labels the measurement outcomes corresponding to the post-measured states $\tilde{\rho}^k$ occurring with probability $p_k$ $\left(\sum_{k}p_k=1\right)$, where  $\tilde{\rho}^k_A=\text{Tr}_B[\tilde{\rho}^k]$. Depending on the possible partitions in $A$, the entanglement measure $E$ computed over $A$ post-measurement can be either a bipartite~\cite{verstraete2004,verstraete2004a,popp2005}, or a multipartite~\cite{sadhukhan2017} measure. Apart from successfully characterizing entanglement in GHZ and GHZ-like states such as the graph~\cite{hein2006} and the stabilizer states~\cite{amaro2018,amaro2020a}, localizable entanglement and related ideas have been immensely useful in defining the correlation length in one-dimensional (1D) quantum spin models~\cite{verstraete2004,verstraete2004a,popp2005,jin2004}, in characterizing quantum phase transitions in the cluster-Ising~\cite{skrovseth2009,smacchia2011} and cluster-XY models~\cite{montes2012} in terms of entanglement, and in entanglement percolation through quantum networks~\cite{acin2007}.

The measurement on the subsystem $B$ completely decouples $B$ from $A$, leading to
\begin{eqnarray}
\rho\rightarrow\tilde{\rho}=\tilde{\rho}_A\otimes\tilde{\rho}_B, 
\end{eqnarray}
implying a complete loss in entanglement over all qubits belonging to the different subsystems $A$ and $B$. A natural question that arises is \emph{whether and how the entanglement $\langle E_A\rangle$ localized on $A$ via measurements on $B$ depends on the entanglement that is lost during the same measurement process}. More specifically, considering a bipartition $A_1:A_2$ of $A$ (i.e., an overall tripartition $A_1:A_2:B$ of the multi-qubit system, see Fig.~\ref{fig:fig1}), and a bipartite entanglement measure $E$, we ponder on the following question: \emph{Does any relation exist between the bipartite entanglement $\langle E_A\rangle \equiv\langle E_{A_1A_2}\rangle$ localized over the subsystem $A$ via single-qubit projection measurements on all qubits in $B$, and the entanglement over different bipartitions prior to the measurement, namely, $E_{A_1A_2:B}$, $E_{A_1:A_2B}$, and $E_{A_2:A_1B}$, that are lost during the measurement process leading to $\langle E_{A_1A_2}\rangle$?} On one hand, answer to this question may give rise to constraints on the entanglement localizable over the subsystem of a multi-qubit system in terms of the entanglement present in the system prior to measurement -- a situation that is of fundamental interest from the perspective of complete characterization of the system via entanglement. On the other hand, such a study may also aid in estimating the localizable entanglement prior to the measurement via the information on the bipartite entanglement present in the system. The latter is advantageous from a practical point of view, specially in situations where performing the measurements and optimizing over all possible measurements may turn out to be difficult.     

A few results exist in this direction. Note that the monotonicity of $E$~\cite{horodecki2009,popescu1997,vedral1998,horodecki2001a,vidal2000} implies that 
\begin{eqnarray}
\langle E_{A_1A_2}\rangle\leq \min\left[E_{A_1:A_2B},E_{A_2:A_1B}\right].
\label{eq:localizable_inequality}
\end{eqnarray}
It has been shown that the inequality~(\ref{eq:localizable_inequality}) can be tightened to an equality in the case of  asymptotic pure state distillation~\cite{smolin2005,yang2009}. At the single copy level, it has also been shown that for all three-qubit pure states with each of $A_1, A_2$, and $B$ being a qubit, the inequality~(\ref{eq:localizable_inequality}) becomes an equality via some measurement on $B$ for a specific choice of entanglement measure~\cite{pollock2021}. In the case of multi-qubit pure states, investigation on the relation between localizable multipartite entanglement $\langle E_{A}\rangle$ and the multipartite entanglement, as quantified by the multiparty entanglement measure $E$, over the state prior to the measurement has also been made~\cite{sadhukhan2017}. However, we are still far from a systematic and complete understanding of the problem in the case of arbitrary multi-qubit quantum states.  

In this paper, we study the interplay between $\langle E_{A_1A_2}\rangle$, $E_{A_1:A_2B}$, $E_{A_2:A_1B}$, and $E_{A_1A_2:B}$ in multi-qubit systems, where each of the partitions $A_1$, $A_2$, and $B$ may consist of multiple qubits. We start with the investigation of a number of paradigmatic $N$-qubit pure states, including the generalized GHZ states~\cite{greenberger1989,dur2000}, the generalized W states~\cite{dur2000,sende2003}, Dicke states~\cite{dicke1954,bergmann2013,lucke2014,kumar2017}, and generalized Dicke states~\cite{sadhukhan2017}, and analytically derive bounds of $\langle E_{A_1A_2}\rangle$ in terms of $E_{A_1:A_2B}$, $E_{A_2:A_1B}$, and $E_{A_1A_2:B}$. In the cases of arbitrary pure states of $N$-qubits, we numerically investigate whether it is always possible to exceed the loss in bipartite entanglement via localization performed through measurement. We extend our investigation to the pure states subjected to the single-qubit phase flip noise~\cite{nielsen2010,holevo2012} of Markovian~\cite{yu2009} and non-Markovian~\cite{Daffer2004,Shrikant2018,Gupta2020} type, and discuss the modifications of the bounds obtained for pure states due to the presence of noise. We also look into the ground states obtained from one-dimensional (1D) interacting quantum spin models, both in the presence and absence of disorder~\cite{de_dominicis_giardina_2006}. More specifically, we consider the 1D transverse-field XY model~\cite{Lieb1961,Barouch1970,Barouch1971,Barouch1971a,dutta1996,sachdev_2011} and the 1D XXZ model in an external field~\cite{Yang1966,Yang1966a,Yang1966b,Langari1998,Mikeska2004,Giamarchi2004}, where disorder can be present in the strength of the field. We numerically demonstrate a cubic dependence of the localizable entanglement on the entanglement lost during measurement in the ground states of the ordered and the disordered models, and demonstrate that the relation is robust against the presence of disorder in the field-strength.

The rest of the paper is organized as follows. In Sec.~\ref{sec:pure_states}, we formally define the localizable entanglement, and compute it, along with the bipartite entanglements present in the un-measured states, in the case of paradigmatic multi-qubit pure states. We also present the numerical data corresponding to arbitrary $N$-qubit pure states, and discuss the implications of the data. The modifications of the results for the pure state due to subjecting the states to single-qubit Markovian and non-Markovian phase flip channels are discussed in Sec.~\ref{sec:noise}. The study of the localizable and the lost entanglement in the ground states of ordered and disordered 1D quantum spin models can be found in Sec.~\ref{sec:spin_model}. Sec.~\ref{sec:conclude} presents the outlook and concluding remarks.

\section{Multi-qubit pure states}
\label{sec:pure_states}

In this section, we explore a number of paradigmatic multi-qubit pure states, and discuss the correlation between the localizable entanglement and the bipartite entanglement that is lost due to measurement. Let us take an $N$-qubit system, $S$, where the qubits are labelled as $1,2,\cdots,N$. As discussed in Sec.~\ref{sec:intro}, we consider a tripartition $A_1:A_2:B$ of the system, and perform single-qubit rank-$1$ projection measurements on all qubits in $B$. Without any loss in generality, we assume that $B$ holds $n(<N-1)$ qubits, and $A=A_1\cup A_2$ consists of the rest $N-n$ qubits. We label the qubits in $B$ as $1,2,\cdots,n$, and the qubits in $A$ as $n+1,n+2,\cdots,N-1,N$. In this situation, $\langle E_{A_1A_2}\rangle$ (Eq.~\ref{eq:loce_def}) takes the form
\begin{eqnarray}
\langle E_{A_1A_2}\rangle=\max \sum_{k=0}^{2^{n}-1}p_k E(\tilde{\rho}^k_{A_1A_2}),
\label{eq:bipartite_loce}
\end{eqnarray}
where $\tilde{\rho}^k_{A_1A_2}=\text{Tr}_B\left[\tilde{\rho}^k\right]$, $\tilde{\rho}^k=\left[M^k\rho M^{k\dagger}\right]/p_k$, and $p_k =\text{Tr}\left[ M^k\rho M^{k\dagger}\right]$, with $M^k$ being the measurement operation corresponding to the measurement outcome $k$, and the maximization is performed over the set of all possible single-qubit rank-$1$ projection measurements on all qubits in $B$. Note that the maximum value of $\langle E_{A_1A_2}\rangle$ can not exceed the  maximum value of the chosen entanglement measure, $E_{max}$, implying $\langle E_{A_1A_2}\rangle\leq E_{max}$.

The measurement operators $\{M^k\}$ can be written as $M^k=I_{A_1A_2}\otimes P^k_B$, with the projectors on the qubits in $B$ corresponding to the measurement outcome $k$ given by $P^k_B=\ket{\tilde{b}_k}\bra{\tilde{b}_k}$, where $k_i=0,1$ and $k\equiv k_{1}k_{2}\cdots k_{n}$ is a multi-index. Here, $\ket{\tilde{b}_k} = \otimes_{\forall i\in B}\ket{b_{k_i}}$, and 
\begin{eqnarray}
\ket{b_0}_i&=& \cos\frac{\theta_i}{2}\ket{0}+\text{e}^{\text{i}\phi_i}\sin\frac{\theta_i}{2}\ket{1},\nonumber \\ 
\ket{b_1}_i&=& \sin\frac{\theta_i}{2}\ket{0}-\text{e}^{\text{i}\phi_i}\cos\frac{\theta_i}{2}\ket{1},
\end{eqnarray}
on all qubits $i\in B$, where $\theta_i,\phi_i\in\mathbb{R}$, $0\leq\theta_i\leq \pi$, $0\leq \phi\leq 2\pi$. Unless otherwise stated, we maintain these notations in all subsequent calculations. Also, we use $A$ and $A_1A_2$ interchangeably, since $A\equiv A_1\cup A_2=A_1A_2$.

\JK{The maximum value of $\langle E_{A_1A_2}\rangle$ as well as the optimal measurement basis on $B$ providing this value depends on the choice of $E$. Therefore, a discussion on the choice of $E$ is in order here. Note that the analytical determination of bounds reported in this as well as in the subsequent sections depend on the computability of the entanglement measure in terms of the state parameters. Also, a computable measure for mixed states is desired as required in the case of noisy systems, discussed in Sec.~\ref{sec:noise}. In view of these, we choose negativity~\cite{peres1996,horodecki1996,vidal2002,zyczkowski1998,lee2000} (see Appendix~\ref{app:negativity} for a brief definition) as the entanglement measure for demonstration of the results, although it is important to remind ourselves that negativity, by definition, does not take into account entangled states with positive partial transpose. Therefore it can not provide the full picture in $\mathbb{C}^{d_1}\otimes\mathbb{C}^{d_2}$ scenario except the case of $\mathbb{C}^2\otimes\mathbb{C}^2$ and $\mathbb{C}^2\otimes\mathbb{C}^3$~\cite{Horodecki1997,Dur2000a,Dur2000b}. We have also tested our results for other entanglement measures, such as logarithmic negativity~\cite{plenio2005} and von-Neumann entropy~\cite{bennett1996,bennett1996a,horodecki1996,horodecki2009,guhne2009} (see Appendix~\ref{app:negativity} for brief definitions), and have found the results to be qualitatively valid. We comment on the subtle differences for different measures as we discuss the specific results in this section  and in the subsequent sections.}



\subsection{Generalized GHZ states}
\label{subsec:gghz_states}

We start with the $N$-qubit generalized GHZ (gGHZ) state~\cite{greenberger1989}, given by
\begin{eqnarray}
\kstate{gGHZ} &=& a_0\ket{0}^{\otimes N}+a_1\ket{1}^{\otimes N},
\end{eqnarray}
where $a_0$ and $a_1$ are complex numbers, i.e.,  $a_0,a_1\in\mathbb{C}$, and the state is normalized, implying $|a_0|^2+|a_1|^2=1$. For the gGHZ state, we present the following proposition. 

\propose{I} \emph{For any tripartition $A_1:A_2:B$ of an $N$-qubit gGHZ state,}
\begin{eqnarray}
\langle E_{A_1A_2}\rangle=E_{A_1A_2:B}=E_{A_1:A_2B}=E_{A_2:A_1B}.
\end{eqnarray}

\begin{proof} Partial transposition of $\rho=\kstate{gGHZ}\bstate{gGHZ}$ with  respect to the subsystems $B$ leads to
\begin{eqnarray}
\rho^{T_B}&=&|a_0|^2(\ket{0}\bra{0})^{\otimes N}+|a_1|^2(\ket{1}\bra{1})^{\otimes N}\nonumber \\ 
&&+ a_0a_1^* (\ket{0}\bra{1})^{\otimes N-n}(\ket{1}\bra{0})^{\otimes n} \nonumber \\
&&+ a_0^*a_1 (\ket{1}\bra{0})^{\otimes N-n}(\ket{0}\bra{1})^{\otimes n},
\end{eqnarray}
with non-zero eigenvalues $|a_0|^2,|a_1|^2,\pm|a_0||a_1|$. Therefore the entanglement between partition $A_1A_2$ and partition $B$, as quantified by negativity, is given by 
\begin{eqnarray}
E_{A_1A_2:B} &=& 2|a_0|\sqrt{1-|a_0|^2}. 
\label{eq:gghz_neg}
\end{eqnarray}

We now compute localizable entanglement, $\langle E_{A_1A_2}\rangle$, for the gGHZ state. Application of the measurement operator $M^k$ corresponding to the measurement outcome $k$ on $\kstate{gGHZ}$ leads to 
\begin{eqnarray}
M^k\kstate{gGHZ} &=& \kstate{gGHZ$_k$}_{A_1A_2}\otimes\ket{\tilde{b}_k}_B
\end{eqnarray}
where 
\begin{eqnarray}
\kstate{gGHZ$_k$} &=& \frac{1}{\sqrt{p_k}}\left(a_0f^k_0\ket{0}^{\otimes (N-n)}+a_1f^k_1\ket{1}^{\otimes (N-n)}\right),\nonumber\\
p_k &=& |a_0|^2|f^k_0|^2+|a_1|^2|f^k_1|^2.
\end{eqnarray}
Here, 
\begin{eqnarray}
f^k_0 &=&\prod_{i\in B}f^{k_i}_0=\prod_{i\in B}\bracket{b_{k_i}}{0}_i,\nonumber\\
f^k_1 &=&\prod_{i\in B}f^{k_i}_1=\prod_{i\in B}\bracket{b_{k_i}}{1}_i   
\end{eqnarray}
are functions of $2n$ real parameters $\{\theta_i,\phi_i\}$, $i=1,\cdots,n$. The negativity, $E_{A_1:A_2}^k$, can now be computed for each post-measured states $\kstate{gGHZ$_k$}$ on $A_1A_2$ as   
\begin{eqnarray}
E_{A_1:A_2}^k &=&  \frac{2|a_0|\sqrt{1-|a_0|^2}|f^k_0||f^k_1|}{p_k}. 
\end{eqnarray}
The localizable entanglement across the bipartition $A_1:A_2$ of $A$ is, therefore, 
\begin{eqnarray}
\langle E_{A_1:A_2}\rangle &=& 2|a_0|\sqrt{1-|a_0|^2}\left[\max\sum_{k=0}^{2^{n}-1}|f^k_0||f^k_1|\right].\nonumber\\ 
\label{eq:final_expression_gghz}
\end{eqnarray}

In order to perform the maximization, we note that $|f^k_{0,1}|$, and subsequently $p_k$ and $E_{A_1:A_2}^k$, are independent of $\{\phi_i\}$, $i=1,\cdots,n$, thereby reducing the maximization problem to one involving $n$ real parameters, $\{\theta_i\}$, $i=1,\cdots,n$. Moreover, since  $|f^k_{0,1}|\geq 0$, and  
\begin{eqnarray}
|f_0^k||f^k_1|&=& \frac{1}{2^{n}}\prod_{\forall i\in B}\sin\theta_i, 
\end{eqnarray}
one obtains
\begin{eqnarray}
\underset{\theta_i}{\max}\sum_{k=0}^{2^{n}-1}|f^k_0||f^k_1|&=&\frac{1}{2^{n}}\sum_{k=0}^{2^{n}-1}\underset{\theta_i}{\max}\prod_{\forall i\in B}\sin\theta_i\nonumber\\
&=& 1,
\end{eqnarray}
where the maximization takes place for $\theta_i=\frac{\pi}{2}$ $\forall i\in B$, implying that a $\sigma^x$ measurement on all qubits in $B$ is optimal for obtaining the maximum $\langle E_{A_1A_2}\rangle$ as  
\begin{eqnarray}
\langle E_{A_1A_2}\rangle &=& 2|a_0|\sqrt{1-|a_0|^2}.
\end{eqnarray}
It is easy to see from the symmetry of the gGHZ state that $E_{A_1A_2:B}=E_{A_1:A_2B}=E_{A_2:A_1B}$, leading to
\begin{eqnarray}
\langle E_{A_1A_2}\rangle=E_{A_1A_2:B}=E_{A_1:A_2B}=E_{A_2:A_1B}.
\label{eq:equality_gGHZ}
\end{eqnarray}
Hence the proof.
\end{proof}

Therefore, in the case of the $N$-qubit gGHZ states, it is not possible to exceed the bipartite entanglement present in the state prior to the measurement via localizable entanglement. Note also that for the $N$-qubit gGHZ state, (\ref{eq:localizable_inequality}) is an equality. \JK{Also, one can use von Neumann entropy to compute entanglment in the case of the pure gGHZ states, where Eq.~(\ref{eq:equality_gGHZ}) is found to be unaltered.}

\begin{figure*}
    \centering
    \includegraphics[width=0.7\textwidth]{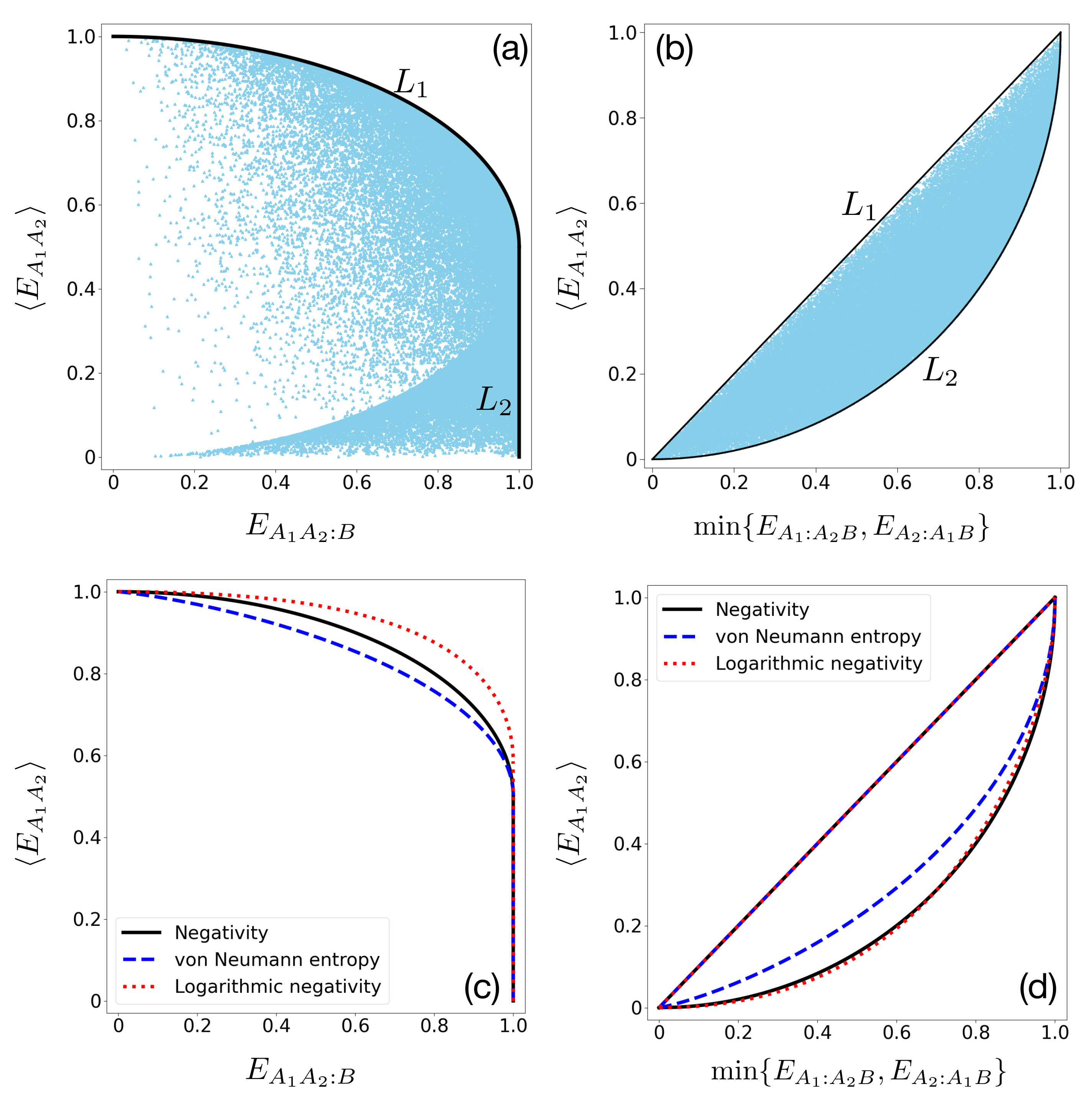}
    \caption{\textbf{Generalized W states.} Scatter plot of a sample of $10^7$ Haar-uniformly generated $3$-qubit gW states on the \textbf{(a)} $\left(E_{A_1A_2:B},\langle E_{A_1A_2}\rangle\right)$ plane and the \textbf{(b)} $\left(\min\{E_{A_1:A_2B},E_{A_2:A_1B}\},\langle E_{A_1A_2}\rangle\right)$ plane, where negativity is used as the entanglement measure. The lines $L_1$ and $L_2$ on the $\left(E_{A_1A_2:B},\langle E_{A_1A_2}\rangle\right)$ plane represent respectively the family of states following Eq.~(\ref{eq:gw_upper_bound}), and $E_{A_1A_2:B}=1$. On the other hand, the lines $L_1$ and $L_2$ on the $\left(\min\{E_{A_1:A_2B},E_{A_2:A_1B}\},\langle E_{A_1A_2}\rangle\right)$ plane represent the family of states obeying Eqs.~(\ref{eq:upper_bound_2}) and (\ref{eq:gw_lower_bound}) respectively. \JK{The bounds for logarithmic negativity and von Neumann entropy are shown in \textbf{(c)} and \textbf{(d)} along with the same for negativity for comparison (see Sec.~\ref{subsec:gw_states}, Note 3 for a discussion). All quantities plotted are dimensionless, except the entanglement and localizable entanglement computed using von Neumann entropy as the entanglement measure, which are in ebits.}}
    \label{fig:fig2}
\end{figure*}

\subsection{Generalized W states}
\label{subsec:gw_states}

Next, we focus on the $N$-qubit generalized W states~\cite{dur2000,sende2003}, given by 
\begin{eqnarray}
\kstate{gW} &=& \sum_{i=1}^N a_i\ket{0}^{\otimes (i-1)}\ket{1}_i\ket{0}^{\otimes (N-i)},
\label{eq:n_qubit_gw}
\end{eqnarray}
where $a_i\in\mathbb{C}$ $\forall i\in\{1,2,\cdots,N\}$, satisfying the normalization condition $\sum_{i=1}^N|a_i|^2=1$. Let us first consider a subset of the $N$-qubit gW states given in Eq.~(\ref{eq:n_qubit_gw}) where the coefficients $a_i$ are real, i.e.,  $a_i\in\mathbb{R}$. We first focus on the correlation between $\langle E_{A_1A_2}\rangle$ and $E_{A_1A_2:B}\equiv E_{AB}$. Negativity over any bipartition $AB$ of the system is given by\footnote{For gW states with real coefficients, this can be obtained by observing the patterns of the negative eigenvalues of the partially transposed density matrix of the smaller systems. Eq.~(\ref{eq:gW_negativity}) is also verified numerically for large gW states with real coefficients as well as complex coefficients, where in the case of the latter, $a_i^2$ is replaced with $|a_i|^2$.} 
\begin{eqnarray}
 E_{A_1A_2:B} &=& 2\left|\left[\left(\sum_{i=n+1}^{n+m} a_i^2+\sum_{i=n+m+1}^{N} a_i^2\right)\sum_{i=1}^n a_i^2\right]^{\frac{1}{2}}\right|,\nonumber\\ 
 \label{eq:gW_negativity}
\end{eqnarray}
where we have assumed that qubits $1,2,\cdots,n$ constitute the subsystem $B$, and qubits $n+1,n+2,\cdots,N$ form the subsystem $B$. On the other hand, computation of $\langle E_{A_1A_2}\rangle$ involves application of $M^k$, $k=0,1,\cdots,2^{n}-1$, on the $n$ qubits in $B$ of $\kstate{gW}$ (see Sec.~\ref{subsec:gghz_states}), leading to
\begin{eqnarray}
M^k\kstate{gW} =\ket{\tilde{b}_k}_B \otimes \ket{\psi_k}_{A_1A_2} 
\end{eqnarray} 
with 
\begin{eqnarray}
\ket{\psi_k} &=&c^{k}_{0} \ket{0}^{\otimes(N-n)} \nonumber\\ &&+ \sum_{i=1}^{N-n} c^{k}_{i}\ket{0}^{\otimes (i-1)}\ket{1}_i\ket{0}^{\otimes (N-n-i)},
\label{eq:post_measured_state}
\end{eqnarray}
where $c^k_0=f_0^k/\sqrt{p_k}$, and $c^k_i=f^k a_{n+i}/\sqrt{p_k}$, $i=1,2,\cdots,N-n$, with $p_k$ being the probability of obtaining the measurement outcome $k$, such that $\sum_{i=0}^{N-n} |c_i^k|^2=1$ ensuring normalization. Note further that among the coefficients $c_i^k$, $i=0,1,\cdots,N-n$, only $c_0^k$ is complex (see Appendix~\ref{app:two_measurement} for explicit examples with the cases of single- and two-qubit measurements).  Using this, negativity in the state $\ket{\psi_k}$ over the bipartition $A_1:A_2$ can be written for all $k$ as\footnote{This expression for negativity of the states of the form in Eq.~(\ref{eq:post_measured_state}) has also been obtained analytically for smaller systems, and has been verified numerically for larger systems.} 
\begin{eqnarray}
E_{A_1A_2}^k&=& \frac{2(f^k)^2}{p_k}\left|\left[\left(\sum_{i=n+1}^{n+m} a_i^2\right)\sum_{n+m+1}^{N} a_i^2\right]^{\frac{1}{2}}\right|.
\label{eq:post-measured_negativity}
\end{eqnarray}
Here, we have assumed, without any loss in generality, that the subsystem $A_1$ ($A_2$) is constituted of the qubits $n+1,n+2,\cdots,n+m$ (qubits $n+m+1,n+m+2,\cdots,N$). Therefore, 
\begin{eqnarray}
\langle E_{A_1A_2}\rangle &=& \sum_k p_k E^k_{A_1A_2} \nonumber \\ 
&=& 2\left|\left[\left(\sum_{i=n+1}^{n+m} a_i^2\right)\sum_{n+m+1}^{N} a_i^2\right]^{\frac{1}{2}}\right| \sum_{k}(f^k)^2. 
\label{eq:loc_neg_intermediate}
\end{eqnarray}
For small values of $n$, it can analytically be shown that the factor $\sum_{k=0}^{2^n-1} (f^k)^2=1$\footnote{See Appendix~\ref{app:two_measurement} for cases upto $n=2$}, while our numerical investigation shows this to be true for high values of $n$ also. This leads to 
\begin{eqnarray}
 \langle E_{A_1A_2}\rangle &=& 2\left|\left[\left(\sum_{i=n+1}^{n+m} a_i^2\right)\sum_{n+m+1}^{N} a_i^2\right]^{\frac{1}{2}}\right|.
 \label{eq:loc_neg}
\end{eqnarray}

We now propose the following for an $N$-qubit gW state with real coefficients.

\propose{II} \emph{In the space $\left(E_{A_1A_2:B},\langle E_{A_1A_2}\rangle\right)$, the localizable entanglement $\langle E_{A_1A_2}\rangle$ of an $N$-qubit normalized gW state with real coefficients is upper bounded by the line 
\begin{eqnarray}
\langle E_{A_1A_2}\rangle = \frac{1}{2}\left(1+\sqrt{1-E_{A_1A_2:B}^2}\right),
\label{eq:gw_upper_bound}
\end{eqnarray}
where $E_{A_1A_2:B}$ is the bipartite entanglement over the bipartition $A_1A_2:B$ in the state prior to measurement on all the qubits in $B$.}



\begin{proof}
For ease of calculation, we focus on $\langle E_{A_1A_2}\rangle^2$ given by (using Eq.~(\ref{eq:loc_neg}) and normalization of the gW state)
\begin{eqnarray}
\langle E_{A_1A_2}\rangle^2=4\left(1-\sum_{i=1}^n a_i^2-\sum_{i=n+1}^{n+m} a_i^2\right)\sum_{i=n+1}^{n+m}a_i^2,
\label{eq:le_square_n}
\end{eqnarray} 
which, for fixed $a=\sum_{i=1}^{n}a_i^2$, is a single-parameter function $F(x)=4x(1-a-x)$ of $x$, where $x\equiv \sum_{i=n+1}^{n+m}a_i^2$. The function $F(x)$ has the maximum value $(1-a)^2$, occurring at $x=(1-a)/2$. This implies $\sum_{i=n+1}^{n+m}a_i^2=(1-a)/2$ for the maximum of $\langle E_{A_1A_2}\rangle=1-a$. Note also from Eq.~(\ref{eq:gW_negativity}) that $E_{A_1A_2:B}=2\sqrt{a(1-a)}$. Eliminating  $a$ and subsequently solving for $\langle E_{A_1A_2}\rangle$, one obtains Eq.~(\ref{eq:gw_upper_bound}). 
\end{proof}

The following Corollaries can be obtained straightforwardly from Proposition I. 

\corollary{II.1} \emph{The family of gW states with real coefficients that satisfy Eq.~(\ref{eq:gw_upper_bound}) are given by
\begin{eqnarray}
\sum_{i=n+1}^{n+m}a_i^2=\sum_{i=n+m+1}^N a_i^2=\frac{1}{2}\left(1-\sum_{i=1}^n a_i^2\right). 
\label{eq:gw_constraint}
\end{eqnarray}
}

\begin{proof}
For a fixed value of $\sum_{i=1}^n a_i^2$, the maximization condition for $\langle E_{A_1A_2}\rangle$ is given by $\sum_{i=n+1}^{n+m}a_i^2=\frac{1}{2}\left(1-\sum_{i=1}^n a_i^2\right)$. From the normalization of the gW states, Eq.~(\ref{eq:gw_constraint}) follows.  
\end{proof}

\corollary{II.2} \emph{For the family of gW states given by Eq.~(\ref{eq:gw_constraint}), $E_{A_1:A_2B}=E_{A_2:A_1B}$.}

\begin{proof}
Similar to Eq.~(\ref{eq:gW_negativity}), entanglement over the bipartitions  $A_2:A_1B$ and $A_1:A_2B$ can be written as 
\begin{eqnarray}
\label{eq:a1a2b}
E_{A_1:A_2B} &=& 2\left|\left[\left(\sum_{i=1}^{n}a_i^2+\sum_{i=n+m+1}^{N}a_i^2\right)\sum_{i=n+1}^{n+m} a_i^2\right]^{\frac{1}{2}}\right|,\nonumber\\ 
\end{eqnarray}
and 
\begin{eqnarray}
\label{eq:a2a1b}
E_{A_2:A_1B} &=& 2\left|\left[\left(\sum_{i=1}^{n}a_i^2+\sum_{i=n+1}^{n+m}a_i^2\right)\sum_{i=n+m+1}^N a_i^2\right]^{\frac{1}{2}}\right|,\nonumber\\ 
\end{eqnarray}
respectively. Clearly, for $\sum_{i=n+1}^{n+m}a_i^2=\sum_{i=n+m+1}^N a_i^2$, $E_{A_1:A_2B}=E_{A_2:A_1B}$.  
\end{proof}

\begin{figure*}
    \centering
    \includegraphics[width=0.7\textwidth]{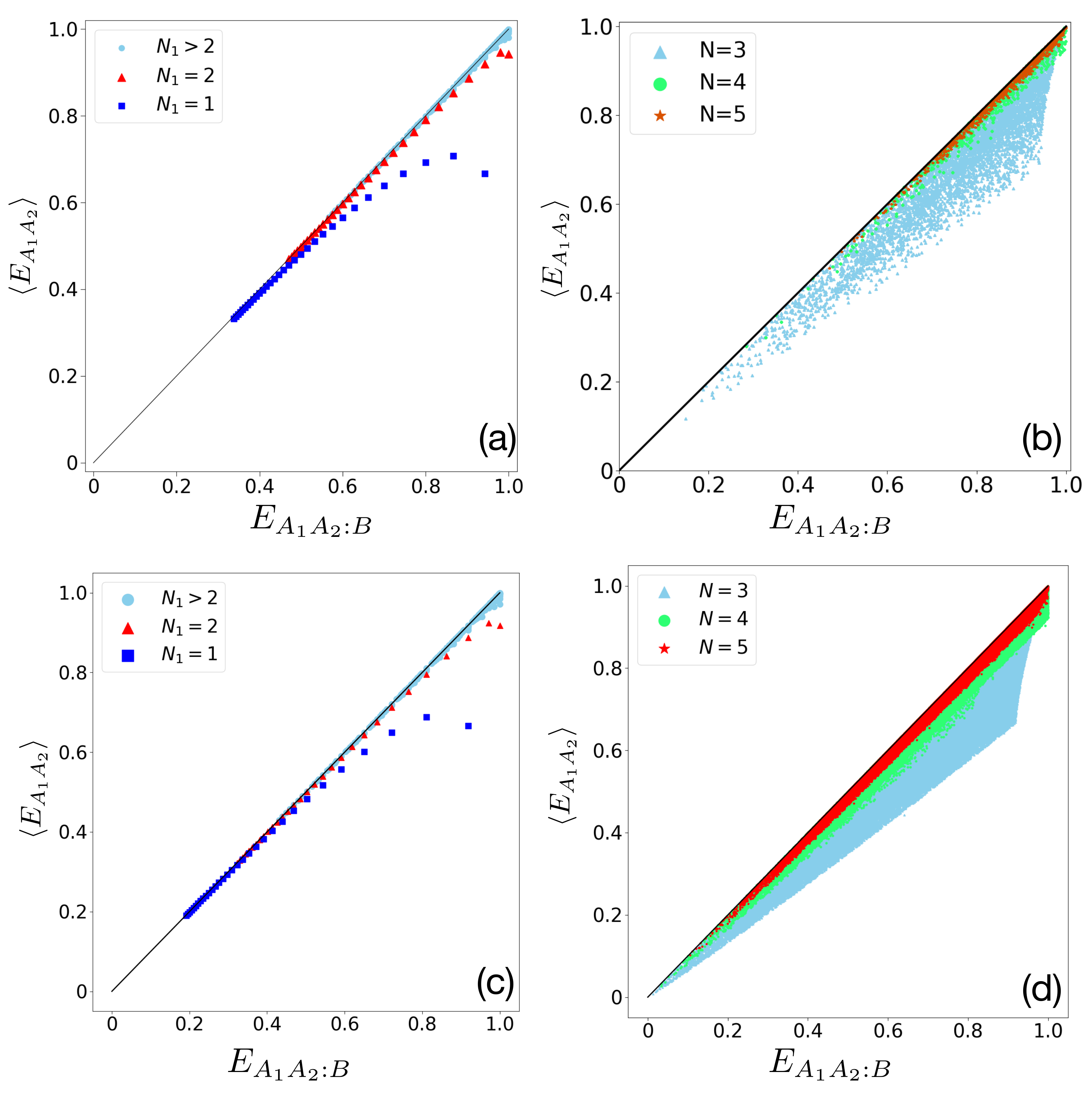}
    \caption{\textbf{Dicke states.} \JK{\textbf{(a), (c)} Scatter plots of $\ket{D(N,N_1)}$ for $N\leq 35$ on the $\left(E_{A_1A_2:B},\langle E_{A_1A_2}\rangle\right)$ plane, with all possible values of $N_1=1,\cdots,N-1$, using \textbf{(a)} negativity and \textbf{(c)} von Neumann entropy as entanglement measures. \textbf{(b), (d).} Scatter plot of Haar-uniformly generated gD states of $N=3$, $N=4$, and $N=5$ qubits on the $\left(E_{A_1A_2:B},\langle E_{A_1A_2}\rangle\right)$ plane, where each of the subsystems $B$ and $A_1$ is constituted of one qubit only. For each value of $N$, a sample of $10^5$ states are used. The chosen entanglement measures are \textbf{(b)} negativity and \textbf{(d)} von Neumann entropy. All quantities plotted are dimensionless, except the entanglement and localizable entanglement computed using von Neumann entropy as the entanglement measure, which are in ebits.}}
    \label{fig:fig3}
\end{figure*}

\propose{III} \emph{In $\left(\min\{E_{A_1:A_2B},E_{A_2:A_1B}\},\langle E_{A_1A_2}\rangle\right)$ space, the localizable entanglement $\langle E_{A_1A_2}\rangle$ of an $N$-qubit normalized gW state with real coefficients is upper-bounded by the line 
\begin{eqnarray}
 \langle E_{A_1A_2}\rangle = \min\{E_{A_1:A_2B},E_{A_2:A_1B}\},
 \label{eq:upper_bound_2}
\end{eqnarray}
and lower-bounded by the line 
\begin{eqnarray}
\langle E_{A_1A_2}\rangle^2-2\langle E_{A_1A_2}\rangle+\left(\min\{E_{A_1:A_2B},E_{A_2:A_1B}\}\right)^2=0,\nonumber\\
\label{eq:gw_lower_bound}
\end{eqnarray} 
where $E_{A_1:A_2B}$ $\left(E_{A_2:A_1B}\right)$ is the bipartite entanglement over the bipartition $A_1:A_2B$ $\left(A_2:A_1B\right)$ in the state prior to measurement on all the qubits in $B$.}

\begin{proof}
The upper bound follows from the monotonicity of $E$ (see Eq.~(\ref{eq:localizable_inequality})). On the other hand, note that from Eqs.~(\ref{eq:a1a2b}) and (\ref{eq:a2a1b}),  
\begin{eqnarray}
E_{A_1:A_2B}^2 &=& 4\left(\sum_{i=n+1}^{n+m} a_i^2\right)\sum_{i=1}^n a_i^2+\langle E_{A_1A_2}\rangle^2,
\label{eq:compare_2}
\end{eqnarray}
and 
\begin{eqnarray}
E_{A_2:A_1B}^2 &=& 4\left(\sum_{i=n+m+1}^N a_i^2\right)\sum_{i=1}^n a_i^2+\langle E_{A_1A_2}\rangle^2,
\label{eq:compare_1}
\end{eqnarray}
respectively, where we have used Eq.~(\ref{eq:loc_neg}). 
Let us proceed by assuming $E_{A_2:A_1B}\geq E_{A_1:A_2B}$, implying $\min\{E_{A_1:A_2B},E_{A_2:A_1B}\}=E_{A_1:A_2B}$, and  $\sum_{i=n+m+1}^N a_i^2\geq\sum_{i=n+1}^{n+m}a_i^2$. For a fixed $a=\sum_{i=n+1}^{n+m}a_i^2$, $\langle E_{A_1A_2}\rangle$ is minimum if $\sum_{i=n+m+1}^Na_i^2$ is minimum, leading to $\sum_{i=n+m+1}^Na_i^2=a$, and subsequently $\langle E_{A_1A_2}\rangle\geq 2a$ and $E_{A_1:A_2B}=2\sqrt{a(1-a)}$. Eliminating $a$, we obtain the equation of the lower bound as
\begin{eqnarray}
\langle E_{A_1A_2}\rangle^2-2\langle E_{A_1A_2}\rangle+E_{A_1:A_2B}^2=0.
\label{eq:lb_1}
\end{eqnarray}
Similarly, assuming $E_{A_1:A_2B}\geq E_{A_2:A_1B}$, one can also prove that 
\begin{eqnarray}
\langle E_{A_1A_2}\rangle^2-2\langle E_{A_1A_2}\rangle+E_{A_2:A_1B}^2=0.
\label{eq:lb_2}
\end{eqnarray}
Eqs.~(\ref{eq:lb_1}) and (\ref{eq:lb_2}) lead to Eq.~(\ref{eq:gw_lower_bound}).
\end{proof}

\note{1} Note here that Propositions II and III, and the related Corollaries are proved for the subclass of gW states with real coefficients, and it is therefore logical to ask whether the same results apply to the gW states with complex coefficients. \JK{While analytical calculation is difficult for a generic $N$-qubit gW state due to increase in the number of state parameters, for demonstration, we perform the calculation for the gW state with $N=3$, and find Propositions II and III and the Corollaries to be unchanged (see Appendix~\ref{app:three-gW}). We also numerically check the applicability of these results for $N$-qubit gW states with with arbitrary $N$ and complex coefficients, and find them to be valid.} Therefore, for generic gW states, Propositions II, III, and Corollaries II.1 and II.2 can be straightforwardly updated by replacing $a_i^2$ with $|a_i|^2$ for all $i=1,\cdots,N$. For demonstrations of these results, see Figs.~\ref{fig:fig2}(a)-(b). Specific examples of the family of states described in Eq.~(\ref{eq:gw_constraint}) can be found in Appendix~\ref{app:examples}.  

\note{2} Note in Fig.~\ref{fig:fig2}(a) that apart from the line given by Eq.~(\ref{eq:gw_upper_bound}), the Haar uniformly~\cite{bengtsson_zyczkowski_2006,banerjee2020} chosen gW states are also bounded by the lines (a) $E_{A_1A_2:B}=0$, (b) $\langle E_{A_1A_2}\rangle=0$, and (c) $E_{A_1A_2:B}=1$, which are not shown explicitly in the figure, and which are obtained from the fact that $0\leq E_{A_1A_2:B},\langle E_{A_1A_2:B}\rangle\leq 1$. It is worthwhile to note that the family of gW states, for which $E_{A_1A_2:B}=1$, are given by  
\begin{eqnarray}
 4\sum_{i=1}^n a_i^2 = 1/\sum_{i=n+1}^N a_i^2.
 \label{eq:unit_entanglement}
\end{eqnarray}
The significance of these states will be clear in Sec.~\ref{subsec:noisy_gw}. For examples of such states, see Appendix~\ref{app:examples}. 

\JK{\note{3} We point out here that one can also use other entanglement measures, such as the logarithmic negativity, or the von Neumann entropy, to demonstrate the above results. With a change in the entanglement measure, the family of gW states providing the bounds (see Eqs.~(\ref{eq:gw_constraint}) and (\ref{eq:unit_entanglement})) remain unchanged, although the functional form of the dependence of $\langle E_{A_1A_2}\rangle$ on $E_{A_1A_2:B}$ (see Eq.~(\ref{eq:gw_upper_bound})) and $\min\{E_{A_1:A_2B},E_{A_2:A_1B}\}$ (see Eq.~(\ref{eq:gw_lower_bound})) corresponding to the bounds change. In Figs.~\ref{fig:fig2}(c)-(d), we have demonstrated this pictorially using negativity, logarithmic negativity, and von Neumann entropy as the entanglement measures, although we refrain from writing the equivalent equations corresponding to Eqs.~(\ref{eq:gw_upper_bound}) and (\ref{eq:gw_lower_bound}) for these measures to keep the text uncluttered. Note, however, that the bound (\ref{eq:localizable_inequality}) remains unaltered with a change in entanglement measures.}

\subsection{Dicke states}
\label{subsec:dicke}

We now consider the class of \emph{symmetric} states that remain invariant under permutation of parties. More specifically, we focus on the Dicke states~\cite{dicke1954,bergmann2013,lucke2014,kumar2017} of $N$-qubits, where $N_0$ qubits are in the ground state $\ket{0}$, and the rest $N_1=N-N_0$ qubits are in the excited state $\ket{1}$.  A Dicke state with $N_1$ excited qubits can be written as 
\begin{eqnarray}
 \ket{D(N,N_1)} = \frac{1}{\sqrt{\genfrac{()}{}{0pt}{}{N}{N_1} }}\sum_{i}\mathcal{P}_i\left(\ket{0}^{\otimes N-N_1}\ket{1}^{\otimes N_1}\right),  
\end{eqnarray} 
where for a fixed $N$, $N_1=0,1,2,\cdots,N$. Here, $\{\mathcal{P}_i\}$ is the set of all possible permutations of $N_0$ ($N_1$) qubits at ground (excited) state, such that $N_0+N_1=N$. Note that $\ket{D(N,0)}$ and $\ket{D(N,N)}$ are product states, while $\ket{D_1}$ and $\ket{D_{N-1}}$ are identical to $N$-qubit W states, or its local unitary equivalents, the results for which can be determined as special cases of the gW states discussed in Sec.~\ref{subsec:gw_states}. Note also that in the case of $n=m<N/2$ $(n=m<(N-1)/2)$ where $N$ is even (odd), with $m$ being the size of the subsystem $A_1$, symmetry of the Dicke states under qubit permutations suggests that $E_{A_1:A_2B}=E_{A_1A_2:B}$. We focus on the scenario where $n=1$, and without any loss in generality, measure on qubit $1$. Negativity for Dicke states are computable over 1:rest bipartition as ~\cite{chen2020} (see also~\cite{sanpera1998,rana2013})
\begin{eqnarray}
 E_{A_1A_2:B} = \underset{i,j,i\neq j}{\max}\frac{1}{\genfrac{()}{}{0pt}{}{N}{N_1}}\sqrt{\genfrac{()}{}{0pt}{}{N-1}{N_1-i}\genfrac{()}{}{0pt}{}{N-1}{N_1-j}},
\end{eqnarray}
where $B$ is constituted of one (measured) qubit, and $i,j=0,1$. \JK{On the other hand, using von Neumann entropy as the entanglement measure~\cite{Moreno2018}, 
\begin{eqnarray}
\label{dickentr}
 E_{A_1A_2:B} =
-\frac{N-N_{1}}{N}\log\frac{N-N_{1}}{N}-\frac{N_{1}}{N}\log\frac{N_{1}}{N}.
\end{eqnarray}}

In the case of $\ket{D(N,N_1)}$, the normalized post-measured states are given by 
\begin{eqnarray}
 \ket{D} &=& \frac{1}{\sqrt{p}}\Big[\sqrt{\frac{N-N_1}{N}}\cos\frac{\theta}{2}D(N-1,N_1)\nonumber\\
 &&+\sqrt{\frac{N_1}{N}}\text{e}^{-\text{i}\phi}\sin\frac{\theta}{2}D(N-1,N_1-1)\Big],\\
 \ket{D_\perp} &=& \frac{1}{\sqrt{p_\perp}}\Big[\sqrt{\frac{N-N_1}{N}}\sin\frac{\theta}{2}D(N-1,N_1)\nonumber\\
 &&-\sqrt{\frac{N_1}{N}}\text{e}^{-\text{i}\phi}\cos\frac{\theta}{2}D(N-1,N_1-1)\Big], 
\end{eqnarray}
with 
\begin{eqnarray}
 p &=& \cos^2\frac{\theta}{2}-\frac{N_1}{N}\cos\theta,\\ 
 p_\perp &=& \sin^2\frac{\theta}{2}+\frac{N_1}{N}\cos\theta.
\end{eqnarray}
Determination of the general form of any entanglement measure over a bipartition of states of the form $\ket{D},\ket{D_\perp}$, and the subsequent analytical optimization is a difficult task. However, our numerical analysis suggests that the optimization of localizable negativity, in the current situation, always takes place in the $\sigma^z$ basis. Using this information, $\ket{D},\ket{D_\perp}$ and $p,p_\perp$ become \begin{eqnarray} 
 \ket{D} &=& \ket{D(N-1,N_1)},\nonumber\\
 \ket{D_\perp}&=&\ket{D(N-1,N_1-1)},
\end{eqnarray}
and 
\begin{eqnarray}
 p=(N-N_1)/N;\;p_\perp=N_1/N,  
\end{eqnarray}
respectively,  resulting in\small 
\begin{eqnarray}
 \langle E_{A_1A_2}\rangle &=& \underset{i,j,i\neq j}{\max}\frac{N-N_1}{N\genfrac{()}{}{0pt}{}{N-1}{N_1}}\sqrt{\genfrac{()}{}{0pt}{}{N-2}{N_1-i}\genfrac{()}{}{0pt}{}{N-2}{N_1-j}}\nonumber\\
 && + \underset{i,j,i\neq j}{\max}\frac{N_1}{N\genfrac{()}{}{0pt}{}{N-1}{N_1-1}}\sqrt{\genfrac{()}{}{0pt}{}{N-2}{N_1-1-i}\genfrac{()}{}{0pt}{}{N-2}{N_1-1-j}}.\nonumber\\ 
\end{eqnarray}\normalsize
when negativity is used as entanglement measure. \JK{Using von Neumann entropy instead, one obtains
\begin{eqnarray}
\langle E_{A_{1}A_{2}}\rangle &=& -M_0[(1-M_1)\log_2\left(1-M_1\right)\nonumber\\ &&+M_1\log_2 M_1]
-(1-M_0)[M_2\log_2 M_2\nonumber\\ &&+(1-M_2)\log_2(1-M_2)], 
\end{eqnarray}\normalsize
with $M_0=(N-N_1)/N$, $M_1=N_1/(N-1)$ and $M_2=(N-N_1)/(N-1)$. It is easy to numerically check that $E_{A_1A_2:B} \geq \langle E_{A_1A_2}\rangle$, which approaches equality as $N$ increases in the case of both negativity and von Neumann entropy. Note also that the variation of $\langle E_{A_1A_2}\rangle$ with $E_{A_1A_2:B}$ is non-monotonic for lower values of $N_1$, resulting in a larger difference between $\langle E_{A_1A_2}\rangle$ and $E_{A_1A_2:B}$ at higher $N$. However, these features disappear as $N_1$ increases. See Figs.~\ref{fig:fig3}(a) and (c).}

\paragraph{Generalized Dicke states.} Using the Dicke states, one can define an $N$-qubit permutation-symmetric state in the form of a \emph{generalized} Dicke (gD) state~\cite{sadhukhan2017}, as 
\begin{eqnarray}
\ket{D(N)} &=& \sum_{N_1=0}^N a_{N_1}\ket{D(N,N_1)},
\end{eqnarray}
where $a_{N_1}\in\mathbb{C}$, and $\sum_{N_1=0}^N |a_{N_1}|^2=1$. Due to a large number of state parameters, analytical calculation is difficult for gD states. However, as in the case of the Dicke states,  the permutation symmetry can be used here also to have $E_{A_1:A_2B}=E_{A_1A_2:B}$ in the case of $n=m<N/2$ $(n=m<(N-1)/2)$ for even (odd) $N$. This implies that it is sufficient to look into the relation between $\langle E_{A_1A_2}\rangle$ and $\min\{E_{A_1:A_2B},E_{A_2:A_1B}\}$, which is given by~(\ref{eq:localizable_inequality}).  Moreover, our numerical results suggest that $\min\{E_{A_1:A_2B},E_{A_2:A_1B}\}=E_{A_1:A_2B}$ for all Haar uniformly generated gD states, which leads to the upper bound of $\langle E_{A_1A_2}\rangle$ as $\langle E_{A_1A_2}\rangle \leq E_{A_1A_2:B}$ \JK{(see Figs.~\ref{fig:fig3}(b) and (d))}. It is clear from the scatter diagrams that as $N$ increases, the difference between $E_{A_1A_2:B}$ and $\langle E_{A_1A_2}\rangle$ decreases, and considerably larger fraction of states are found to obey $\langle E_{A_1A_2}\rangle=E_{A_1A_2:B}$ in the situation where each of the subsystems $B$ and $A_1$ holds only one qubit.         

\subsection{Arbitrary multi-qubit pure states}
\label{subsec:pure_arbit}

\begin{figure*}
    \centering
    \includegraphics[width=\textwidth]{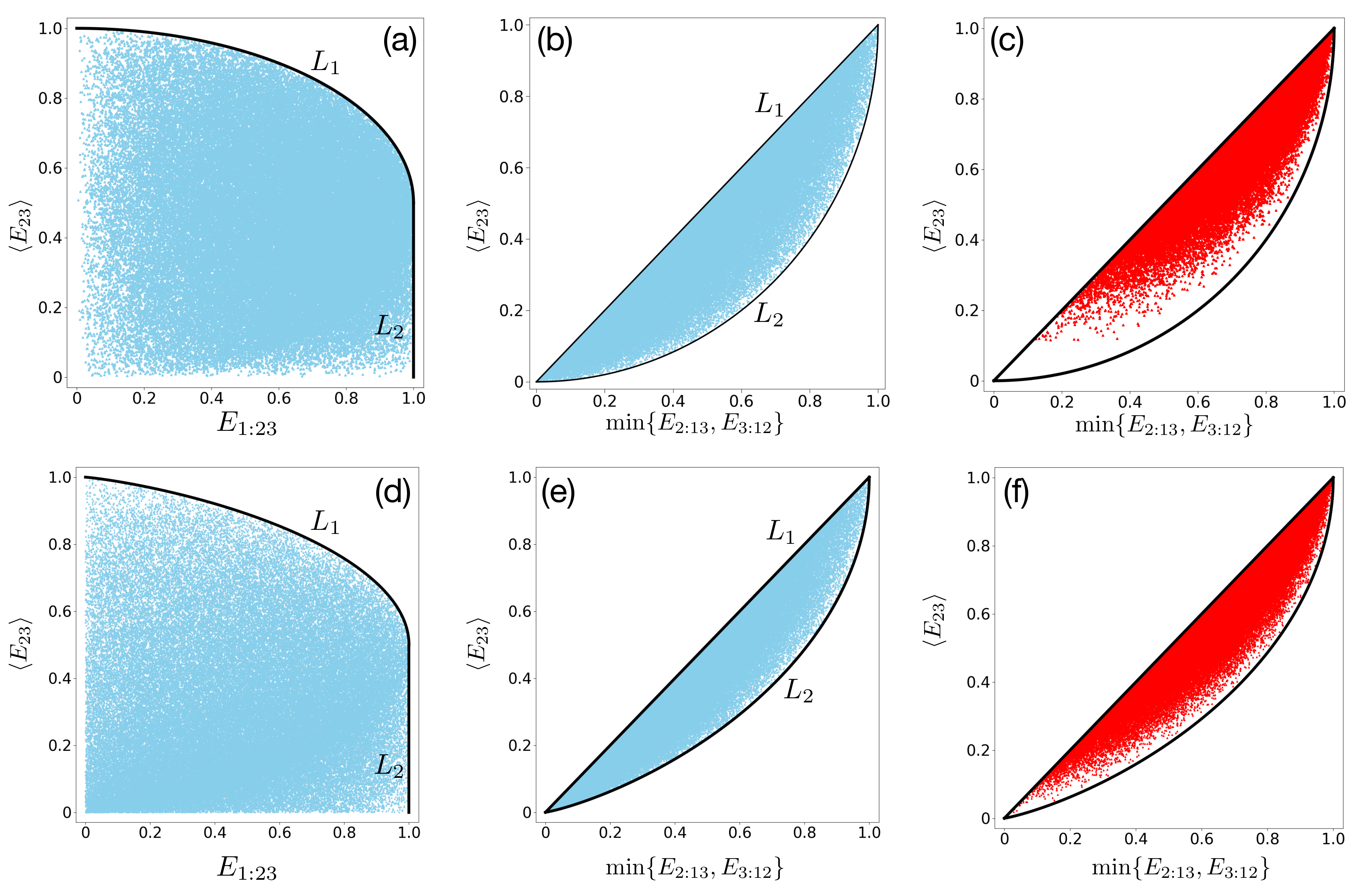}
    \caption{\textbf{GHZ and W-class states.} \JK{Scatter plot of a sample of $10^7$ Haar-uniformly generated three-qubit W-class states on the \textbf{(a)} $\left(E_{1:23},\langle E_{23}\rangle\right)$ and the \textbf{B.} $\left(\min\{E_{2:13},E_{3:12}\},\langle E_{23}\rangle\right)$ plane. The lines $L1$ and $L2$ bounding the states correspond to Eq.~(\ref{eq:w_upper_bound}) and $E_{1:23}=1$ respectively. \textbf{(c)} A sample of $10^7$ Haar-uniformly generated GHZ-class states are also found to be bound by the same upper and lower bounds as proposed in Proposition III. In \textbf{(a)-(c)}, negativity is used as entanglement measure, while the corresponding figures for von Neumann entropy are given in \textbf{(d)-(f)}. All quantities plotted are dimensionless, except the entanglement and localizable entanglement computed using von Neumann entropy as the entanglement measure, which are in ebits.}}
    \label{fig:fig4}
\end{figure*}

In this section, we investigate the question posed in Sec.~\ref{sec:intro} for arbitrary states of systems with arbitrary number of qubits. 

\paragraph{Three-qubit systems.} While analytical calculation of the relevant quantities is difficult for large number of qubits, in relation to Proposition II, some analytical results can be derived for the three-qubit systems $(N=3)$ with $n=m=1$. Note that the three-qubit gW states are a subset of the three-qubit W class states, which, together with the three-qubit GHZ class states, form the complete set of three-qubit pure states~\cite{dur2000}.
We explore whether the bound in Proposition II also applies to the three-qubit W class states given by~\cite{dur2000} 
\begin{eqnarray}
\ket{\psi_{\text{W}}} &=& a_0\ket{000}+a_1\ket{100}+a_2\ket{010}+a_3\ket{001},
\label{eq:wclass}
\end{eqnarray}
where $\sum_{i=0}^3|a_i|^2=1$ and $a_i\in\mathbb{C}$, $i=0,1,2,3$. For ease of discussion, we consider the qubits $1$, $2$, and $3$ to be subsystems $B$, $A_1$, and $A_2$, respectively. Also, we first consider the subclass of the W-class states with real coefficients only, i.e., $a_i\in\mathbb{R}$, $i=0,1,2,3$, and numerically verify that the negativity over different bipartitions of the three-qubit W-class states are given by 
\begin{eqnarray}
 \label{eq:w_negativity_1}
 E_{1:23} &=& 2 \left|a_{1}\sqrt{a_{2}^{2} + a_{3}^{2}}\right|,\\
  \label{eq:w_negativity_2}
 E_{2:13} &=& 2 \left|a_{2}\sqrt{a_{1}^{2} + a_{3}^{2}}\right|,\\
 E_{3:12} &=& 2 \left|a_{3}\sqrt{a_{1}^{2} + a_{2}^{2}}\right|.
 \label{eq:w_negativity_3}
\end{eqnarray}
Further, performing local projection measurements in the basis $\{\ket{b_0},\ket{b_1}\}$ on the qubit $1$, one obtains the post-measured states on qubits $2$ and $3$ to be in the same form as in Eq.~(\ref{eq:post_measured_state}) with $f_0^k$ and $f^k$ $(k=0,1)$ given in Appendix~\ref{app:two_measurement}. Similar to the three-qubit gW states, negativity of the post-measured states are independent of the measurement-basis, and are given by $2|a_2||a_3|$. This leads to 
\begin{eqnarray}
\langle E_{23}\rangle &=& 2|a_2||a_3|.
\end{eqnarray}
We are now in a position to present the following proposition for the W class states of three qubits. 

\propose{IV} \emph{In the space $\left(E_{1:23},\langle E_{23}\rangle\right)$, the localizable entanglement $\langle E_{23}\rangle$ of a $3$-qubit normalized W-class state with real coefficients is upper bounded by the line 
\begin{eqnarray}
\langle E_{23}\rangle = \frac{1}{2}\left(1+\sqrt{1-E_{1:23}^2}\right),
\label{eq:w_upper_bound}
\end{eqnarray}
where $E_{1:23}$ is the bipartite entanglement over the bipartition $1:23$ in the state prior to measurement on the qubit $1$.}

\begin{proof}
Similar to the case of the gW states, we can write $\langle E_{23}\rangle^2$ as
\begin{eqnarray}
\langle E_{23}\rangle^2 &=& 4a_2^2(1-a_0^2-a_1^2-a_2^2). 
\end{eqnarray}
For a fixed value of $a_0^2+a_1^2=a^2$, the maximum of $\langle E_{23}\rangle^2$, and therefore of $\langle E_{23}\rangle$ occurs at $a_2^2=(1-a^2)/2$, the maximum value of $\langle E_{23}\rangle$ being  $(1-a^2)$. Also, $E_{1:23}=2\left|a_1\sqrt{1-a^2}\right|$. Note that the maximum value of $\langle E_{23}\rangle$ as well as $E_{1:23}$ have two free parameters, $a_0$ and $a_1$, constrained by $a_0^2+a_1^2$ being a constant, $a^2$. Eliminating $a_1$ from $\langle E_{23}\rangle$ and $E_{1:23}$, followed by solving for $\langle E_{23}\rangle$ leads to 
\begin{eqnarray}
 \langle E_{23}\rangle &=& \frac{1}{2}\left[(1-a_0^2)+\sqrt{(1-a_0^2)^2-E_{1:23}^2}\right],
 \label{eq:three_qubit_w_class_form_1}
\end{eqnarray}
where $0\leq a_0^2\leq a^2$. Further maximization w.r.t. $a_0$ implies $a_0=0$, leading to Eq.~(\ref{eq:w_upper_bound}).  
\end{proof} 

Similar to the three-qubit gW states, the following Corollaries originate from Proposition IV.

\corollary{IV.1} \emph{The family of W class states with real coefficients that satisfy Eq.~(\ref{eq:w_upper_bound}) are given by
\begin{eqnarray}
a_2^2=a_3^2=(1-a_1^2)/2.
\label{eq:w_constraint}
\end{eqnarray}
}

\begin{proof}
The proof of this Corollary follows from the maximization condition of $\langle E_{23}\rangle$, and the normalization of the W-class state. 
\end{proof}

\corollary{IV.2} \emph{For the family of W-class states given by Eq.~(\ref{eq:w_constraint}), $E_{2:13}=E_{3:12}$.}

\begin{proof}
The proof of this corollary follows from identifying the states satisfying Eq.~(\ref{eq:w_constraint}) as the three-qubit gW states, and from the proof of Corollary II.2. 
\end{proof}

\propose{V} \emph{In $\left(\min\{E_{2:13},E_{3:12}\},\langle E_{23}\rangle\right)$ space, the localizable entanglement $\langle E_{23}\rangle$ of a three-qubit normalized W-class state with real coefficients is upper-bounded by the line 
\begin{eqnarray}
 \langle E_{23}\rangle = \min\{E_{2:13},E_{3:12}\},
 \label{eq:w_upper_bound_2}
\end{eqnarray}
and lower-bounded by the line 
\begin{eqnarray}
\langle E_{23}\rangle^2-2\langle E_{23}\rangle+\left(\min\{E_{2:13},E_{3:12}\}\right)^2=0,
\label{eq:w_lower_bound}
\end{eqnarray} 
where $E_{2:13}$ $\left(E_{3:12}\right)$ is the bipartite entanglement over the bipartition $2:13$ $\left(3:12\right)$ in the state prior to measurement on the qubit $1$.}

\begin{proof}
Similar to the proof of Proposition III, the upper bound follows from the monotonicity of $E$ (see Eq.~(\ref{eq:localizable_inequality})). To prove the lower bound, we start by assuming $E_{2:13}\geq E_{3:12}$ and $\min\{E_{2:13},E_{3:12}\}=E_{3:12}$, which, by virtue of Eqs.~(\ref{eq:w_negativity_2}) and (\ref{eq:w_negativity_3}), implies $a_2^2\geq a_3^2$. For a fixed $a_3$ and $a_0$, $\langle E_{23}\rangle$ is minimum if $a_2$ is minimum, leading to $a_2=a_3$, and subsequently $\langle E_{23}\rangle\geq 2a_3^2$. On the other hand, exploiting normalization of the W-class state,  $E_{3:12}=2\left|a_3\sqrt{1-a_0^2-a_3^2}\right|$. Eliminating $a_3$ from $E_{3:12}$ and the minimum of $\langle E_{23}\rangle$, we obtain
\begin{eqnarray}
 \langle E_{23}\rangle^2+E_{3:12}^2-2\langle E_{23}\rangle(1-a_0^2)=0. 
\end{eqnarray}
Similar to Proposition V, it can be shown that $\langle E_{23}\rangle$ attains a minimum for $a_0=0$, leading to Eq.~(\ref{eq:w_lower_bound}).
\end{proof}

\note{4} Similar to the case of $N$-qubit gW states, in this case also, we numerically verify that the Propositions IV and V remain valid in the case of generic three-qubit states from W class with complex coefficients. The bounds in the case of three-qubit W class states are demonstrated in Fig.~\ref{fig:fig4}(a)-(b).

\note{5} We also investigate the three-qubit GHZ class states, given by 
\begin{eqnarray}
\ket{\psi_{\text{GHZ}}} &=& \sum_{i=0}^7c_i\ket{\phi_i},
\end{eqnarray}
with $c_i\in\mathbb{C}$ and $\{\ket{\phi_i};i=0,1,\cdots,7\}$ being the standard product basis in the Hilbert space of three qubits. While analytical investigation is difficult due to a large number of parameters involved in these states, our numerical analysis  does not provide any evidence of the  existence of an upper bound of $\langle E_{23}\rangle$ on the $\left(E_{1:23},\langle E_{23}\rangle\right)$ plane. However, our investigation involving a sample of $10^7$ Haar-uniformly generated GHZ-class states did not find any example that violates the lower bound proposed in Proposition V (see Fig.~\ref{fig:fig4}(c) for a demonstration). While this does not analytically prove the validity of the lower bound, this implies that Proposition IV \emph{ has the potential to distinguish between the three-qubit W class states from the GHZ class states}.

\paragraph{Larger systems.} In order to study the correlation between localizable entanglement  $\langle E_{A_1A_2}\rangle$ and the bipartite entanglement lost due to the measurement, namely, $E_{A_1A_2:B}$, $E_{A_1:A_2B}$, and $E_{A_2:A_1B}$ in the case of pure states on $N$-qubits, we look into how the states are distributed on the $\left(E_{A_1A_2:B},\langle E_{A_1A_2}\rangle\right)$ and the $\left(\min\left[E_{A_1:A_2B},E_{A_2:A_1B}\right],\langle E_{A_1A_2}\rangle\right)$ spaces.  To investigate this, we define the followings.
\begin{eqnarray}
\label{eq:delta1}
 \delta_1 &=& \langle E_{A_1A_2}\rangle - E_{AB},\\ 
 \delta_2 &=& \langle E_{A_1A_2}\rangle - \min\{E_{A_1B:A_2},E_{A_2B:A_1}\}.
 \label{eq:delta2}
\end{eqnarray}
Note that $\delta_1\geq 0$ $\left(\delta_2\geq 0\right)$ implies $\langle E_{A_1A_2}\rangle \geq E_{A_1A_2:B}$ $\left(E_{A_1A_2}\rangle \geq \min\{E_{A_1B:A_2},E_{A_2B:A_1}\}\right)$, which is representative of a situation where one can, on average, localize at least $E_{A_1A_2:B}$ $\left(\min\{E_{A_2:A_1B},E_{A_1:A_2B}\}\right)$ amount of entanglement via local projection measurements on the qubits in $B$. The percentages of $N$-qubit states for which $\delta_1> 0$ are included in Table~\ref{tab:percentage_table} for different combinations of $N$ and $m$, keeping $n=1$, where in each case, a sample of $10^5$ Haar-uniformly generated pure states are considered. Clearly, the percentage of states for which $\delta_1>0$ increases overall with the increase in the number of qubits, implying that the number of states for which $\langle E_{A_1A_2}\rangle>E_{A_1A_2:B}$ are more for larger systems. However, the percentage of states for which $\delta_1=0$, up to our numerical accuracy, is negligibly small for all cases of $(N,n,m)$.   On the other hand, as expected, $\delta_2>0$ does not occur for any multi-qubit pure states, as it would imply a violation of~(\ref{eq:localizable_inequality}). Moreover, for a very small fraction of states, $\delta_2=0$. This fraction overall increases with an increase in the number of qubits in the system. 

\JK{\note{6} We point out here that the analysis presented in Sec.~\ref{subsec:pure_arbit} can also be carried out using entanglement measures other than negativity, where no qualitative results are changed except Eqs.~(\ref{eq:three_qubit_w_class_form_1}) and (\ref{eq:w_lower_bound}). While we refrain from writing the corresponding equivalent equations, we pictorially demonstrate this in Fig.~\ref{fig:fig4}(d)-(f) using von Neumann entropy. Similar features are also found for logarithmic negativity. Note also that the data included in Table~\ref{tab:percentage_table} are specific to the choice of negativity as an entanglement measure.}

\begin{figure*}
    \centering
    \includegraphics[width=0.7\textwidth]{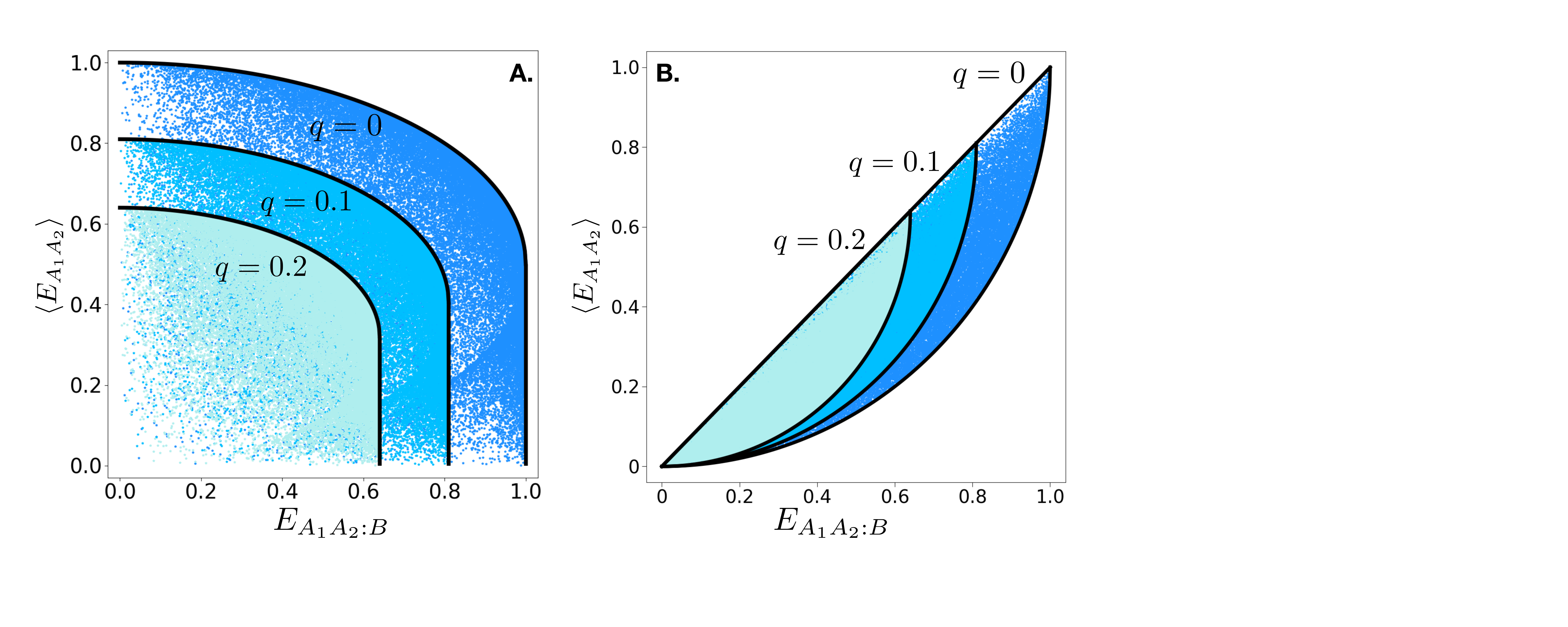}
    \caption{\textbf{Generalized W states under Markovian phase-flip channels.} Modification of boundaries of gW states proposed in Propositions II and III on the \textbf{(a)} $\left(E_{A_1A_2:B},\langle E_{A_1A_2}\rangle\right)$ and the \textbf{(b)} $\left(\min\{E_{A_1:A_2B},E_{A_2:A_1B}\},\langle E_{A_1A_2}\rangle\right)$ plane due to Markovian phase-flip noise of noise strength $q=0,0.1,0.2$ (see also Fig.~\ref{fig:fig2}), using negativity as the entanglement measure. The modified boundaries are given in Propositions VI and VII. The trivial boundaries corresponding to zero entanglement lines are not shown. All quantities plotted in all figures are dimensionless.}
    \label{fig:fig5}
\end{figure*}

\section{Quantum states under phase-flip noise}
\label{sec:noise}

It is now logical to ask whether and how the results reported in Sec.~\ref{sec:pure_states} are modified if the multi-qubit system is subjected to noise~\cite{nielsen2010}. In this paper, we consider a situation where qubits of the multi-qubit system are sent through independent phase-flip channels~\cite{nielsen2010,holevo2012}. These phase-flip channels can be either Markovian~\cite{yu2009}, or non-Markovian~\cite{Daffer2004,Shrikant2018,Gupta2020}. Using the Kraus operator representation, the evolution of the quantum state $\rho_0=\ket{\psi}\bra{\psi}$ under these phase-flip channels are given by 
\begin{eqnarray}
 \rho = \sum_{\alpha}K_\alpha \rho_0 K_\alpha^\dagger,
\end{eqnarray}
such that $\sum_{\alpha}K_\alpha^\dagger K_\alpha = I$, $I$ being the identity operator in the Hilbert space of the multi-qubit system, and Kraus operators $\{K_\alpha\}$, takes the form $K_\alpha=\sqrt{p_\alpha}K^\prime_\alpha$ with
\begin{eqnarray}
 K^\prime_\alpha =\bigotimes_{i=1}^N K^\prime_{\alpha_i};\;p_\alpha = \prod_{i=1}^N p_{\alpha_i}. 
\end{eqnarray}
Here, the index $\alpha\equiv \alpha_1\alpha_2\cdots\alpha_N$ is interpreted as a multi-index, $\sum_{\alpha_i}p_{\alpha_i}=1$, and $K^\prime_{\alpha_i}$ are the single-qubit Kraus operators, the form of which depends on the type of noise under consideration. In the case of the phase-flip noise, $\alpha_i\in\{0,1\}$, $K^\prime_{\alpha_i=0} = I_i$, and $K^\prime_{\alpha_i=1} = \sigma^z_i$, with
\begin{eqnarray}
p_{\alpha_i=0} &=& 1-\frac{q}{2},\;p_{\alpha_i=1} = \frac{q}{2}
\end{eqnarray}
in the Markovian case~\cite{yu2009}, and 
\begin{eqnarray}
p_{\alpha_i=0} &=& \left(1-\frac{q}{2}\right)\left(1-\frac{\alpha q}{2}\right),\nonumber\\
p_{\alpha_i=1} &=& \left[1+\alpha\left(1-\frac{q}{2}\right)\right]\frac{q}{2}
\end{eqnarray}
in the non-Markovian case~\cite{Daffer2004,Shrikant2018,Gupta2020}. Here,  $q$ is the \emph{noise strength} ($0\leq q\leq 1$), and $\alpha$ is the non-Markovianity parameter ($0\leq \alpha\leq 1$). For ease of discussion, from now onward, we denote entanglement in the noiseless scenario with a superscript ``$0$". For example, $\langle E_{A_1A_2}\rangle^0$ and $E_{AB}^0$ denote the localizable entanglement over the subsystem $A$ and the bipartite entanglement over the bipartition $A:B$ of the system in the case of $\rho_0$. \JK{Also, in this section, we use negativity as an entanglement measure to demonstrate all our results.}  

\subsection{Generalized GHZ states}
\label{subsec:noisy_gghz}

We start our discussions with the $N$-qubit gGHZ state subjected to phase-flip noise on all qubits. In this situation, we prove the following proposition. 

\propose{VI} \emph{For any tripartition $A_1:A_2:B$ of an $N$-qubit gGHZ state under uncorrelated phase-flip channel on all qubits,
\begin{eqnarray}
\langle E_{A_1A_2}\rangle=E_{A_1A_2:B}=E_{A_1:A_2B}=E_{A_2:A_1B},
\label{eq:pf_gghz_bound}
\end{eqnarray}
irrespective of whether the noise is Markovian, or non-Markovian.}

\begin{proof}
The $N$-qubit gGHZ state, under the Markovian phase-flip noise on all qubits, takes the form
\begin{eqnarray}
 \rho&=&\left(|a_0|^2(\ket{0}\bra{0})^{\otimes N}+|a_1|^2(\ket{1}\bra{1})^{\otimes N}\right)\nonumber \\ 
&&+ (1-q)^N\left(a_0a_1^* (\ket{0}\bra{1})^{\otimes N}+ a_0^*a_1 (\ket{1}\bra{0})^{\otimes N}\right).
\end{eqnarray}
Partial transposition of $\rho$ with  respect to the subsystems $B$ leads to
\begin{eqnarray}
\rho^{T_B}&=&\left(|a_0|^2(\ket{0}\bra{0})^{\otimes N}+|a_1|^2(\ket{1}\bra{1})^{\otimes N}\right)\nonumber \\ 
&&+ (1-q)^Na_0a_1^* (\ket{0}\bra{1})^{\otimes N-n}(\ket{1}\bra{0})^{\otimes n} \nonumber \\
&&+ (1-q)^Na_0^*a_1 (\ket{1}\bra{0})^{\otimes N-n}(\ket{0}\bra{1})^{\otimes n},
\end{eqnarray}
with non-zero eigenvalues $ |a_0|^2,|a_1|^2,\pm (1-q)^N |a_0||a_1|$. Therefore the entanglement between partition $A$ and partition $B$, as quantified by negativity~\cite{peres1996,horodecki1996,vidal2002,zyczkowski1998,lee2000}, is given by 
\begin{eqnarray}
E_{A_1A_2:B} &=& 2(1-q)^N|a_0|\sqrt{1-|a_0|^2}\nonumber\\ &=& (1-q)^N E_{A_1A_2:B}^0. 
\label{eq:gghz_neg_noise}
\end{eqnarray}

To calculate the localizable entanglement over the subsystem $A$ with bipartition $A_1:A_2$, we proceed as in the case of Proposition I, and write the post-measured states on $A$ as 
\begin{eqnarray}
 \tilde{\rho}_A^k=\text{Tr}_B\left[\left(M^k\rho M^{k\dagger}\right)/{p_k}\right],
\end{eqnarray}
which, written explicitly, takes the form
\begin{eqnarray}
 \tilde{\rho}^k_A &=& \frac{1}{p_k}\Big[\Big\{|a_0|^2 |f_{0}^{k}|^2 (\ket{0}\bra{0})^{\otimes N-n} \nonumber\\ &&+ |a_1|^2 |f_{1}^{k}|^2 (\ket{1}\bra{1})^{\otimes N-n}\Big\} \nonumber\\ &&+ (1-q)^N \big\{a_0 a_1^* f_{0}^{k} f_{1}^{k*} (\ket{0}\bra{1})^{\otimes N-n}\nonumber\\ &&+a_0^* a_1 f_{1}^{k} f_{0}^{k*} (\ket{1}\bra{0})^{\otimes N-n})\big\}\Big],
\end{eqnarray}
with
\begin{eqnarray}
 p_k &=& \left(|a_0|^2|f^k_0|^2+|a_1|^2|f^k_1|^2\right).
\end{eqnarray}
Partial transposition of  $\tilde{\rho}^k_A$ over any bipartition $A_1:A_2$, and subsequent calculation of negativity followed by the optimization of average negativity over $A_1:A_2$ yields
\begin{eqnarray}
 \langle E_{A_1:A_2}\rangle = 2(1-q)^N| a_0|\sqrt{1-|a_0|^2}\left[\max\sum_{k=0}^{2^n-1}|f^k_0||f^k_1|\right],\nonumber\\
\end{eqnarray}
as in Eq.~(\ref{eq:final_expression_gghz}). The maximization is similar to that shown in the proof of Proposition I, leading to 
\begin{eqnarray}
 \langle E_{A_1:A_2}\rangle&=&(1-q)^N\langle E_{A_1A_2} \rangle^0 = E_{A_1A_2:B}.
\end{eqnarray}
Also, from the symmetry of $\rho$, $E_{A_1A_2:B}=E_{A_1:A_2B}=E_{A_2:A_1B}$, leading to Eq.~(\ref{eq:pf_gghz_bound}).

\begin{figure*}
    \centering
    \includegraphics[width=0.8\textwidth]{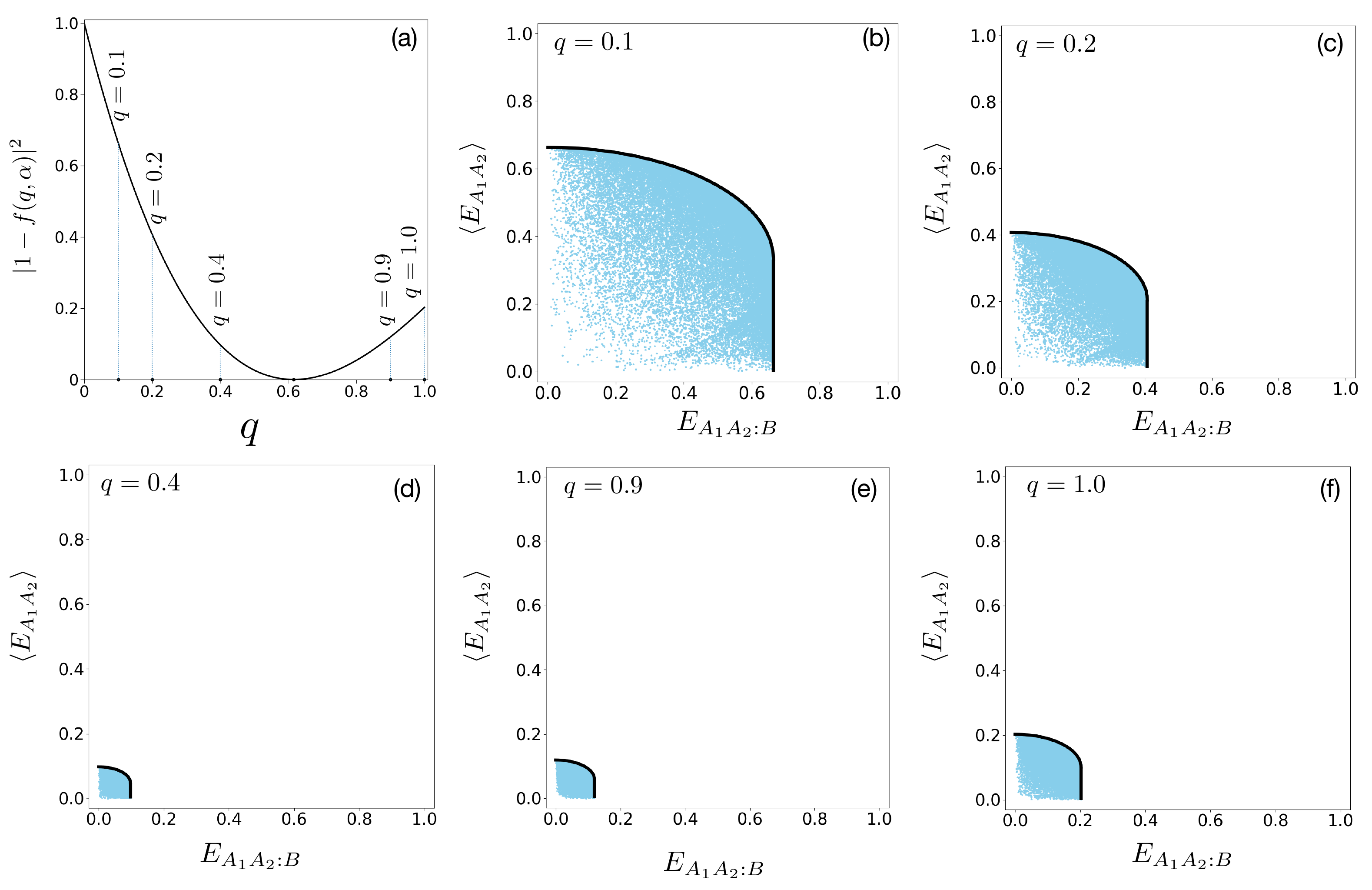}
    \caption{\textbf{Generalized W states under non-Markovian phase-flip noise.} \textbf{(a)} The variation of $|1-f(q,\alpha)|^2$ as a function of $q$ for $\alpha=0.9$, demonstrating the variation of entanglement with $q$ (see Eqs.~(\ref{eq:gw_non_markovian_q_1})-(\ref{eq:gw_non_markovian_q_4})),  which exhibits monotonic decay followed by a vanishing at $q_c$ (Eq.~(\ref{eq:critical_noise})), and a subsequent revival for $q>q_c$, where negativity is used as entanglement measure. In the entire range $0\leq q\leq 1$, five values of noise strengths, namely \textbf{(b)} $q=0.1$, \textbf{(c)} $q=0.2$, \textbf{(d)} $q=0.4$, \textbf{(e)} $q=0.9$, and \textbf{(f)} $q=1.0$ are chosen, and the modification of the boundaries on the four-qubit gW states, as given in Proposition II, with varying $q$ is demonstrated. The region accessible by the gW states on the  $\left(E_{A_1A_2:B},\langle E_{A_1A_2}\rangle\right)$ decreases at first, vanishes at $q=q_c$, and then revives again. All quantities plotted in all figures are dimensionless.}
    \label{fig:fig6}
\end{figure*}

Same line of calculations would follow for the non-Markovian phase-flip channel, leading to 
\begin{eqnarray}
 E_{A_1A_2:B} &=&  \left|1-f(q,\alpha)\right|^NE_{A_1A_2:B}^0,\nonumber\\
 \langle E_{A_1A_2}\rangle &=&  \left|1-f(q,\alpha)\right|^N \langle E_{A_1A_2}\rangle^0, 
\end{eqnarray}
with 
\begin{eqnarray}
 f(q,\alpha)=q\left\{1+\alpha\left(1-\frac{q}{2}\right)\right\},
 \label{eq:non_markovian_noise_factor}
\end{eqnarray}
which implies $\langle E_{A_1A_2}\rangle=E_{A_1A_2:B}$. Similar proof follows for $E_{A_1:A_2B}$, and $E_{A_2:A_1B}$ also, resulting in Eq.~(\ref{eq:pf_gghz_bound}) for the non-Markovian phase-flip channel. Hence the proof.
\end{proof}

\note{7} A comparative discussion on the variation of entanglement with $q$ in the cases of the Markovian and the non-Markovian phase-flip channels is in order here. Note that in the former case, entanglement decays monotonically with $q$, as indicated from the $(1-q)^N$ dependence, while the decay fastens exponentially with increasing number of qubits. It also indicates that entanglement vanishes asymptotically with increasing $q$, attaining zero value only at $q=1$. Similar features are also present in the case of the non-Markovian channel, except one where in contrast to entanglement vanishing only at $q=1$ in the former case, entanglement vanishes at a finite critical $q$ in the latter, given by 
\begin{eqnarray}
 q_c=\frac{1}{\alpha}\left(1+\alpha-\sqrt{1+\alpha^2}\right).
 \label{eq:critical_noise}
\end{eqnarray}
For $q>q_c$, entanglement revives again. Note that $q_c$ is a monotonically decreasing function of $\alpha$, which,  in the limit $\alpha\rightarrow 0$ (the Markovian limit), goes to $1$.

\subsection{Generalized W states}
\label{subsec:noisy_gw}

We now focus on the $N$-qubit gW states under the phase-flip noise on all qubits. Analytical investigation of $\langle E_{A_1A_2}\rangle$ as well as bipartite entanglement over the unmeasured state in such cases is difficult due to the increasing number of state as well as optimization parameters. However, our numerical investigation suggests that irrespective of the tripartition $A_1:A_2:B$ of the $N$-qubit system, the optimization of $\langle E_{A_1A_2}\rangle$ always takes place via $\sigma^z$ measurement on all qubits $i\in B$. This result can be utilized to determine the dependence of $\langle E_{A_1A_2}\rangle$ and $E_{A_1A_2:B}$ on the noise strength $q$, and extend the Propositions II and III, as follows. 

\paragraph{Markovian phase-flip channels.} In the case of the Markovian phase-flip channels, we obtain\footnote{These expressions are determined for systems of small sizes, and are verified numerically for larger systems with different combinations of $(N,n,m)$.}
\begin{eqnarray}
\label{eq:gw_markovian_q_1}
\langle E_{A_1A_2}\rangle &=& (1-q)^2\langle E_{A_1A_2}\rangle^0,\\ 
E_{A_1A_2:B} &=& (1-q)^2 E^0_{A_1A_2:B},
\label{eq:gw_markovian_q_2}
\end{eqnarray}
Given these results, we present the following Proposition.

\propose{VII} \emph{In the space $\left(E_{A_1A_2:B},\langle E_{A_1A_2}\rangle\right)$, the localizable entanglement $\langle E_{A_1A_2}\rangle$ of an $N$-qubit normalized gW state subjected to Markovian phase-flip channels of the same strength, $q$, on all qubits is bounded by the lines
\begin{eqnarray}
\langle E_{A_1A_2}\rangle = \frac{1}{2}\left[(1-q)^2+\sqrt{(1-q)^4-E_{A_1A_2:B}^2}\right],
\label{eq:gw_upper_bound_noise}
\end{eqnarray}
and 
\begin{eqnarray}
 E_{A_1A_2:B}=(1-q)^2,
 \label{eq:other_line}
\end{eqnarray}
where $E_{A_1A_2:B}$ is the bipartite entanglement over the bipartition $A_1A_2:B$ in the state prior to measurement on all the qubits in $B$.}

\note{8} It is worthwhile to note that the line given in Eq.~(\ref{eq:other_line}) corresponds to the family of states, described by Eq.~(\ref{eq:unit_entanglement}), subjected to the single-qubit phase-flip channels on all qubits. 

\note{9} Proposition VI implies that the area on the $\left(E_{A_1A_2:B},\langle E_{A_1A_2}\rangle\right)$ plane confining the noisy gW states shrinks with increasing noise strength $q$, and vanishes at $q=1$. This is demonstrated in Fig.~\ref{fig:fig5}(a).

Noting that $E_{A_1:A_2B}$ and $E_{A_2:A_1B}$ also have similar dependence on $q$ as $E_{A_1A_2:B}$, Proposition III can be extended to the case of gW states under phase-flip noise, as follows.

\propose{VIII} \emph{In $\left(\min\{E_{A_1:A_2B},E_{A_2:A_1B}\},\langle E_{A_1A_2}\rangle\right)$ space, the localizable entanglement $\langle E_{A_1A_2}\rangle$ of an $N$-qubit normalized gW state subjected to the phase-flip channel of strength $q$ on all qubits is upper-bounded by the line 
\begin{eqnarray}
 \langle E_{A_1A_2}\rangle = \min\{E_{A_1:A_2B},E_{A_2:A_1B}\},
 \label{eq:pf_upper_bound_2}
\end{eqnarray}
and lower-bounded by the line 
\begin{eqnarray}
&& \langle E_{A_1A_2}\rangle^2-2(1-q)^2\langle E_{A_1A_2}\rangle\nonumber\\&&+\left(\min\{E_{A_1:A_2B},E_{A_2:A_1B}\}\right)^2=0,
\label{eq:pf_gw_lower_bound}
\end{eqnarray} 
where $E_{A_1:A_2B}$ $\left(E_{A_2:A_1B}\right)$ is the entanglement over the bipartition $A_1:A_2B$ $\left(A_2:A_1B\right)$ in the state prior to measurement on all the qubits in $B$.}

\begin{figure*}
    \centering
    \includegraphics[width=0.7\textwidth]{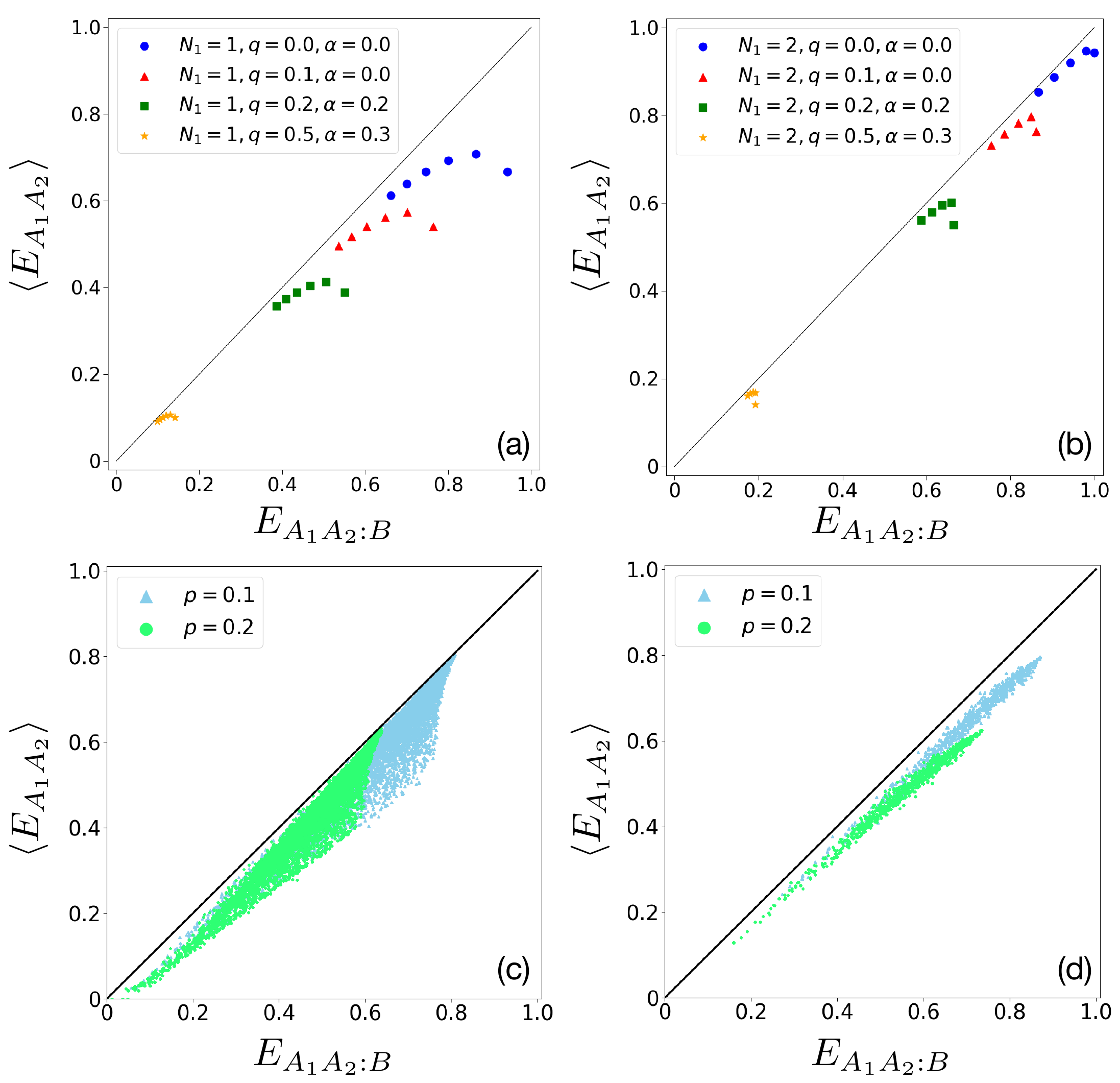}
    \caption{\textbf{Dicke and generalized Dicke states under phase-flip noise.} \textbf{(a), (b)} Scatter plots of $N$-qubit Dicke states, for upto $N=8$, under Markovian and non-Markovian phase-flip noise are shown on the   $\left(E_{A_1A_2:B},\langle E_{A_1A_2}\rangle\right)$ planes for \textbf{(a)} $N_1=1$, and \textbf{(b)} $N_1=2$. \textbf{(c), (d)} On the other hand, scatter plots of samples of $10^7$ gD states with \textbf{(c)} $N=3$ and \textbf{(d)} $N=4$ are shown on the $\left(E_{A_1A_2:B},\langle E_{A_1A_2}\rangle\right)$ plane. In all plots, negativity is used as entanglement measure. All quantities plotted in all figures are dimensionless.}
    \label{fig:fig7}
\end{figure*}

\paragraph{Non-Markovian phase-flip channels.} A similar approach can also be taken in the case of the non-Markovian channels, where entanglement has the following dependence on $q$ and $\alpha$:  
\begin{eqnarray}
\label{eq:gw_non_markovian_q_1}
\langle E_{A_1A_2}\rangle &=& \left|1-f(q,\alpha)\right|^2\langle E_{A_1A_2}\rangle^0,\\ 
E_{A_1A_2:B} &=& \left|1-f(q,\alpha)\right|^2 E^0_{A_1A_2:B},\\
\label{eq:gw_non_markovian_q_2}
E_{A_1:A_2B} &=& \left|1-f(q,\alpha)\right|^2 E^0_{A_1:A_2B},\\
\label{eq:gw_non_markovian_q_3}
E_{A_2:A_1B} &=& \left|1-f(q,\alpha)\right|^2 E^0_{A_2:A_1B},
\label{eq:gw_non_markovian_q_4}
\end{eqnarray}
with $f(q,\alpha)$  given in Eq.~(\ref{eq:non_markovian_noise_factor}). Using these, one can straightforwardly obtain the results on the bounds on the gW states when subjected to non-Markovian phase-flip channels, by replacing the $(1-q)^2$ factors with $\left|1-f(q,\alpha)\right|^2$. To keep the text uncluttered, we refrain from writing these Propositions explicitly. 

Similar to the Markovian case, for a fixed value of $\alpha$, the area on the $E_{A_1A_2:B}-\langle E_{A_1A_2}\rangle$ plane confining the noisy gW states shrinks with increasing $q$ in the case of the non-Markovian phase flip channel also.  However, in contrast to the Markovian case, the area vanishes at a critical noise strength $q_c$, given in Eq.~(\ref{eq:critical_noise}), and then revives again for $q>q_c$ (see Fig.~\ref{fig:fig6} for a demonstration). It is worthwhile to note that the decay of entanglement in the case of the gW states under Markovian and non-Markovian phase-flip channel is independent of the number of qubits in the system, as opposed to the case of the gGHZ states, where the dependence is exponential in $N$.

\subsection{Numerical results}

The complexity of the states obtained via applying single-qubit phase-flip noise to all qubits of a multi-qubit states prevents analytical investigation into the relation between the localizable end the destroyed entanglement in most of the cases. In this subsection, we discuss the numerical results obtained for the mixed states generated via subjecting three-qubit W-class states, $N$-qubit Dicke and gD states, and arbitrary $N$-qubit pure states to Markovian and non-Markovian phase-flip channels.

\paragraph{Three-qubit states.} Our numerical investigation of the three-qubit W-class states subjected to Markovian as well as non-Markovian phase-flip channels indicate that the variations of the localizable entanglement $\langle E_{23}\rangle$ as well as the bipartite entanglement lost during measurement, namely, $E_{1:23}$, $E_{2:13}$, and $E_{3:12}$, are identical to that described in Eqs.~(\ref{eq:gw_markovian_q_1})-(\ref{eq:gw_markovian_q_2}) (Markovian) and Eqs.~(\ref{eq:gw_non_markovian_q_1})-(\ref{eq:gw_non_markovian_q_4}) (non-Markovian).
This implies that the mixed states obtained from the three-qubit W-class states, for a specific noise strength $q$, are bounded in a similar way as described for the gW states under phase-flip noise. On the other hand, while a similar numerical analysis is difficult for the GHZ class states due to the increased number of parameters, we observe that similar to the pure GHZ-class states discussed in Note 4 (see also Fig.~\ref{fig:fig4}(c).), the bounds proposed in Eqs.~(\ref{eq:pf_upper_bound_2})-(\ref{eq:pf_gw_lower_bound}) hold also for three-qubit GHZ-class states subjected to the Markovian phase-flip noise channels.

\begin{table} 
\begin{tabular}{|p{0.025\linewidth}|p{0.025\linewidth}|p{0.025\linewidth}|p{0.8\linewidth}|}
\hline 
   $N$ & $n$ & $m$ & Fraction of  states (in $\%$) with $\delta_1>0$ \\
\hline 
 &  &  & 
\begin{tabular}{p{1.20cm}|p{1.145cm}|p{1.155cm}|p{1.175cm}|p{1.155cm}}
  $q=0.0$ & $q=0.1$ & $q=0.2$ & $q=0.3$ & $q=0.4$ 
\end{tabular} \\
\hline 
3 & 1 & 1 & 
\begin{tabular}{c|c|c|c|c}
  $23.343\%$ & $23.608\%$ & $23.097\%$ & $21.530\%$ & $18.890\%$ 
\end{tabular} \\
\hline 
4 & 1 & 1 & 
\begin{tabular}{c|c|c|c|c}
  $38.704\%$ & $20.461\%$ & $15.757\%$ & $15.514\%$ & $15.336\%$
\end{tabular} \\
\hline
5 & 1 & 1 & 
\begin{tabular}{c|c|c|c|c}
  $44.920\%$ & $14.213\%$ & $12.942\%$ & $10.531\%$ & $07.487\%$ 
\end{tabular} \\
\hline
5 & 1 & 2 & 
\begin{tabular}{c|c|c|c|c}
  $100.00\%$ & $100.00\%$ & $100.00\%$ & $100.00\%$ & $93.402\%$ 
\end{tabular} \\
\hline
\end{tabular}  
\caption{Variations of the fraction of Haar uniformly generated three-, four-, and five-qubit states for which $\delta_1>0$, as a function of the noise parameter $q$, which assumes five values, $q=0.0,0.1,0.2,0.3,$ and $0.4$, from left to right along the row for a specific combination of $N$, $n$, and $m$, in the case of the Markovian phase-flip noise, where negativity is used as entanglement measure. In all cases, we have kept $n=1$, where $n$ and $m$ being the sizes of $B$ and $A_1$ respectively. For the three-qubit system, only GHZ class states are considered. For each case, the percentages are determined from a sample size of $10^5$ Haar-uniformly generated states. }
\label{tab:percentage_table}
\end{table}

\paragraph{Dicke states and generalized Dicke states under noise.} We also numerically investigate the $N$-qubit Dicke and gD states subjected to Markovian and non-Markovian phase-flip channels. Fig.~\ref{fig:fig7}(a)-(b) depict the scatter plots of the Dicke states up to $N=8$, and $N_1=1,2$, where Markovian and non-Markovian phase-flip channel is applied to all qubits. It is clear from the figures that similar to the case of the Pure Dicke states (see Sec.~\ref{subsec:pure_arbit}), (a) the variation of $\langle E_{A_1A_2}\rangle$ with $E_{A_1A_2:B}$ remains non-monotonic, and (b) with increasing $N_1$, the states tend to the $\langle E_{A_1A_2}\rangle=E_{A_1A_2:B}$ line. Also, increasing the non-Markovianity factor $\alpha$ generally tends to lower values of $\langle E_{A_1A_2}\rangle$ and $E_{A_1A_2:B}$. 

In the case of the gD states, our numerical investigation suggests that while the upper bound $\langle E_{A_1A_2}\rangle \leq E_{A_1A_2:B}$ remains valid even in the presence of phase-flip noise irrespective of whether it is Markovian, or non-Markovian. However, with increasing number of qubits, as in the case of the pure gD states, more states are concentrated close to the $\langle E_{A_1A_2}\rangle = E_{A_1A_2:B}$ line, although the number of states for which $\langle E_{A_1A_2}\rangle = E_{A_1A_2:B}$ diminishes drastically. This is demonstrated in Fig.~\ref{fig:fig7}(c)-(d) for gD states with $N=3,4$ qubits, respectively, under Markovian phase-flip noise. The results remain qualitatively the same even in the presence of non-Markovian phase-flip noise also. 

\paragraph{Arbitrary pure states under noise.} It is important to note that the numerical investigation for arbitrary $N$-qubit pure states subjected to phase-flip channels is resource-intensive even for a small number of qubits due to the optimization involved in the computation of localizable entanglement, as the number of optimization parameter increases with increasing $n$, the size of the measured subsystem $B$. In this paper, we restrict ourselves in reporting data for which $B$ is constituted of one qubit only. Similar to the pure states of $N$-qubits described in Sec.~\ref{subsec:pure_arbit}, we focus on $\delta_1$ and $\delta_2$ (Eqs.~(\ref{eq:delta1})-(\ref{eq:delta2})) also for the mixed states obtained by subjecting $N$-qubit arbitrary pure states to Markovian phase-flip channels. As expected, for all investigated values of $q$, no states are found for which $\delta_2\geq 0$, implying a violation of inequality (\ref{eq:localizable_inequality}), which is similar to the case of the pure states (see Sec.~\ref{subsec:pure_arbit}). On the other hand, the percentage of states for which $\delta_1>0$ are tabulated in Table~\ref{tab:percentage_table} for different noise strengths in the case of the Markovian phase-flip channel. It is clear from the table that (a) for a fixed $N$ with $n=1$, the number of states for which $\delta_1>0$ overall decreases with increasing $q$, and (b) for a specific $q$ value $>0$, such states overall decreases in number with increasing $N$, as long as $n$ and $m$ are fixed at $1$.   

\JK{It is important to note that the effect of single-qubit Pauli noise on localizable entanglement has been explored in literature~\cite{banerjee2020,Banerjee2022}, and a set of hierarchies between the localizable entanglement over a specific subsystem in a multiqubit state is proposed~\cite{Banerjee2022} in situations when local noise acts on either the subparts or on all the qubits of the whole system.  In contrast, our work probes the multiparty systems subjected to single-qubit Pauli noise on all qubits via the bounds on localizable entanglement imposed by the entanglement present in the noisy system prior to measurement.  We once again stress here that while we have used negativity as entanglement measure to discuss the results in Sec.~\ref{sec:noise}, using other computable bipartite entanglement measures for mixed states, eg. logarithmic negativity, yields qualitatively similar results, with changes only in the functional dependence of $\langle E_{A_1A_2}\rangle$ on $E_{A_1A_2:B}$  and $\min\{E_{A_1:A_2B},E_{A_2:A_1B}\}$ corresponding to the bounds, and  in the data presented in Table~\ref{tab:percentage_table}, which exclusively correspond to negativity.}

\section{Interacting 1D quantum spin models}
\label{sec:spin_model}


It is natural to ask whether the bounds discussed in Sec.~\ref{sec:pure_states} for the pure states also exist in the ground states of paradigmatic quantum spin Hamiltonians.  In order to investigate this, we focus on the one-dimensional (1D) quantum spin chains with $N$ spin-$1/2$ particles, governed by a Hamiltonian given by~\cite{Mikeska2004} 
\begin{eqnarray}
H&=&\sum_{i=1}^{N}\Bigg[\frac{J^{xy}_{i,i+1}}{4}\Big\{(1+\gamma)\sigma^{x}_{i}\sigma^{x}_{i+1}+(1-\gamma)\sigma^{y}_{i}\sigma^{y}_{i+1}\Big\}\nonumber\\ &&+\frac{J^{zz}_{i,i+1}}{4}\sigma^{z}_{i}\sigma^{z}_{i+1}+\frac{h_{i}}{2}\sigma^{z}_{i}\Bigg].
\label{eq:hamiltonian}
\end{eqnarray}
In Eq.~(\ref{eq:hamiltonian}), $\sigma$'s are Pauli operators, $\gamma$ is the $xy$ anisotropy parameter, $h_{i}$ is local magnetic field strength corresponding to spin $i$, and $J^{xy}_{i,i+1}$ $\left(J^{zz}_{i,i+1}\right)$ represents the nearest-neighbor $xy$ $(zz)$ interaction strengths. Also, we assume periodic boundary condition (PBC) in the system, implying $\sigma_{N+1}^{x,y,z}\equiv \sigma_1^{x,y,z}$. A number of  paradigmatic 1D quantum spin models can be represented by different special cases of $H$. In this paper we are interested in two of them, namely, (a) transverse-field XY model (TXY) $\left(0< \gamma\leq 1,J^{zz}_{i,i+1}=0\right)$~\cite{Lieb1961,Barouch1970,Barouch1971,Barouch1971a,dutta1996,sachdev_2011} (note that the transverse-field Ising model (TIM)~\cite{Pfeuty1970,dutta1996,sachdev_2011} is a special case of the TXY model with $\gamma=1$), and (b) $XXZ$ model with magnetic field (XXZ) $\left(\gamma=0\right)$~\cite{Yang1966,Yang1966a,Yang1966b,Langari1998,Mikeska2004,Giamarchi2004}.

\subsection{Ordered quantum spin models}
\label{subsec:ordered}

\begin{figure*}
    \centering
    \includegraphics[width=\linewidth]{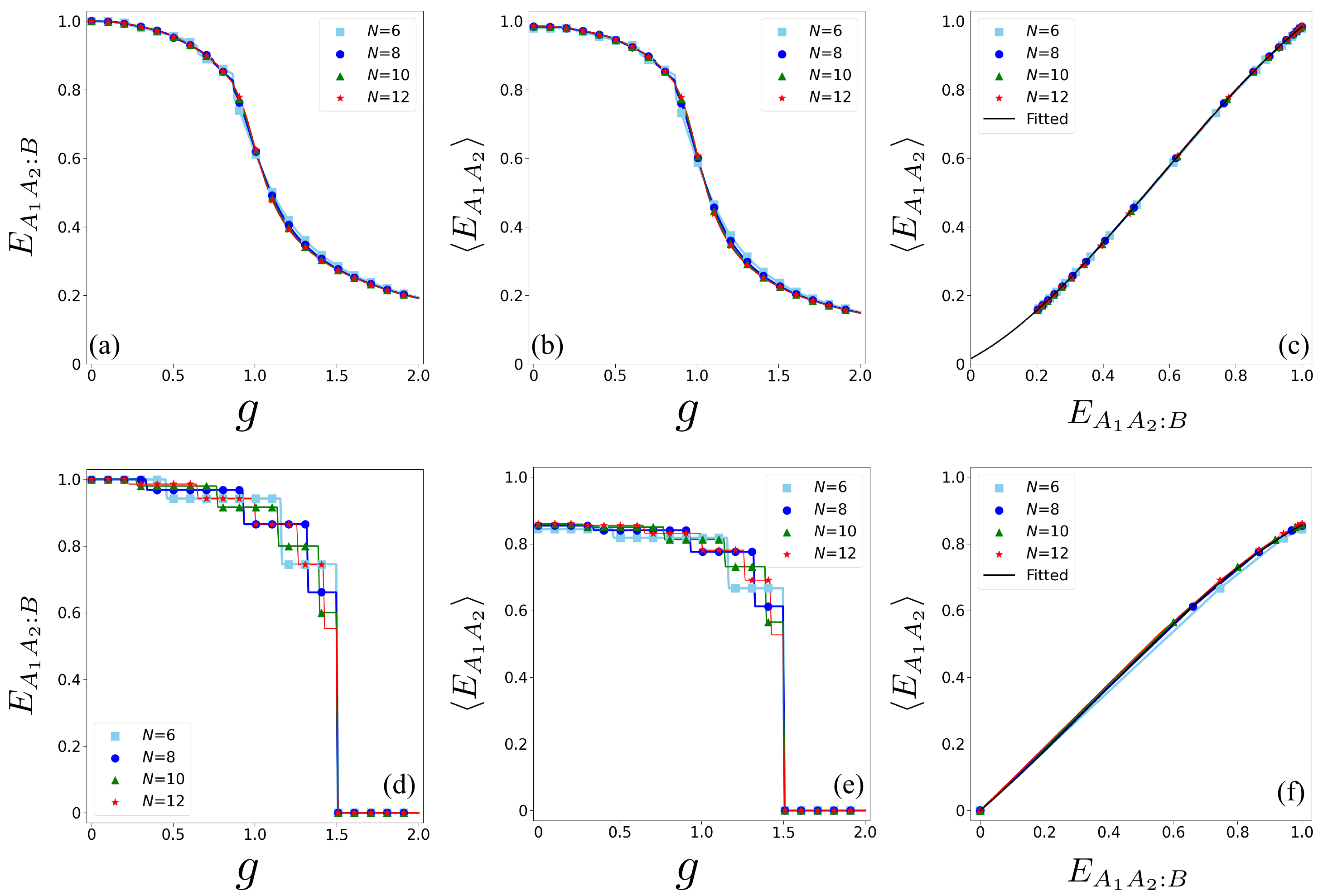}
    \caption{\textbf{Ordered 1D quantum spin models.} \JK{Variations of $E_{A_1A_2:B}$ (\textbf{(a), (d)}) and $\langle E_{A_1A_2}\rangle$ (\textbf{(b), (e)}) as functions of $g$ for the TXY model (\textbf{(a), (b)}) and the XXZ model (\textbf{(d), (e)}) with different system sizes.} \textbf{(c), (f)} Scatter plots of the ground states in 1D TXY model \textbf{(c)}, and 1D XXZ model \textbf{(f)} in an external field on the $\left(E_{A_1A_2:B},\langle E_{A_1A_2}\rangle\right)$ plane The variations of $\langle E_{A_1A_2}\rangle$ as a function of $E_{A_1A_2:B}$ are fitted to Eq.~(\ref{eq:xy_ordered_fit}) for both the 1D TXY model and the 1D XXZ model. In the case of the TXY model, $\gamma$ is taken to be $0.5$, while for the XXZ model, $\Delta=0.5$. In this figure, we have used negativity to quantify entanglement. All quantities plotted in all figures are dimensionless.}
    \label{fig:fig8}
\end{figure*}

In the case of the \emph{ordered} quantum spin models where order exists in all spin-spin couplings as well as the field strengths, we assume $J^{xy}_{i,i+1}=J^{xy}>0$, $J^{zz}_{i,i+1}=J^{zz}>0$, and $h_{i}=h>0$ for all $i=1,2,\cdots,N$. In such models, we numerically investigate the correlation between $\langle E_{A_1A_2}\rangle$ and  $\{E_{A_1A_2:B}$, $E_{A_1B:A_2}, E_{A_2B:A_1}\}$ in the ground state, which is obtained via numerical diagonalization of $H$. We first consider the ordered AFM TXY model, and define $g=h/J^{xy}$ as the dimensionless field-strength. The model exhibits a quantum phase transition from an antiferromagnetic (AFM) phase $(g<1)$ to a paramagnetic (PM) phase $(g>1)$ at $g_c=1$, for all values of $\gamma>0$~\cite{Barouch1970,Barouch1971,Barouch1971a,sachdev_2011,dutta1996,osterloh2002,amico2008,osborne2002}. Figs.~\ref{fig:fig8}(a) and (b) depict the variations of $E_{A_1A_2:B}$ and $\langle E_{A_1A_2}\rangle$, respectively, with negativity quantifying entanglement, as functions of $g$. The shape of the curve changes from convex to concave at $g=1$, indicating that the absolute value of the first derivative of entanglement w.r.t. $g$ exhibits a sharp peak at $g=1$, signalling the quantum phase transition. Fig.~\ref{fig:fig8}(c) presents the scatter plot of $\langle E_{A_1A_2\rangle}-E_{A_1A_2:B}$ corresponding to the ordered TXY model for different values of $N$ using negativity, where for each $N$, values of $n,m=1$. This, along with the PBC ensures that $E_{A_1A_2:B}=E_{A_1:A_2B}$, while our numerical findings suggest that for the ground states of the TXY model with $n=m=1$, $\min\{E_{A_1:A_2B},E_{A_2:A_1B}\} = E_{A_1:A_2B}$ for all values of $N$.  This implies that it is sufficient to investigate the dependence of $\langle E_{A_1A_2}\rangle$ on $E_{A_1A_2:B}$. It is evident from the figure that $\langle E_{A_1A_2}\rangle$ is positively correlated with $E_{A_1A_2:B}$. Irrespective of the value of $N$, the data suggests a cubic dependence of $\langle E_{A_1A_2}\rangle$ on $E_{A_1A_2:B}$, given by 
\begin{eqnarray}
\langle E_{A_1A_2}\rangle &=&  \lambda_3 E_{A_1A_2:B}^3+\lambda_2 E_{A_1A_2:B}^2\nonumber\\&&+\lambda_1 E_{A_1A_2:B}+\lambda_0, 
\label{eq:xy_ordered_fit}
\end{eqnarray}
where the values of $\lambda_{0,1,2,3}$ can be obtained by fitting the numerical data to Eq.~(\ref{eq:xy_ordered_fit}) (see Table.~\ref{tab:fitting_parameters}). 
Note that at $g\rightarrow\infty$, both $E_{A_1A_2:B}$ and $\langle E_{A_1A_2}\rangle$ tend to vanish as the ground state of the TXY model becomes fully polarized.  Therefore one expects $\lambda_0=0$ in Eq.~(\ref{eq:xy_ordered_fit}). However,  we fit only the numerical data up to $g=2$ to Eq.~(\ref{eq:xy_ordered_fit}), which results in a small non-zero value of $\lambda_0$. Our numerical investigation also suggests that the form of Eq.~(\ref{eq:xy_ordered_fit}) remains invariant with a change in the values of the $xy$ anisotropy parameter $\gamma$.  \JK{Also, the qualitative results as well as the form (\ref{eq:xy_ordered_fit}) is invariant with a change in the entanglement measures, with a change only in the values of the fitting parameters. See Table.~\ref{tab:fitting_parameters}.}

\begin{figure*}
    \centering
    \includegraphics[width=\linewidth]{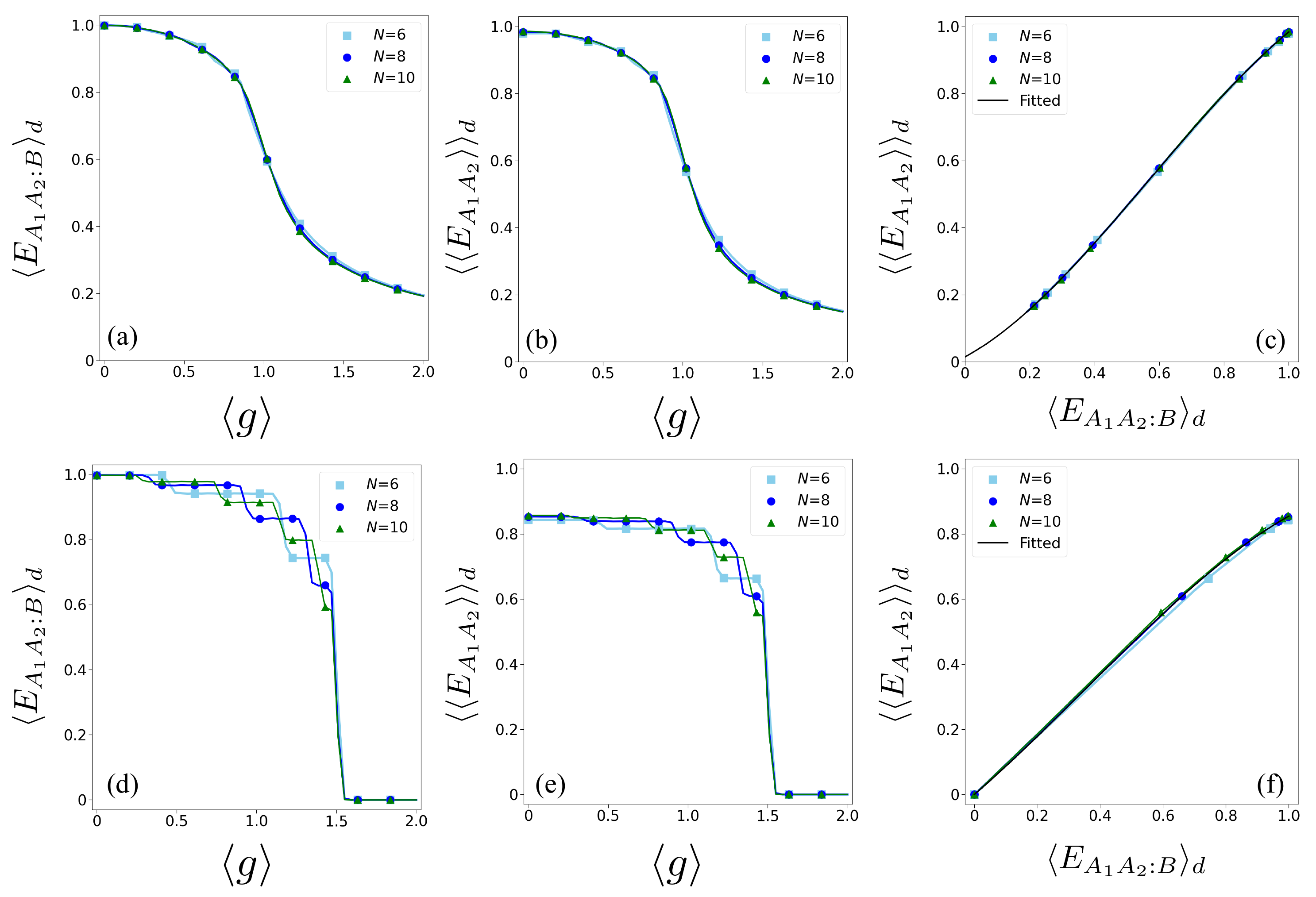}
    \caption{\textbf{Disordered 1D quantum spin models.} \JK{Variations of $\langle E_{A_1A_2:B}\rangle_d$ (\textbf{(a), (d)}) and $\langle\langle E_{A_1A_2}\rangle\rangle_d$ (\textbf{(b), (e)}) as functions of $\langle g\rangle$ for the TXY model (\textbf{(a), (b)}) and the XXZ model (\textbf{(d), (e)}) with different system sizes.} \textbf{(c), (f)} Scatter plots of the ground states in 1D TXY model \textbf{(c)}, and 1D XXZ model \textbf{(f)} in an external field on the $\left(\langle E_{A_1A_2:B}\rangle_d,\langle\langle E_{A_1A_2}\rangle\rangle_d\right)$ plane. The variations of $\langle\langle E_{A_1A_2}\rangle\rangle_d$ as a function of $\langle E_{A_1A_2:B}\rangle_d$ are fitted to Eq.~(\ref{eq:xy_ordered_fit}) for both the 1D TXY model and the 1D XXZ model. In the case of the TXY model, $\gamma$ is taken to be $0.5$, while for the XXZ model, $\Delta=0.5$. For all cases, $\sigma_g=$ 0.05. In this figure, we have used negativity to quantify entanglement. All quantities plotted in all figures are dimensionless.}
    \label{fig:fig9}
\end{figure*}

In the case of the XXZ model in a magnetic field along the $z$ direction with PBC, we assume $J^{zz}=\Delta J^{xy}$, where $\Delta$ signifies the $z$-anisotropy parameter, and denote the dimensionless field-strength by $g=h/J^{xy}$. For $-1\leq \Delta\leq 1$, the model undergoes a quantum phase transition from the XY phase to the ferromagnetic (FM) phase at the critical field strength $g_c=\pm(1+\Delta)$~\cite{Mikeska2004}. Similar to the TXY model, we fix $\Delta$ and investigate the correlation between $\langle E_{A_1A_2}\rangle$ and $E_{A_1A_2:B}$. The data for the variations of $E_{A_1A_2:B}$ and $\langle E_{A_1A_2}\rangle$ with $g$ using negativity are presented in Figs.~\ref{fig:fig8}(d) and (e), respectively, for $\Delta=0.5$, where the quantum phase transition is indicated ar $g_c=1.5$ via a sharp fall in the value of entanglement. Fig.~\ref{fig:fig8}(f) depicts the scatter plot of the points $\left(E_{A_1A_2}\rangle,E_{A_1A_2:B}\right)$ with negativity as entanglement measure, where the fitted curve has the same form as in Eq.~(\ref{eq:xy_ordered_fit}).
Note that in the case of the XXZ model in an external field, both $E_{A_1A_2:B}$ and $\langle E_{A_1A_2}\rangle$ reduces to zero for $g>1.5$. Therefore, the $(E_{A_1A_2:B},\langle E_{A_1A_2}\rangle)$  data includes the point $(0,0)$, leading to $\lambda_0=0$ in the fitted curve.  It is also worthwhile to note that in the case of the XXZ model also, the overall relation between $\langle E_{A_1A_2}\rangle$ and $E_{A_1A_2:B}$ remains unaltered with a change in the value of $\Delta$ within the mentioned region $-1\leq \Delta \leq 1$. \JK{All these results for the XXZ model qualitatively remain the same if the choice of entanglement measure is changed, with only a change in the fitting parameters (see Table.~\ref{tab:fitting_parameters})}. It is, therefore, clear from our numerical analysis that each of these 1D models can be classified by the variations of $\langle E_{A_1A_2}\rangle$ with $E_{A_1A_2:B}$, irrespective of the values of the system parameters as well as the system-size. 

\begin{table*}
    \JK{
    \centering
    \begin{tabular}{c}
    TXY model \\ 
    \begin{tabular}{|c|c|c|c|}
    \hline 
    Parameters & Negativity & Logarithmic Negativity & von Neumann entropy \\
    \hline
      $\lambda_0$   & 0.015  & 0.058 & -0.0064 \\
    \hline 
       $\lambda_1$  & 0.511  & 0.225 & 0.735 \\
    \hline 
       $\lambda_2$  & 1.087  & 1.516 & 0.703 \\ 
    \hline 
       $\lambda_3$  & -0.629  & -0.812 & -0.454 \\
    \hline   
    \end{tabular} \\
    XXZ model \\
    \begin{tabular}{|c|c|c|c|}
    \hline 
    Parameters & Negativity & Logarithmic Negativity & von Neumann entropy \\
    \hline
      $\lambda_0$   & 2.799$\times 10^{-5}$  & 0 & 0\\
    \hline 
       $\lambda_1$  & 0.828  & 1.207$\pm$0.001 & 0.786 \\
    \hline 
       $\lambda_2$  & 0.387$\pm$0.002  & -0.648$\pm$0.008 & 0.611$\pm$0.001 \\ 
    \hline 
       $\lambda_3$  & -0.359  & 0.331$\pm$0.002 & -0.612 \\
    \hline   
    \end{tabular}    
    \end{tabular} 
    \caption{Fitting parameters corresponding to Eq.~(\ref{eq:xy_ordered_fit}) for the ordered TXY and XXZ model, with negativity, logarithmic negativity, and von Neumann entropy as entanglement measures. We have reported the error in the fitting iff the absolute value of the error $\geq 10^{-3}$.}
    \label{tab:fitting_parameters}
    }
\end{table*}

\subsection{Disordered quantum spin models}
\label{subsec:disorder}

In a \emph{disordered} quantum spin model~\cite{de_dominicis_giardina_2006}, the values of a relevant system parameter, such as $g$, are chosen from a Gaussian distribution, $P(g)$, of fixed mean, $\langle g\rangle$, and fixed standard deviation, $\sigma_g$. Here, $\sigma_g$ represents the \emph{strength} of the disorder, and each random value of $g$ represents a \emph{random parameter configuration} of the quantum spin model, describing a \emph{random realization} of the system. For each random realization of the system, the quantity of interest, $Q(g)$, can be computed. A subsequent \emph{quenched} average of $Q(g)$ over a statistically large number of random realizations is given by 
\begin{eqnarray}
 \langle Q\rangle_d &=& \int P(g) Q(g)dg,
\end{eqnarray}
where the subscript $d$ represents a quenched average, and $\langle Q\rangle$ is effectively a function of $\langle g\rangle$ and $\sigma_g$. Note that the corresponding \emph{ordered} result can be obtained as a special case at $\sigma_g=0$. Note also that disorder can, in principle, be present in a number of system parameters. In this paper, however, we confine ourselves in situations where only one chosen system parameter is disordered.   

We start with the TXY model, choosing the field-strength $h$ (and hence $g$, where $J^{xy}$ is constant) to be the disordered system parameter. The quenched averaged localizable entanglement, $\langle\langle E_{A_1A_2}\rangle\rangle_d$, and the bipartite entanglement, $\langle E_{A_1A_2:B}\rangle_d$, quantified using negativity, in the ground state of the system are numerically computed for different values of the disorder strength $\sigma_{g}$, and are plotted against $\langle g\rangle$ in Figs.~\ref{fig:fig9}(a) and (b). In each of these computations on an $N$-qubit system, both $n$ and $m$ are taken to be $1$, similar to the ordered scenario. Fig.~\ref{fig:fig9}(c), on the other hand, represents the scatter plots of the ground state of the 1D disordered TXY model on the $\langle\langle E_{A_1A_2}\rangle\rangle_d$-$\langle E_{A_1A_2:B}\rangle_d$ plane using negativity, where the numerical data is found to be well-fitted to Eq.~(\ref{eq:xy_ordered_fit}) with   $\lambda_0=0.014$, $\lambda_1=0.520$, $\lambda_2=1.065$, and $\lambda_3=-0.617$, where errors $\leq 10^{-3}$ are neglected.  This  indicates that the relation between $\langle E_{A_1A_2}\rangle$  and $E_{A_1A_2:B}$ for the TXY model is qualitatively robust against disorder in the field. Also, varying $\sigma_g$ in the range $0.01\leq \sigma_g\leq 0.1$ only changes the fitting parameters negligibly. Similar analysis is performed for the $XXZ$ model in an external field to arrive at a similar conclusion, where the data corresponding to negativity is presented in Figs.~\ref{fig:fig9}(d)-(f). \JK{Also, similar to the ordered case, all of these results remain qualitatively unchanged with a change in the entanglement measure.}

\section{Conclusions and outlook}
\label{sec:conclude}

In this paper, we investigate dependence of the \emph{gain} in the entanglement through localization over a group of qubits in a multi-qubit system via single-qubit projection measurements on the rest of the qubits on the amount of \emph{loss} in bipartite entanglement during these measurements. We probe a number of paradigmatic $N$-qubit pure states, namely, the generalized GHZ, generalized W, Dicke, and the generalized Dicke states. We derive analytical bounds for the generalized GHZ and the generalized W states. We show that the gain is always equal to the loss in the former, while in the latter, lower and upper bounds of localizable entanglement can be derived in terms of the bipartite entanglement present in the system prior to the measurement process. In the case of the Dicke and the generalized Dicke states, a combination of analytical and numerical investigations reveal that the localizable entanglement tend to be equal to a component of the lost bipartite entanglement when the number of qubits increases. Modifications of these results, when the system is subjected to single-qubit Markovian and non-Markovian phase-flip channels, are also discussed. We extend our study to the ground states of the 1D quantum spin models, namely, the transverse-field XY model and the XXZ model in an external field, and numerically demonstrate a cubic dependence of the localizable entanglement over the bipartite entanglement in the ground state prior to measurement, where measurement is restricted to one qubit only, and the entanglement is always computed in the 1:rest bipartition. This dependence is shown to be qualitatively robust even in the presence of disorder in the field strength.  

A number of possible avenues of future research emerge from this study. Within the orbit of the results reported in this paper, it is important to understand how the results obtained in the case of the pure states are modified when different types of noise that are commonly occurring in experiments~\cite{schindler2013,bermudez2017}, such as the bit-flip, depolarizing, and amplitude-damping~\cite{nielsen2010,holevo2012} noise are present in the system. However, investigating such noise channels may present new challenges in deriving the appropriate bounds, if any, on localizable entanglement. Also, in the case of the 1D quantum spin systems considered in this paper, it would be interesting to see the effect of the presence of disorder in the spin-spin interaction strengths along with a disordered field-strength. Besides, a plethora of quantum spin models are important from the perspective of quantum information theory~\cite{amico2008,dechiara2018}, and it is interesting to investigate whether a specific relation between the localized and the lost entanglement, similar to the one in the case of the models described in this study, exist in the ground states of these models in the presence and absence of disordered interactions, as well as in situations where the system is made open by allowing an interaction with the environment~\cite{breuer2002}.  

\acknowledgements 

We acknowledge the use of QIClib ( \href{https://github.com/titaschanda/QIClib}{https://github.com/titaschanda/QIClib}) -- a modern C++ library for general purpose quantum information processing and quantum computing.

\appendix

\section{Entanglement measures}
\label{app:negativity}

\JK{Here we define the entanglement measures used in this paper. The amount of entanglement between two partitions $A$ and $B$ of a bipartite quantum state $\rho_{AB}$ can be quantified by a bipartite entanglement measure~\cite{horodecki2009,guhne2009}. In this paper, for both pure as well as mixed states, we focus on negativity~\cite{peres1996,horodecki1996,zyczkowski1998,vidal2002,lee2000} as a bipartite entanglement measure, which is defined as   
\begin{eqnarray} 
E_{A:B}^{neg}=||\rho_{AB}^{T_{B}}||-1,
\end{eqnarray}
which corresponds to the absolute value of the sum of negative eigenvalues, $\lambda$, of $\rho_{AB}^{T_{B}}$, given by 
\begin{eqnarray}
E_{A:B}^{neg}=2\left|\sum_{\lambda_i<0}\lambda_i\right|.
\end{eqnarray}
Here, $||\varrho|| = \mbox{Tr}\sqrt{\varrho^{\dagger}\varrho}$ is the trace norm of the density operator $\varrho$, computed as the sum of the singular values of $\varrho$. The matrix $\rho_{AB}^{T_{B}}$ is obtained by performing partial transposition of the density matrix $\rho_{AB}$ with respect to the subsystem  $B$. Since we only focus on negativity throughout this paper, we discard the superscript from $E^{neg}_{A:B}$, and denote the negativity between the partitions $A$ and $B$ by $E_{A:B}$. One can also define the \emph{logarithmic negativity}~\cite{plenio2005} $\mathcal{L}_{A:B}$ over the same partitions $A:B$ as 
\begin{eqnarray}
\mathcal{L}_{A:B}=\log_{2}(E_{A:B}+1).
\end{eqnarray}}

\JK{In the case of a pure bipartite state $\rho_{AB}$, the entnglement over the bipartition $A:B$ can also be quantified by the von-Neumann entropy~\cite{bennett1996,bennett1996a,horodecki1996,horodecki2009,guhne2009} of the reduced density matrix $\rho_A=\text{Tr}_B(\rho_{AB})$ as 
\begin{eqnarray}
    S=-\rho_A\log_2\rho_A = -\sum_{\lambda}\lambda\log_2\lambda,
\end{eqnarray}
where $\{\lambda\}$ are the eigenvalues of $\rho_A$.}

\section{Single- and two-qubit measurements}
\label{app:two_measurement}

We first consider the single-qubit measurement ($n=1$) in an $N$-qubit gW state, where (see Eq.~(\ref{eq:post_measured_state}))
\begin{eqnarray}
 f_0^0 &=& a_1\text{e}^{-\text{i}\phi}\sin\frac{\theta}{2},\; f_0^1 = -a_1\text{e}^{-\text{i}\phi}\cos\frac{\theta}{2},
 \label{eq:single_measure_coeff_0}
\end{eqnarray}
\begin{eqnarray}
 f^0 &=& \cos\frac{\theta}{2},\; f^1 = \sin\frac{\theta}{2},
 \label{eq:single_measure_coeff_1}
\end{eqnarray}
 and
\begin{eqnarray}
 p_0 &=& a_1^2\sin^2\frac{\theta}{2}+ \cos^2\frac{\theta}{2}\sum_{i=1}^{N-1}a_{1+i}^2,\nonumber\\ 
 p_1 &=& a_1^2\cos^2\frac{\theta}{2}+ \sin^2\frac{\theta}{2}\sum_{i=1}^{N-1}a_{1+i}^2,
\end{eqnarray}
such that $\langle\psi_k|\psi_k\rangle=1$ for $k=0,1$. On the other hand, in the case of two-qubit projection measurements ($n=2$) on an $N$-qubit gW state, the post-measured states are of the form given in Eq.~(\ref{eq:post_measured_state}), with the coefficients $f^k_0$'s and $f^k$'s as  
\begin{widetext}
\begin{eqnarray}
f_0^0 &=& a_1\text{e}^{-\text{i}\phi_1}\sin\frac{\theta_1}{2}\cos\frac{\theta_2}{2}+a_2\text{e}^{-\text{i}\phi_2}\cos\frac{\theta_1}{2}\sin\frac{\theta_2}{2},\; f_0^1 = a_1\text{e}^{-\text{i}\phi_1}\sin\frac{\theta_1}{2}\sin\frac{\theta_2}{2}-a_2\text{e}^{-\text{i}\phi_2}\cos\frac{\theta_1}{2}\cos\frac{\theta_2}{2},\nonumber\\ 
f_0^2 &=& -a_1\text{e}^{-\text{i}\phi_1}\cos\frac{\theta_1}{2}\cos\frac{\theta_2}{2}+a_2\text{e}^{-\text{i}\phi_2}\sin\frac{\theta_1}{2}\sin\frac{\theta_2}{2},\; f_0^3 =-a_1\text{e}^{-\text{i}\phi_1}\cos\frac{\theta_1}{2}\sin\frac{\theta_2}{2}-a_2\text{e}^{-\text{i}\phi_2}\sin\frac{\theta_1}{2}\cos\frac{\theta_2}{2},
\end{eqnarray}
and 
\begin{eqnarray}
f^0 &=& \cos\frac{\theta_1}{2}\cos\frac{\theta_2}{2},\; f^1 = \cos\frac{\theta_1}{2}\sin\frac{\theta_2}{2},\; f^2 = \sin\frac{\theta_1}{2}\cos\frac{\theta_2}{2},\; f^3 = \sin\frac{\theta_1}{2}\sin\frac{\theta_2}{2},
\end{eqnarray} 
 The probabilities of obtaining the measurement outcome $k=0,1,2,3$ are given by
\begin{eqnarray}
 p_0 &=& a_1^2 \sin^2\frac{\theta_1}{2}\cos^2\frac{\theta_2}{2} + a_2^2 \cos^2\frac{\theta_1}{2}\sin^2\frac{\theta_2}{2}+\cos^2\frac{\theta_1}{2}\cos^2\frac{\theta_2}{2}\sum_{i=1}^{N-2}a_{2+i}^2,\nonumber\\ 
 p_1 &=& a_1^2 \sin^2\frac{\theta_1}{2}\sin^2\frac{\theta_2}{2}+ a_2^2 \cos^2\frac{\theta_1}{2}\cos^2\frac{\theta_2}{2}+\cos^2\frac{\theta_1}{2}\sin^2\frac{\theta_2}{2}\sum_{i=1}^{N-2}a_{2+i}^2,\nonumber\\
 p_2 &=& a_1^2 \cos^2\frac{\theta_1}{2}\cos^2\frac{\theta_2}{2}+ a_2^2 \sin^2\frac{\theta_1}{2}\sin^2\frac{\theta_2}{2}+\sin^2\frac{\theta_1}{2}\cos^2\frac{\theta_2}{2}\sum_{i=1}^{N-2}a_{2+i}^2,\nonumber\\
 p_3 &=& a_1^2 \cos^2\frac{\theta_1}{2}\sin^2\frac{\theta_2}{2}+ a_2^2 \sin^2\frac{\theta_1}{2}\cos^2\frac{\theta_2}{2}+\sin^2\frac{\theta_1}{2}\sin^2\frac{\theta_2}{2}\sum_{i=1}^{N-2}a_{2+i}^2.
\end{eqnarray}
\end{widetext}

For a three-qubit state belonging to the W class (see Eq.~(\ref{eq:wclass})), measurement on qubit $1$ can be described in a way similar to that for the three-qubit gW state, with $f_0^k$ $(k=0,1)$ given by
\begin{eqnarray}
 f_0^0 &=& a_0\cos\frac{\theta}{2}+a_1\text{e}^{-\text{i}\phi}\sin\frac{\theta}{2},\nonumber\\
 f_0^1 &=& a_0\sin\frac{\theta}{2}-a_1\text{e}^{-\text{i}\phi}\cos\frac{\theta}{2},
\end{eqnarray}
and $f^k$ $(k=0,1)$ given by 
\begin{eqnarray}
f^0&=&\cos\frac{\theta}{2},\;\; f^1=\sin\frac{\theta}{2},
\end{eqnarray}
and 
\begin{eqnarray}
p_0 &=& \left|a_0\cos\frac{\theta}{2}+a_1\text{e}^{-\text{i}\phi}\sin\frac{\theta}{2}\right|^2+(a_2^2+a_3^2)\cos^2\frac{\theta}{2},\nonumber\\
p_1&=&\left|a_0\sin\frac{\theta}{2}-a_1\text{e}^{-\text{i}\phi}\cos\frac{\theta}{2}\right|^2+(a_2^2+a_3^2)\sin^2\frac{\theta}{2}.\nonumber\\
\end{eqnarray}

\section{Three-qubit gW states with complex coefficients}
\label{app:three-gW}

\JK{Here, we present the crucial steps for computing the localizable entanglement over qubits $2$ and $3$ in a three-qubit gW state $\ket{\psi}=a_1\ket{100}+a_2\ket{010}+a_3\ket{001}$ (Eq.~(\ref{eq:n_qubit_gw}) with $N=3$) with complex coefficients $a_1,a_2$, and $a_3$, via a measurement on qubit $1$. The post-measured states $\ket{\psi_k}$ (Eq.~(\ref{eq:post_measured_state})) are given by $\ket{\psi_k}=c_0^k\ket{00}+c_1^k\ket{01}+c_2^k\ket{10}$, $k=1,2$,  such that $|c_0^k|^2+|c_1^k|^2+|c_2^k|^2=1$, where the forms of $c_i^k$, $i=0,1,2$, $k=1,2$ are given in Sec.~\ref{subsec:gw_states}. The state $\ket{\psi_k}\bra{\psi_k}$, upon partial transposition, becomes 
\begin{eqnarray}
\begin{bmatrix}
|c^k_0|^2 & c_1^kc_0^{k*} & c_0^kc_2^{k*} & c_1^kc_2^{k*} \\
c_0^kc_1^{k*}  & |c_1^k|^2 & 0 & 0 \\
c_2^kac_0^{k*} & 0  &  |c_2^k|^2 & 0 \\
c_2^kc_1^{k*} & 0 & 0 & 0 \\
\end{bmatrix},
\end{eqnarray} 
having the eigenvalues 
\begin{eqnarray}
    \lambda_0&=& -|c_1^k||c_2^k|, \nonumber\\
    \lambda_1&=& |c_1^k||c_2^k|, \nonumber\\
    \lambda_2&=& \frac{1}{2}(1-\sqrt{1-4|c_1^k|^2|c_2^k|^2}), \nonumber\\
    \lambda_3&=& \frac{1}{2}(1+\sqrt{1-4|c_1^k|^2|c_2^k|^2}), 
\end{eqnarray}
among which only $\lambda_0$ is negative. Therefore $\langle E_{23}^k\rangle=|c_1^k||c_2^k|$. On the other hand, partial transpose of the state $\ket{\psi}\bra{\psi}$ with respect to the bipartition $1:23$ followed by a diagonalization leads to the non-zero eigenvalues 
 \begin{eqnarray}
     \lambda_0 &=& |a_1|^2,\nonumber\\
     \lambda_1 &=& -|a_1|\sqrt{|a_2|^2 +|a_3|^2},\nonumber\\ 
     \lambda_2 &=& |a_1|\sqrt{|a_2|^2 +|a_3|^2},\nonumber\\
     \lambda_3 &=& |a_2|^2 + |a_3|^2,
 \end{eqnarray}
leading to $E_{1:23}=|a_1|\sqrt{|a_2|^2 +|a_3|^2}$. Replacing $c_i^k$ as per definition in terms of $a_i$ in $\langle E_{23}\rangle$ and using $E_{1:23}$, one can arrive at the Propositions II and III and the corresponding Corollaries.}

\section{Specific examples}
\label{app:examples}

In this section, we demonstrate a number of results discussed in Secs.~\ref{sec:pure_states} and \ref{sec:noise} using specific examples. 

\paragraph{Upper bound for gW states.} Consider the cases of $N=3,4$, for which the family of states providing the upper bound of $\langle E_{A_1A_2}\rangle$ in the case of gW states with real coefficients can be written as 
\begin{eqnarray}
 \ket{\Psi_3} &=& a\ket{100}+\sqrt{\frac{1-a^2}{2}}(\ket{010}+\ket{001}),\\ 
 \ket{\Psi_4} &=& a\ket{1000}+\sqrt{\frac{1-a^2}{2}}\ket{0100}+b\ket{0010}\nonumber\\ &&+\sqrt{\frac{1-a^2-2b^2}{2}}\ket{0001},
\end{eqnarray}
where $a,b$ are real numbers. The behaviour of $\langle E_{A_1A_2}\rangle$ as a function of $a,b$ is shown in Figs.~\ref{fig:fig2}(a) and (b). Also, $\langle E_{A_1A_2}\rangle$ for the state $\ket{\Psi_3}$ provides an upper bound for $\langle E_{A_1A_2} \rangle$ corresponding to all three-qubit gW states with complex coefficients. This is demonstrated by the $\langle E_{A_1A_2}\rangle$ values corresponding to $10^6$ Haar-uniformly generated three-qubit gW states lying below the upper bound.

\paragraph{Maximum bipartite entanglement for gW states.} The line $E_{A_1A_2:B}=1$ corresponds to the family of states given by 
\begin{eqnarray}
 \sum_{i\in B}a_i^2 = \left(4\sum_{i\in A}a_i^2\right)^{-1}.
\end{eqnarray}
Specifically, for a three-qubit system, this family is given by 
\begin{eqnarray}
 \ket{\Phi_3} &=& \sqrt{\frac{1}{2}}\ket{100}+a\ket{010}+\sqrt{\frac{1}{2}-a^2}\ket{001},
\end{eqnarray} 
while for $N=4$, $n=m=1$, 
\begin{eqnarray}
 \ket{\Phi_4} &=& \sqrt{\frac{1}{2}}\ket{1000}+a\ket{0100}+ b\ket{0010}\nonumber \\ &&+\sqrt{\frac{1}{2}-a^2-b^2}\ket{0001}.
\end{eqnarray}

\bibliography{ref}

\begin{thebibliography}{104}%
\makeatletter
\providecommand \@ifxundefined [1]{%
 \@ifx{#1\undefined}
}%
\providecommand \@ifnum [1]{%
 \ifnum #1\expandafter \@firstoftwo
 \else \expandafter \@secondoftwo
 \fi
}%
\providecommand \@ifx [1]{%
 \ifx #1\expandafter \@firstoftwo
 \else \expandafter \@secondoftwo
 \fi
}%
\providecommand \natexlab [1]{#1}%
\providecommand \enquote  [1]{``#1''}%
\providecommand \bibnamefont  [1]{#1}%
\providecommand \bibfnamefont [1]{#1}%
\providecommand \citenamefont [1]{#1}%
\providecommand \href@noop [0]{\@secondoftwo}%
\providecommand \href [0]{\begingroup \@sanitize@url \@href}%
\providecommand \@href[1]{\@@startlink{#1}\@@href}%
\providecommand \@@href[1]{\endgroup#1\@@endlink}%
\providecommand \@sanitize@url [0]{\catcode `\\12\catcode `\$12\catcode
  `\&12\catcode `\#12\catcode `\^12\catcode `\_12\catcode `\%12\relax}%
\providecommand \@@startlink[1]{}%
\providecommand \@@endlink[0]{}%
\providecommand \url  [0]{\begingroup\@sanitize@url \@url }%
\providecommand \@url [1]{\endgroup\@href {#1}{\urlprefix }}%
\providecommand \urlprefix  [0]{URL }%
\providecommand \Eprint [0]{\href }%
\providecommand \doibase [0]{http://dx.doi.org/}%
\providecommand \selectlanguage [0]{\@gobble}%
\providecommand \bibinfo  [0]{\@secondoftwo}%
\providecommand \bibfield  [0]{\@secondoftwo}%
\providecommand \translation [1]{[#1]}%
\providecommand \BibitemOpen [0]{}%
\providecommand \bibitemStop [0]{}%
\providecommand \bibitemNoStop [0]{.\EOS\space}%
\providecommand \EOS [0]{\spacefactor3000\relax}%
\providecommand \BibitemShut  [1]{\csname bibitem#1\endcsname}%
\let\auto@bib@innerbib\@empty
\bibitem [{\citenamefont {Horodecki}\ \emph {et~al.}(2009)\citenamefont
  {Horodecki}, \citenamefont {Horodecki}, \citenamefont {Horodecki},\ and\
  \citenamefont {Horodecki}}]{horodecki2009}%
  \BibitemOpen
  \bibfield  {author} {\bibinfo {author} {\bibfnamefont {Ryszard}\ \bibnamefont
  {Horodecki}}, \bibinfo {author} {\bibfnamefont {Pawe{\l}}\ \bibnamefont
  {Horodecki}}, \bibinfo {author} {\bibfnamefont {Micha{\l}}\ \bibnamefont
  {Horodecki}}, \ and\ \bibinfo {author} {\bibfnamefont {Karol}\ \bibnamefont
  {Horodecki}},\ }\bibfield  {title} {\enquote {\bibinfo {title} {{Quantum
  entanglement}},}\ }\href {\doibase 10.1103/RevModPhys.81.865} {\bibfield
  {journal} {\bibinfo  {journal} {Rev. Mod. Phys.}\ }\textbf {\bibinfo {volume}
  {81}},\ \bibinfo {pages} {865--942} (\bibinfo {year} {2009})}\BibitemShut
  {NoStop}%
\bibitem [{\citenamefont {G{\"u}hne}\ and\ \citenamefont
  {T{\'o}th}(2009)}]{guhne2009}%
  \BibitemOpen
  \bibfield  {author} {\bibinfo {author} {\bibfnamefont {Otfried}\ \bibnamefont
  {G{\"u}hne}}\ and\ \bibinfo {author} {\bibfnamefont {G{\'e}za}\ \bibnamefont
  {T{\'o}th}},\ }\bibfield  {title} {\enquote {\bibinfo {title} {Entanglement
  detection},}\ }\href
  {http://www.sciencedirect.com/science/article/pii/S0370157309000623}
  {\bibfield  {journal} {\bibinfo  {journal} {Phys. Rep.}\ }\textbf {\bibinfo
  {volume} {474}},\ \bibinfo {pages} {1--75} (\bibinfo {year}
  {2009})}\BibitemShut {NoStop}%
\bibitem [{\citenamefont {Bennett}\ \emph {et~al.}(1993)\citenamefont
  {Bennett}, \citenamefont {Brassard}, \citenamefont {Cr\'epeau}, \citenamefont
  {Jozsa}, \citenamefont {Peres},\ and\ \citenamefont
  {Wootters}}]{bennett1993}%
  \BibitemOpen
  \bibfield  {author} {\bibinfo {author} {\bibfnamefont {Charles~H.}\
  \bibnamefont {Bennett}}, \bibinfo {author} {\bibfnamefont {Gilles}\
  \bibnamefont {Brassard}}, \bibinfo {author} {\bibfnamefont {Claude}\
  \bibnamefont {Cr\'epeau}}, \bibinfo {author} {\bibfnamefont {Richard}\
  \bibnamefont {Jozsa}}, \bibinfo {author} {\bibfnamefont {Asher}\ \bibnamefont
  {Peres}}, \ and\ \bibinfo {author} {\bibfnamefont {William~K.}\ \bibnamefont
  {Wootters}},\ }\bibfield  {title} {\enquote {\bibinfo {title} {Teleporting an
  unknown quantum state via dual classical and einstein-podolsky-rosen
  channels},}\ }\href {\doibase 10.1103/PhysRevLett.70.1895} {\bibfield
  {journal} {\bibinfo  {journal} {Phys. Rev. Lett.}\ }\textbf {\bibinfo
  {volume} {70}},\ \bibinfo {pages} {1895--1899} (\bibinfo {year}
  {1993})}\BibitemShut {NoStop}%
\bibitem [{\citenamefont {Bouwmeester}\ \emph {et~al.}(1997)\citenamefont
  {Bouwmeester}, \citenamefont {Pan}, \citenamefont {Mattle}, \citenamefont
  {Eibl}, \citenamefont {Weinfurter},\ and\ \citenamefont
  {Zeilinger}}]{bouwmeester1997}%
  \BibitemOpen
  \bibfield  {author} {\bibinfo {author} {\bibfnamefont {Dik}\ \bibnamefont
  {Bouwmeester}}, \bibinfo {author} {\bibfnamefont {Jian-Wei}\ \bibnamefont
  {Pan}}, \bibinfo {author} {\bibfnamefont {Klaus}\ \bibnamefont {Mattle}},
  \bibinfo {author} {\bibfnamefont {Manfred}\ \bibnamefont {Eibl}}, \bibinfo
  {author} {\bibfnamefont {Harald}\ \bibnamefont {Weinfurter}}, \ and\ \bibinfo
  {author} {\bibfnamefont {Anton}\ \bibnamefont {Zeilinger}},\ }\bibfield
  {title} {\enquote {\bibinfo {title} {Experimental quantum teleportation},}\
  }\href {\doibase 10.1038/37539} {\bibfield  {journal} {\bibinfo  {journal}
  {Nature}\ }\textbf {\bibinfo {volume} {390}},\ \bibinfo {pages} {575--579}
  (\bibinfo {year} {1997})}\BibitemShut {NoStop}%
\bibitem [{\citenamefont {Bennett}\ and\ \citenamefont
  {Wiesner}(1992)}]{bennett1992}%
  \BibitemOpen
  \bibfield  {author} {\bibinfo {author} {\bibfnamefont {Charles~H.}\
  \bibnamefont {Bennett}}\ and\ \bibinfo {author} {\bibfnamefont {Stephen~J.}\
  \bibnamefont {Wiesner}},\ }\bibfield  {title} {\enquote {\bibinfo {title}
  {Communication via one- and two-particle operators on einstein-podolsky-rosen
  states},}\ }\href {\doibase 10.1103/PhysRevLett.69.2881} {\bibfield
  {journal} {\bibinfo  {journal} {Phys. Rev. Lett.}\ }\textbf {\bibinfo
  {volume} {69}},\ \bibinfo {pages} {2881--2884} (\bibinfo {year}
  {1992})}\BibitemShut {NoStop}%
\bibitem [{\citenamefont {Mattle}\ \emph {et~al.}(1996)\citenamefont {Mattle},
  \citenamefont {Weinfurter}, \citenamefont {Kwiat},\ and\ \citenamefont
  {Zeilinger}}]{mattle1996}%
  \BibitemOpen
  \bibfield  {author} {\bibinfo {author} {\bibfnamefont {Klaus}\ \bibnamefont
  {Mattle}}, \bibinfo {author} {\bibfnamefont {Harald}\ \bibnamefont
  {Weinfurter}}, \bibinfo {author} {\bibfnamefont {Paul~G.}\ \bibnamefont
  {Kwiat}}, \ and\ \bibinfo {author} {\bibfnamefont {Anton}\ \bibnamefont
  {Zeilinger}},\ }\bibfield  {title} {\enquote {\bibinfo {title} {Dense coding
  in experimental quantum communication},}\ }\href {\doibase
  10.1103/PhysRevLett.76.4656} {\bibfield  {journal} {\bibinfo  {journal}
  {Phys. Rev. Lett.}\ }\textbf {\bibinfo {volume} {76}},\ \bibinfo {pages}
  {4656--4659} (\bibinfo {year} {1996})}\BibitemShut {NoStop}%
\bibitem [{\citenamefont {Sen(De)}\ and\ \citenamefont
  {Sen}(2011)}]{sende2010}%
  \BibitemOpen
  \bibfield  {author} {\bibinfo {author} {\bibfnamefont {Aditi}\ \bibnamefont
  {Sen(De)}}\ and\ \bibinfo {author} {\bibfnamefont {Ujjwal}\ \bibnamefont
  {Sen}},\ }\bibfield  {title} {\enquote {\bibinfo {title} {Quantum advantage
  in communication networks},}\ }\href@noop {} {\bibfield  {journal} {\bibinfo
  {journal} {Phys. News}\ }\textbf {\bibinfo {volume} {40}},\ \bibinfo {pages}
  {17--32} (\bibinfo {year} {2011})},\ \Eprint
  {http://arxiv.org/abs/arXiv:1105.2412} {arXiv:1105.2412} \BibitemShut
  {NoStop}%
\bibitem [{\citenamefont {Ekert}(1991)}]{ekert1991}%
  \BibitemOpen
  \bibfield  {author} {\bibinfo {author} {\bibfnamefont {Artur~K.}\
  \bibnamefont {Ekert}},\ }\bibfield  {title} {\enquote {\bibinfo {title}
  {Quantum cryptography based on bell's theorem},}\ }\href {\doibase
  10.1103/PhysRevLett.67.661} {\bibfield  {journal} {\bibinfo  {journal} {Phys.
  Rev. Lett.}\ }\textbf {\bibinfo {volume} {67}},\ \bibinfo {pages} {661--663}
  (\bibinfo {year} {1991})}\BibitemShut {NoStop}%
\bibitem [{\citenamefont {Jennewein}\ \emph {et~al.}(2000)\citenamefont
  {Jennewein}, \citenamefont {Simon}, \citenamefont {Weihs}, \citenamefont
  {Weinfurter},\ and\ \citenamefont {Zeilinger}}]{jennewein2000}%
  \BibitemOpen
  \bibfield  {author} {\bibinfo {author} {\bibfnamefont {Thomas}\ \bibnamefont
  {Jennewein}}, \bibinfo {author} {\bibfnamefont {Christoph}\ \bibnamefont
  {Simon}}, \bibinfo {author} {\bibfnamefont {Gregor}\ \bibnamefont {Weihs}},
  \bibinfo {author} {\bibfnamefont {Harald}\ \bibnamefont {Weinfurter}}, \ and\
  \bibinfo {author} {\bibfnamefont {Anton}\ \bibnamefont {Zeilinger}},\
  }\bibfield  {title} {\enquote {\bibinfo {title} {Quantum cryptography with
  entangled photons},}\ }\href {\doibase 10.1103/PhysRevLett.84.4729}
  {\bibfield  {journal} {\bibinfo  {journal} {Phys. Rev. Lett.}\ }\textbf
  {\bibinfo {volume} {84}},\ \bibinfo {pages} {4729--4732} (\bibinfo {year}
  {2000})}\BibitemShut {NoStop}%
\bibitem [{\citenamefont {Hubeny}(2015)}]{hubeny2015}%
  \BibitemOpen
  \bibfield  {author} {\bibinfo {author} {\bibfnamefont {Veronika~E}\
  \bibnamefont {Hubeny}},\ }\bibfield  {title} {\enquote {\bibinfo {title} {The
  {AdS}/{CFT} correspondence},}\ }\href {\doibase
  10.1088/0264-9381/32/12/124010} {\bibfield  {journal} {\bibinfo  {journal}
  {Classical Quant. Grav.}\ }\textbf {\bibinfo {volume} {32}},\ \bibinfo
  {pages} {124010} (\bibinfo {year} {2015})}\BibitemShut {NoStop}%
\bibitem [{\citenamefont {Pastawski}\ \emph {et~al.}(2015)\citenamefont
  {Pastawski}, \citenamefont {Yoshida}, \citenamefont {Harlow},\ and\
  \citenamefont {Preskill}}]{pastawski2015}%
  \BibitemOpen
  \bibfield  {author} {\bibinfo {author} {\bibfnamefont {Fernando}\
  \bibnamefont {Pastawski}}, \bibinfo {author} {\bibfnamefont {Beni}\
  \bibnamefont {Yoshida}}, \bibinfo {author} {\bibfnamefont {Daniel}\
  \bibnamefont {Harlow}}, \ and\ \bibinfo {author} {\bibfnamefont {John}\
  \bibnamefont {Preskill}},\ }\bibfield  {title} {\enquote {\bibinfo {title}
  {Holographic quantum error-correcting codes: toy models for the bulk/boundary
  correspondence},}\ }\href {\doibase 10.1007/JHEP06(2015)149} {\bibfield
  {journal} {\bibinfo  {journal} {J. High. Energy Phys.}\ }\textbf {\bibinfo
  {volume} {2015}},\ \bibinfo {pages} {149} (\bibinfo {year}
  {2015})}\BibitemShut {NoStop}%
\bibitem [{\citenamefont {Almheiri}\ \emph {et~al.}(2015)\citenamefont
  {Almheiri}, \citenamefont {Dong},\ and\ \citenamefont
  {Harlow}}]{almheiri2015}%
  \BibitemOpen
  \bibfield  {author} {\bibinfo {author} {\bibfnamefont {Ahmed}\ \bibnamefont
  {Almheiri}}, \bibinfo {author} {\bibfnamefont {Xi}~\bibnamefont {Dong}}, \
  and\ \bibinfo {author} {\bibfnamefont {Daniel}\ \bibnamefont {Harlow}},\
  }\bibfield  {title} {\enquote {\bibinfo {title} {Bulk locality and quantum
  error correction in ads/cft},}\ }\href {\doibase 10.1007/JHEP04(2015)163}
  {\bibfield  {journal} {\bibinfo  {journal} {J. High Energy Phys.}\ }\textbf
  {\bibinfo {volume} {4}},\ \bibinfo {pages} {163} (\bibinfo {year}
  {2015})}\BibitemShut {NoStop}%
\bibitem [{\citenamefont {{Jahn}}\ \emph {et~al.}(2017)\citenamefont {{Jahn}},
  \citenamefont {{Gluza}}, \citenamefont {{Pastawski}},\ and\ \citenamefont
  {{Eisert}}}]{jahn2017}%
  \BibitemOpen
  \bibfield  {author} {\bibinfo {author} {\bibfnamefont {A.}~\bibnamefont
  {{Jahn}}}, \bibinfo {author} {\bibfnamefont {M.}~\bibnamefont {{Gluza}}},
  \bibinfo {author} {\bibfnamefont {F.}~\bibnamefont {{Pastawski}}}, \ and\
  \bibinfo {author} {\bibfnamefont {J.}~\bibnamefont {{Eisert}}},\ }\bibfield
  {title} {\enquote {\bibinfo {title} {Holography and criticality in matchgate
  tensor networks},}\ }\href {https://arxiv.org/abs/1711.03109} {\bibfield
  {journal} {\bibinfo  {journal} {arXiv:1711.03109}\ } (\bibinfo {year}
  {2017})}\BibitemShut {NoStop}%
\bibitem [{\citenamefont {Page}\ and\ \citenamefont
  {Wootters}(1983)}]{page1983}%
  \BibitemOpen
  \bibfield  {author} {\bibinfo {author} {\bibfnamefont {Don~N.}\ \bibnamefont
  {Page}}\ and\ \bibinfo {author} {\bibfnamefont {William~K.}\ \bibnamefont
  {Wootters}},\ }\bibfield  {title} {\enquote {\bibinfo {title} {Evolution
  without evolution: Dynamics described by stationary observables},}\ }\href
  {\doibase 10.1103/PhysRevD.27.2885} {\bibfield  {journal} {\bibinfo
  {journal} {Phys. Rev. D}\ }\textbf {\bibinfo {volume} {27}},\ \bibinfo
  {pages} {2885--2892} (\bibinfo {year} {1983})}\BibitemShut {NoStop}%
\bibitem [{\citenamefont {Gambini}\ \emph {et~al.}(2009)\citenamefont
  {Gambini}, \citenamefont {Porto}, \citenamefont {Pullin},\ and\ \citenamefont
  {Torterolo}}]{gambini2009}%
  \BibitemOpen
  \bibfield  {author} {\bibinfo {author} {\bibfnamefont {Rodolfo}\ \bibnamefont
  {Gambini}}, \bibinfo {author} {\bibfnamefont {Rafael~A.}\ \bibnamefont
  {Porto}}, \bibinfo {author} {\bibfnamefont {Jorge}\ \bibnamefont {Pullin}}, \
  and\ \bibinfo {author} {\bibfnamefont {Sebasti\'an}\ \bibnamefont
  {Torterolo}},\ }\bibfield  {title} {\enquote {\bibinfo {title} {Conditional
  probabilities with dirac observables and the problem of time in quantum
  gravity},}\ }\href {\doibase 10.1103/PhysRevD.79.041501} {\bibfield
  {journal} {\bibinfo  {journal} {Phys. Rev. D}\ }\textbf {\bibinfo {volume}
  {79}},\ \bibinfo {pages} {041501} (\bibinfo {year} {2009})}\BibitemShut
  {NoStop}%
\bibitem [{\citenamefont {Moreva}\ \emph {et~al.}(2014)\citenamefont {Moreva},
  \citenamefont {Brida}, \citenamefont {Gramegna}, \citenamefont {Giovannetti},
  \citenamefont {Maccone},\ and\ \citenamefont {Genovese}}]{moreva2014}%
  \BibitemOpen
  \bibfield  {author} {\bibinfo {author} {\bibfnamefont {Ekaterina}\
  \bibnamefont {Moreva}}, \bibinfo {author} {\bibfnamefont {Giorgio}\
  \bibnamefont {Brida}}, \bibinfo {author} {\bibfnamefont {Marco}\ \bibnamefont
  {Gramegna}}, \bibinfo {author} {\bibfnamefont {Vittorio}\ \bibnamefont
  {Giovannetti}}, \bibinfo {author} {\bibfnamefont {Lorenzo}\ \bibnamefont
  {Maccone}}, \ and\ \bibinfo {author} {\bibfnamefont {Marco}\ \bibnamefont
  {Genovese}},\ }\bibfield  {title} {\enquote {\bibinfo {title} {Time from
  quantum entanglement: An experimental illustration},}\ }\href {\doibase
  10.1103/PhysRevA.89.052122} {\bibfield  {journal} {\bibinfo  {journal} {Phys.
  Rev. A}\ }\textbf {\bibinfo {volume} {89}},\ \bibinfo {pages} {052122}
  (\bibinfo {year} {2014})}\BibitemShut {NoStop}%
\bibitem [{\citenamefont {Lambert}\ \emph {et~al.}(2013)\citenamefont
  {Lambert}, \citenamefont {Chen}, \citenamefont {Cheng}, \citenamefont {Li},
  \citenamefont {Chen},\ and\ \citenamefont {Nori}}]{lambert2013}%
  \BibitemOpen
  \bibfield  {author} {\bibinfo {author} {\bibfnamefont {Neill}\ \bibnamefont
  {Lambert}}, \bibinfo {author} {\bibfnamefont {Yueh-Nan}\ \bibnamefont
  {Chen}}, \bibinfo {author} {\bibfnamefont {{Yuan Chung}}\ \bibnamefont
  {Cheng}}, \bibinfo {author} {\bibfnamefont {Che-Ming}\ \bibnamefont {Li}},
  \bibinfo {author} {\bibfnamefont {{Guang Yin}}\ \bibnamefont {Chen}}, \ and\
  \bibinfo {author} {\bibfnamefont {Franco}\ \bibnamefont {Nori}},\ }\bibfield
  {title} {\enquote {\bibinfo {title} {Quantum biology},}\ }\href
  {https://doi.org/10.1038/nphys2474} {\bibfield  {journal} {\bibinfo
  {journal} {Nat. Phys.}\ }\textbf {\bibinfo {volume} {9}},\ \bibinfo {pages}
  {10--18} (\bibinfo {year} {2013})}\BibitemShut {NoStop}%
\bibitem [{\citenamefont {Raimond}\ \emph {et~al.}(2001)\citenamefont
  {Raimond}, \citenamefont {Brune},\ and\ \citenamefont
  {Haroche}}]{raimond2001}%
  \BibitemOpen
  \bibfield  {author} {\bibinfo {author} {\bibfnamefont {J.~M.}\ \bibnamefont
  {Raimond}}, \bibinfo {author} {\bibfnamefont {M.}~\bibnamefont {Brune}}, \
  and\ \bibinfo {author} {\bibfnamefont {S.}~\bibnamefont {Haroche}},\
  }\bibfield  {title} {\enquote {\bibinfo {title} {Manipulating quantum
  entanglement with atoms and photons in a cavity},}\ }\href {\doibase
  10.1103/RevModPhys.73.565} {\bibfield  {journal} {\bibinfo  {journal} {Rev.
  Mod. Phys.}\ }\textbf {\bibinfo {volume} {73}},\ \bibinfo {pages} {565--582}
  (\bibinfo {year} {2001})}\BibitemShut {NoStop}%
\bibitem [{\citenamefont {Prevedel}\ \emph {et~al.}(2009)\citenamefont
  {Prevedel}, \citenamefont {Cronenberg}, \citenamefont {Tame}, \citenamefont
  {Paternostro}, \citenamefont {Walther}, \citenamefont {Kim},\ and\
  \citenamefont {Zeilinger}}]{prevedel2009}%
  \BibitemOpen
  \bibfield  {author} {\bibinfo {author} {\bibfnamefont {R.}~\bibnamefont
  {Prevedel}}, \bibinfo {author} {\bibfnamefont {G.}~\bibnamefont
  {Cronenberg}}, \bibinfo {author} {\bibfnamefont {M.~S.}\ \bibnamefont
  {Tame}}, \bibinfo {author} {\bibfnamefont {M.}~\bibnamefont {Paternostro}},
  \bibinfo {author} {\bibfnamefont {P.}~\bibnamefont {Walther}}, \bibinfo
  {author} {\bibfnamefont {M.~S.}\ \bibnamefont {Kim}}, \ and\ \bibinfo
  {author} {\bibfnamefont {A.}~\bibnamefont {Zeilinger}},\ }\bibfield  {title}
  {\enquote {\bibinfo {title} {Experimental realization of dicke states of up
  to six qubits for multiparty quantum networking},}\ }\href {\doibase
  10.1103/PhysRevLett.103.020503} {\bibfield  {journal} {\bibinfo  {journal}
  {Phys. Rev. Lett.}\ }\textbf {\bibinfo {volume} {103}},\ \bibinfo {pages}
  {020503} (\bibinfo {year} {2009})}\BibitemShut {NoStop}%
\bibitem [{\citenamefont {Barz}(2015)}]{barz2015}%
  \BibitemOpen
  \bibfield  {author} {\bibinfo {author} {\bibfnamefont {Stefanie}\
  \bibnamefont {Barz}},\ }\bibfield  {title} {\enquote {\bibinfo {title}
  {Quantum computing with photons: introduction to the circuit model, the
  one-way quantum computer, and the fundamental principles of photonic
  experiments},}\ }\href {http://stacks.iop.org/0953-4075/48/i=8/a=083001}
  {\bibfield  {journal} {\bibinfo  {journal} {J. Phys. B}\ }\textbf {\bibinfo
  {volume} {48}},\ \bibinfo {pages} {083001} (\bibinfo {year}
  {2015})}\BibitemShut {NoStop}%
\bibitem [{\citenamefont {Leibfried}\ \emph {et~al.}(2003)\citenamefont
  {Leibfried}, \citenamefont {Blatt}, \citenamefont {Monroe},\ and\
  \citenamefont {Wineland}}]{leibfried2003}%
  \BibitemOpen
  \bibfield  {author} {\bibinfo {author} {\bibfnamefont {D.}~\bibnamefont
  {Leibfried}}, \bibinfo {author} {\bibfnamefont {R.}~\bibnamefont {Blatt}},
  \bibinfo {author} {\bibfnamefont {C.}~\bibnamefont {Monroe}}, \ and\ \bibinfo
  {author} {\bibfnamefont {D.}~\bibnamefont {Wineland}},\ }\bibfield  {title}
  {\enquote {\bibinfo {title} {Quantum dynamics of single trapped ions},}\
  }\href {\doibase 10.1103/RevModPhys.75.281} {\bibfield  {journal} {\bibinfo
  {journal} {Rev. Mod. Phys.}\ }\textbf {\bibinfo {volume} {75}},\ \bibinfo
  {pages} {281--324} (\bibinfo {year} {2003})}\BibitemShut {NoStop}%
\bibitem [{\citenamefont {Leibfried}\ \emph {et~al.}(2005)\citenamefont
  {Leibfried}, \citenamefont {Knill}, \citenamefont {Seidelin}, \citenamefont
  {Britton}, \citenamefont {Blakestad}, \citenamefont {Chiaverini},
  \citenamefont {Hume}, \citenamefont {Itano}, \citenamefont {Jost},
  \citenamefont {Langer}, \citenamefont {Ozeri}, \citenamefont {Reichle},\ and\
  \citenamefont {Wineland}}]{leibfried2005}%
  \BibitemOpen
  \bibfield  {author} {\bibinfo {author} {\bibfnamefont {D.}~\bibnamefont
  {Leibfried}}, \bibinfo {author} {\bibfnamefont {E.}~\bibnamefont {Knill}},
  \bibinfo {author} {\bibfnamefont {S.}~\bibnamefont {Seidelin}}, \bibinfo
  {author} {\bibfnamefont {J.}~\bibnamefont {Britton}}, \bibinfo {author}
  {\bibfnamefont {R.~B.}\ \bibnamefont {Blakestad}}, \bibinfo {author}
  {\bibfnamefont {J.}~\bibnamefont {Chiaverini}}, \bibinfo {author}
  {\bibfnamefont {D.~B.}\ \bibnamefont {Hume}}, \bibinfo {author}
  {\bibfnamefont {W.~M.}\ \bibnamefont {Itano}}, \bibinfo {author}
  {\bibfnamefont {J.~D.}\ \bibnamefont {Jost}}, \bibinfo {author}
  {\bibfnamefont {C.}~\bibnamefont {Langer}}, \bibinfo {author} {\bibfnamefont
  {R.}~\bibnamefont {Ozeri}}, \bibinfo {author} {\bibfnamefont
  {R.}~\bibnamefont {Reichle}}, \ and\ \bibinfo {author} {\bibfnamefont
  {D.~J.}\ \bibnamefont {Wineland}},\ }\bibfield  {title} {\enquote {\bibinfo
  {title} {Creation of a six-atom 'schr{\"o}dinger cat' state},}\ }\href
  {\doibase 10.1038/nature04251} {\bibfield  {journal} {\bibinfo  {journal}
  {Nature}\ }\textbf {\bibinfo {volume} {438}},\ \bibinfo {pages} {639--642}
  (\bibinfo {year} {2005})}\BibitemShut {NoStop}%
\bibitem [{\citenamefont {Brown}\ \emph {et~al.}(2016)\citenamefont {Brown},
  \citenamefont {Kim},\ and\ \citenamefont {Monroe}}]{brown2016}%
  \BibitemOpen
  \bibfield  {author} {\bibinfo {author} {\bibfnamefont {Kenneth~R.}\
  \bibnamefont {Brown}}, \bibinfo {author} {\bibfnamefont {Jungsang}\
  \bibnamefont {Kim}}, \ and\ \bibinfo {author} {\bibfnamefont {Christopher}\
  \bibnamefont {Monroe}},\ }\bibfield  {title} {\enquote {\bibinfo {title}
  {Co-designing a scalable quantum computer with trapped atomic ions},}\ }\href
  {https://doi.org/10.1038/npjqi.2016.34} {\bibfield  {journal} {\bibinfo
  {journal} {Nature Phys. J. Quant. Inf.}\ }\textbf {\bibinfo {volume} {2}},\
  \bibinfo {pages} {16034 EP --} (\bibinfo {year} {2016})}\BibitemShut
  {NoStop}%
\bibitem [{\citenamefont {Mandel}\ \emph {et~al.}(2003)\citenamefont {Mandel},
  \citenamefont {Greiner}, \citenamefont {Widera}, \citenamefont {Rom},
  \citenamefont {Hansch},\ and\ \citenamefont {Bloch}}]{mandel2003}%
  \BibitemOpen
  \bibfield  {author} {\bibinfo {author} {\bibfnamefont {Olaf}\ \bibnamefont
  {Mandel}}, \bibinfo {author} {\bibfnamefont {Markus}\ \bibnamefont
  {Greiner}}, \bibinfo {author} {\bibfnamefont {Artur}\ \bibnamefont {Widera}},
  \bibinfo {author} {\bibfnamefont {Tim}\ \bibnamefont {Rom}}, \bibinfo
  {author} {\bibfnamefont {Theodor~W.}\ \bibnamefont {Hansch}}, \ and\ \bibinfo
  {author} {\bibfnamefont {Immanuel}\ \bibnamefont {Bloch}},\ }\bibfield
  {title} {\enquote {\bibinfo {title} {Controlled collisions for multi-particle
  entanglement of optically trapped atoms},}\ }\href {\doibase
  10.1038/nature02008} {\bibfield  {journal} {\bibinfo  {journal} {Nature}\
  }\textbf {\bibinfo {volume} {425}},\ \bibinfo {pages} {937--940} (\bibinfo
  {year} {2003})}\BibitemShut {NoStop}%
\bibitem [{\citenamefont {Bloch}(2005)}]{bloch2005}%
  \BibitemOpen
  \bibfield  {author} {\bibinfo {author} {\bibfnamefont {Immanuel}\
  \bibnamefont {Bloch}},\ }\bibfield  {title} {\enquote {\bibinfo {title}
  {Exploring quantum matter with ultracold atoms in optical lattices},}\ }\href
  {\doibase 10.1088/0953-4075/38/9/013} {\bibfield  {journal} {\bibinfo
  {journal} {J. Phys. B: At. Mol. Opt. Phys.}\ }\textbf {\bibinfo {volume}
  {38}},\ \bibinfo {pages} {S629--S643} (\bibinfo {year} {2005})}\BibitemShut
  {NoStop}%
\bibitem [{\citenamefont {Bloch}\ \emph {et~al.}(2008)\citenamefont {Bloch},
  \citenamefont {Dalibard},\ and\ \citenamefont {Zwerger}}]{bloch2008}%
  \BibitemOpen
  \bibfield  {author} {\bibinfo {author} {\bibfnamefont {Immanuel}\
  \bibnamefont {Bloch}}, \bibinfo {author} {\bibfnamefont {Jean}\ \bibnamefont
  {Dalibard}}, \ and\ \bibinfo {author} {\bibfnamefont {Wilhelm}\ \bibnamefont
  {Zwerger}},\ }\bibfield  {title} {\enquote {\bibinfo {title} {Many-body
  physics with ultracold gases},}\ }\href {\doibase 10.1103/RevModPhys.80.885}
  {\bibfield  {journal} {\bibinfo  {journal} {Rev. Mod. Phys.}\ }\textbf
  {\bibinfo {volume} {80}},\ \bibinfo {pages} {885--964} (\bibinfo {year}
  {2008})}\BibitemShut {NoStop}%
\bibitem [{\citenamefont {Clarke}\ and\ \citenamefont
  {Wilhelm}(2008)}]{clarke2008}%
  \BibitemOpen
  \bibfield  {author} {\bibinfo {author} {\bibfnamefont {John}\ \bibnamefont
  {Clarke}}\ and\ \bibinfo {author} {\bibfnamefont {Frank~K.}\ \bibnamefont
  {Wilhelm}},\ }\bibfield  {title} {\enquote {\bibinfo {title} {Superconducting
  quantum bits},}\ }\href {\doibase 10.1038/nature07128} {\bibfield  {journal}
  {\bibinfo  {journal} {Nature}\ }\textbf {\bibinfo {volume} {453}},\ \bibinfo
  {pages} {1031--1042} (\bibinfo {year} {2008})}\BibitemShut {NoStop}%
\bibitem [{\citenamefont {Barends}\ \emph {et~al.}(2014)\citenamefont
  {Barends}, \citenamefont {Kelly}, \citenamefont {Megrant}, \citenamefont
  {Veitia}, \citenamefont {Sank}, \citenamefont {Jeffrey}, \citenamefont
  {White}, \citenamefont {Mutus}, \citenamefont {Fowler}, \citenamefont
  {Campbell}, \citenamefont {Chen}, \citenamefont {Chen}, \citenamefont
  {Chiaro}, \citenamefont {Dunsworth}, \citenamefont {Neill}, \citenamefont
  {O'Malley}, \citenamefont {Roushan}, \citenamefont {Vainsencher},
  \citenamefont {Wenner}, \citenamefont {Korotkov}, \citenamefont {Cleland},\
  and\ \citenamefont {Martinis}}]{barends2014}%
  \BibitemOpen
  \bibfield  {author} {\bibinfo {author} {\bibfnamefont {R.}~\bibnamefont
  {Barends}}, \bibinfo {author} {\bibfnamefont {J.}~\bibnamefont {Kelly}},
  \bibinfo {author} {\bibfnamefont {A.}~\bibnamefont {Megrant}}, \bibinfo
  {author} {\bibfnamefont {A.}~\bibnamefont {Veitia}}, \bibinfo {author}
  {\bibfnamefont {D.}~\bibnamefont {Sank}}, \bibinfo {author} {\bibfnamefont
  {E.}~\bibnamefont {Jeffrey}}, \bibinfo {author} {\bibfnamefont {T.~C.}\
  \bibnamefont {White}}, \bibinfo {author} {\bibfnamefont {J.}~\bibnamefont
  {Mutus}}, \bibinfo {author} {\bibfnamefont {A.~G.}\ \bibnamefont {Fowler}},
  \bibinfo {author} {\bibfnamefont {B.}~\bibnamefont {Campbell}}, \bibinfo
  {author} {\bibfnamefont {Y.}~\bibnamefont {Chen}}, \bibinfo {author}
  {\bibfnamefont {Z.}~\bibnamefont {Chen}}, \bibinfo {author} {\bibfnamefont
  {B.}~\bibnamefont {Chiaro}}, \bibinfo {author} {\bibfnamefont
  {A.}~\bibnamefont {Dunsworth}}, \bibinfo {author} {\bibfnamefont
  {C.}~\bibnamefont {Neill}}, \bibinfo {author} {\bibfnamefont
  {P.}~\bibnamefont {O'Malley}}, \bibinfo {author} {\bibfnamefont
  {P.}~\bibnamefont {Roushan}}, \bibinfo {author} {\bibfnamefont
  {A.}~\bibnamefont {Vainsencher}}, \bibinfo {author} {\bibfnamefont
  {J.}~\bibnamefont {Wenner}}, \bibinfo {author} {\bibfnamefont {A.~N.}\
  \bibnamefont {Korotkov}}, \bibinfo {author} {\bibfnamefont {A.~N.}\
  \bibnamefont {Cleland}}, \ and\ \bibinfo {author} {\bibfnamefont {John~M.}\
  \bibnamefont {Martinis}},\ }\bibfield  {title} {\enquote {\bibinfo {title}
  {Superconducting quantum circuits at the surface code threshold for fault
  tolerance},}\ }\href {https://doi.org/10.1038/nature13171} {\bibfield
  {journal} {\bibinfo  {journal} {Nature}\ }\textbf {\bibinfo {volume} {508}},\
  \bibinfo {pages} {500} (\bibinfo {year} {2014})}\BibitemShut {NoStop}%
\bibitem [{\citenamefont {Negrevergne}\ \emph {et~al.}(2006)\citenamefont
  {Negrevergne}, \citenamefont {Mahesh}, \citenamefont {Ryan}, \citenamefont
  {Ditty}, \citenamefont {Cyr-Racine}, \citenamefont {Power}, \citenamefont
  {Boulant}, \citenamefont {Havel}, \citenamefont {Cory},\ and\ \citenamefont
  {Laflamme}}]{negrevergne2006}%
  \BibitemOpen
  \bibfield  {author} {\bibinfo {author} {\bibfnamefont {C.}~\bibnamefont
  {Negrevergne}}, \bibinfo {author} {\bibfnamefont {T.~S.}\ \bibnamefont
  {Mahesh}}, \bibinfo {author} {\bibfnamefont {C.~A.}\ \bibnamefont {Ryan}},
  \bibinfo {author} {\bibfnamefont {M.}~\bibnamefont {Ditty}}, \bibinfo
  {author} {\bibfnamefont {F.}~\bibnamefont {Cyr-Racine}}, \bibinfo {author}
  {\bibfnamefont {W.}~\bibnamefont {Power}}, \bibinfo {author} {\bibfnamefont
  {N.}~\bibnamefont {Boulant}}, \bibinfo {author} {\bibfnamefont
  {T.}~\bibnamefont {Havel}}, \bibinfo {author} {\bibfnamefont {D.~G.}\
  \bibnamefont {Cory}}, \ and\ \bibinfo {author} {\bibfnamefont
  {R.}~\bibnamefont {Laflamme}},\ }\bibfield  {title} {\enquote {\bibinfo
  {title} {Benchmarking quantum control methods on a 12-qubit system},}\ }\href
  {\doibase 10.1103/PhysRevLett.96.170501} {\bibfield  {journal} {\bibinfo
  {journal} {Phys. Rev. Lett.}\ }\textbf {\bibinfo {volume} {96}},\ \bibinfo
  {pages} {170501} (\bibinfo {year} {2006})}\BibitemShut {NoStop}%
\bibitem [{\citenamefont {Augusiak}\ \emph {et~al.}(2012)\citenamefont
  {Augusiak}, \citenamefont {Cucchietti},\ and\ \citenamefont
  {Lewenstein}}]{Augusiak2012}%
  \BibitemOpen
  \bibfield  {author} {\bibinfo {author} {\bibfnamefont {R.}~\bibnamefont
  {Augusiak}}, \bibinfo {author} {\bibfnamefont {F.~M.}\ \bibnamefont
  {Cucchietti}}, \ and\ \bibinfo {author} {\bibfnamefont {M.}~\bibnamefont
  {Lewenstein}},\ }\enquote {\bibinfo {title} {Many-body physics from a quantum
  information perspective},}\ in\ \href {\doibase 10.1007/978-3-642-10449-7_6}
  {\emph {\bibinfo {booktitle} {Modern Theories of Many-Particle Systems in
  Condensed Matter Physics}}},\ \bibinfo {editor} {edited by\ \bibinfo {editor}
  {\bibfnamefont {Daniel~C.}\ \bibnamefont {Cabra}}, \bibinfo {editor}
  {\bibfnamefont {Andreas}\ \bibnamefont {Honecker}}, \ and\ \bibinfo {editor}
  {\bibfnamefont {Pierre}\ \bibnamefont {Pujol}}}\ (\bibinfo  {publisher}
  {Springer Berlin Heidelberg},\ \bibinfo {address} {Berlin, Heidelberg},\
  \bibinfo {year} {2012})\ pp.\ \bibinfo {pages} {245--294}\BibitemShut
  {NoStop}%
\bibitem [{\citenamefont {Amico}\ \emph {et~al.}(2008)\citenamefont {Amico},
  \citenamefont {Fazio}, \citenamefont {Osterloh},\ and\ \citenamefont
  {Vedral}}]{amico2008}%
  \BibitemOpen
  \bibfield  {author} {\bibinfo {author} {\bibfnamefont {Luigi}\ \bibnamefont
  {Amico}}, \bibinfo {author} {\bibfnamefont {Rosario}\ \bibnamefont {Fazio}},
  \bibinfo {author} {\bibfnamefont {Andreas}\ \bibnamefont {Osterloh}}, \ and\
  \bibinfo {author} {\bibfnamefont {Vlatko}\ \bibnamefont {Vedral}},\
  }\bibfield  {title} {\enquote {\bibinfo {title} {Entanglement in many-body
  systems},}\ }\href {\doibase 10.1103/RevModPhys.80.517} {\bibfield  {journal}
  {\bibinfo  {journal} {Rev. Mod. Phys.}\ }\textbf {\bibinfo {volume} {80}},\
  \bibinfo {pages} {517--576} (\bibinfo {year} {2008})}\BibitemShut {NoStop}%
\bibitem [{\citenamefont {Chiara}\ and\ \citenamefont
  {Sanpera}(2018)}]{dechiara2018}%
  \BibitemOpen
  \bibfield  {author} {\bibinfo {author} {\bibfnamefont {Gabriele~De}\
  \bibnamefont {Chiara}}\ and\ \bibinfo {author} {\bibfnamefont {Anna}\
  \bibnamefont {Sanpera}},\ }\href {\doibase 10.1088/1361-6633/aabf61}
  {\bibfield  {journal} {\bibinfo  {journal} {Reports on Progress in Physics}\
  }\textbf {\bibinfo {volume} {81}},\ \bibinfo {pages} {074002} (\bibinfo
  {year} {2018})}\BibitemShut {NoStop}%
\bibitem [{\citenamefont {Horodecki}(2001)}]{horodecki2001a}%
  \BibitemOpen
  \bibfield  {author} {\bibinfo {author} {\bibfnamefont {Micha\l{}}\
  \bibnamefont {Horodecki}},\ }\bibfield  {title} {\enquote {\bibinfo {title}
  {Entanglement measures},}\ }\href@noop {} {\bibfield  {journal} {\bibinfo
  {journal} {Quantum Inf. Comput.}\ }\textbf {\bibinfo {volume} {1}},\ \bibinfo
  {pages} {3--26} (\bibinfo {year} {2001})}\BibitemShut {NoStop}%
\bibitem [{\citenamefont {Greenberger}\ \emph {et~al.}(1989)\citenamefont
  {Greenberger}, \citenamefont {Horne},\ and\ \citenamefont
  {Zeilinger}}]{greenberger1989}%
  \BibitemOpen
  \bibfield  {author} {\bibinfo {author} {\bibfnamefont {D.~M.}\ \bibnamefont
  {Greenberger}}, \bibinfo {author} {\bibfnamefont {M.~A.}\ \bibnamefont
  {Horne}}, \ and\ \bibinfo {author} {\bibfnamefont {A.}~\bibnamefont
  {Zeilinger}},\ }\href
  {http://inis.iaea.org/search/search.aspx?orig_q=RN:22064349} {\emph {\bibinfo
  {title} {Bell's theorem, quantum theory and conceptions of the universe}}}\
  (\bibinfo  {publisher} {Kluwer},\ \bibinfo {address} {Netherlands},\ \bibinfo
  {year} {1989})\BibitemShut {NoStop}%
\bibitem [{\citenamefont {Hein}\ \emph {et~al.}(2006)\citenamefont {Hein},
  \citenamefont {D\"{u}r}, \citenamefont {Eisert}, \citenamefont {Raussendorf},
  \citenamefont {Van~den Nest},\ and\ \citenamefont {J.~Briegel}}]{hein2006}%
  \BibitemOpen
  \bibfield  {author} {\bibinfo {author} {\bibfnamefont {M}~\bibnamefont
  {Hein}}, \bibinfo {author} {\bibfnamefont {W}~\bibnamefont {D\"{u}r}},
  \bibinfo {author} {\bibfnamefont {Jens}\ \bibnamefont {Eisert}}, \bibinfo
  {author} {\bibfnamefont {Robert}\ \bibnamefont {Raussendorf}}, \bibinfo
  {author} {\bibfnamefont {M}~\bibnamefont {Van~den Nest}}, \ and\ \bibinfo
  {author} {\bibfnamefont {H}~\bibnamefont {J.~Briegel}},\ }\bibfield  {title}
  {\enquote {\bibinfo {title} {Entanglement in graph states and its
  applications},}\ }\href {https://arxiv.org/abs/quant-ph/0602096} {\bibfield
  {journal} {\bibinfo  {journal} {arXiv:quant-ph/0602096}\ } (\bibinfo {year}
  {2006})}\BibitemShut {NoStop}%
\bibitem [{\citenamefont {Raussendorf}\ \emph {et~al.}(2003)\citenamefont
  {Raussendorf}, \citenamefont {Browne},\ and\ \citenamefont
  {Briegel}}]{raussendorf2003}%
  \BibitemOpen
  \bibfield  {author} {\bibinfo {author} {\bibfnamefont {Robert}\ \bibnamefont
  {Raussendorf}}, \bibinfo {author} {\bibfnamefont {Daniel~E.}\ \bibnamefont
  {Browne}}, \ and\ \bibinfo {author} {\bibfnamefont {Hans~J.}\ \bibnamefont
  {Briegel}},\ }\bibfield  {title} {\enquote {\bibinfo {title}
  {Measurement-based quantum computation on cluster states},}\ }\href {\doibase
  10.1103/PhysRevA.68.022312} {\bibfield  {journal} {\bibinfo  {journal} {Phys.
  Rev. A}\ }\textbf {\bibinfo {volume} {68}},\ \bibinfo {pages} {022312}
  (\bibinfo {year} {2003})}\BibitemShut {NoStop}%
\bibitem [{\citenamefont {Fujii}(2015)}]{fujii2015}%
  \BibitemOpen
  \bibfield  {author} {\bibinfo {author} {\bibfnamefont {Keisuke}\ \bibnamefont
  {Fujii}},\ }\bibfield  {title} {\enquote {\bibinfo {title} {Quantum
  computation with topological codes: from qubit to topological
  fault-tolerance},}\ }\href {https://arxiv.org/abs/1504.01444} {\bibfield
  {journal} {\bibinfo  {journal} {arXiv:1504.01444}\ } (\bibinfo {year}
  {2015})}\BibitemShut {NoStop}%
\bibitem [{\citenamefont {DiVincenzo}\ \emph {et~al.}(1998)\citenamefont
  {DiVincenzo}, \citenamefont {Fuchs}, \citenamefont {Mabuchi}, \citenamefont
  {Smolin}, \citenamefont {Thapliyal},\ and\ \citenamefont
  {Uhlmann}}]{divincenzo1998}%
  \BibitemOpen
  \bibfield  {author} {\bibinfo {author} {\bibfnamefont {D.~P.}\ \bibnamefont
  {DiVincenzo}}, \bibinfo {author} {\bibfnamefont {C.~A.}\ \bibnamefont
  {Fuchs}}, \bibinfo {author} {\bibfnamefont {H.}~\bibnamefont {Mabuchi}},
  \bibinfo {author} {\bibfnamefont {J.~A.}\ \bibnamefont {Smolin}}, \bibinfo
  {author} {\bibfnamefont {A.}~\bibnamefont {Thapliyal}}, \ and\ \bibinfo
  {author} {\bibfnamefont {A.}~\bibnamefont {Uhlmann}},\ }\bibfield  {title}
  {\enquote {\bibinfo {title} {Entanglement of assistance},}\ }\href
  {https://arxiv.org/abs/quant-ph/9803033} {\bibfield  {journal} {\bibinfo
  {journal} {arXiv:quant-ph/9803033}\ } (\bibinfo {year} {1998})}\BibitemShut
  {NoStop}%
\bibitem [{\citenamefont {Verstraete}\ \emph
  {et~al.}(2004{\natexlab{a}})\citenamefont {Verstraete}, \citenamefont
  {Popp},\ and\ \citenamefont {Cirac}}]{verstraete2004}%
  \BibitemOpen
  \bibfield  {author} {\bibinfo {author} {\bibfnamefont {F.}~\bibnamefont
  {Verstraete}}, \bibinfo {author} {\bibfnamefont {M.}~\bibnamefont {Popp}}, \
  and\ \bibinfo {author} {\bibfnamefont {J.~I.}\ \bibnamefont {Cirac}},\
  }\bibfield  {title} {\enquote {\bibinfo {title} {Entanglement versus
  correlations in spin systems},}\ }\href {\doibase
  10.1103/PhysRevLett.92.027901} {\bibfield  {journal} {\bibinfo  {journal}
  {Phys. Rev. Lett.}\ }\textbf {\bibinfo {volume} {92}},\ \bibinfo {pages}
  {027901} (\bibinfo {year} {2004}{\natexlab{a}})}\BibitemShut {NoStop}%
\bibitem [{\citenamefont {Verstraete}\ \emph
  {et~al.}(2004{\natexlab{b}})\citenamefont {Verstraete}, \citenamefont
  {Mart\'{\i}n-Delgado},\ and\ \citenamefont {Cirac}}]{verstraete2004a}%
  \BibitemOpen
  \bibfield  {author} {\bibinfo {author} {\bibfnamefont {F.}~\bibnamefont
  {Verstraete}}, \bibinfo {author} {\bibfnamefont {M.~A.}\ \bibnamefont
  {Mart\'{\i}n-Delgado}}, \ and\ \bibinfo {author} {\bibfnamefont {J.~I.}\
  \bibnamefont {Cirac}},\ }\bibfield  {title} {\enquote {\bibinfo {title}
  {Diverging entanglement length in gapped quantum spin systems},}\ }\href
  {\doibase 10.1103/PhysRevLett.92.087201} {\bibfield  {journal} {\bibinfo
  {journal} {Phys. Rev. Lett.}\ }\textbf {\bibinfo {volume} {92}},\ \bibinfo
  {pages} {087201} (\bibinfo {year} {2004}{\natexlab{b}})}\BibitemShut
  {NoStop}%
\bibitem [{\citenamefont {Popp}\ \emph {et~al.}(2005)\citenamefont {Popp},
  \citenamefont {Verstraete}, \citenamefont {Mart\'{\i}n-Delgado},\ and\
  \citenamefont {Cirac}}]{popp2005}%
  \BibitemOpen
  \bibfield  {author} {\bibinfo {author} {\bibfnamefont {M.}~\bibnamefont
  {Popp}}, \bibinfo {author} {\bibfnamefont {F.}~\bibnamefont {Verstraete}},
  \bibinfo {author} {\bibfnamefont {M.~A.}\ \bibnamefont
  {Mart\'{\i}n-Delgado}}, \ and\ \bibinfo {author} {\bibfnamefont {J.~I.}\
  \bibnamefont {Cirac}},\ }\bibfield  {title} {\enquote {\bibinfo {title}
  {Localizable entanglement},}\ }\href {\doibase 10.1103/PhysRevA.71.042306}
  {\bibfield  {journal} {\bibinfo  {journal} {Phys. Rev. A}\ }\textbf {\bibinfo
  {volume} {71}},\ \bibinfo {pages} {042306} (\bibinfo {year}
  {2005})}\BibitemShut {NoStop}%
\bibitem [{\citenamefont {Sadhukhan}\ \emph {et~al.}(2017)\citenamefont
  {Sadhukhan}, \citenamefont {Roy}, \citenamefont {Pal}, \citenamefont
  {Rakshit}, \citenamefont {Sen(De)},\ and\ \citenamefont
  {Sen}}]{sadhukhan2017}%
  \BibitemOpen
  \bibfield  {author} {\bibinfo {author} {\bibfnamefont {Debasis}\ \bibnamefont
  {Sadhukhan}}, \bibinfo {author} {\bibfnamefont {Sudipto~Singha}\ \bibnamefont
  {Roy}}, \bibinfo {author} {\bibfnamefont {Amit~Kumar}\ \bibnamefont {Pal}},
  \bibinfo {author} {\bibfnamefont {Debraj}\ \bibnamefont {Rakshit}}, \bibinfo
  {author} {\bibfnamefont {Aditi}\ \bibnamefont {Sen(De)}}, \ and\ \bibinfo
  {author} {\bibfnamefont {Ujjwal}\ \bibnamefont {Sen}},\ }\bibfield  {title}
  {\enquote {\bibinfo {title} {Multipartite entanglement accumulation in
  quantum states: Localizable generalized geometric measure},}\ }\href
  {\doibase 10.1103/PhysRevA.95.022301} {\bibfield  {journal} {\bibinfo
  {journal} {Phys. Rev. A}\ }\textbf {\bibinfo {volume} {95}},\ \bibinfo
  {pages} {022301} (\bibinfo {year} {2017})}\BibitemShut {NoStop}%
\bibitem [{\citenamefont {Amaro}\ \emph {et~al.}(2018)\citenamefont {Amaro},
  \citenamefont {M\"{u}ller},\ and\ \citenamefont {Pal}}]{amaro2018}%
  \BibitemOpen
  \bibfield  {author} {\bibinfo {author} {\bibfnamefont {David}\ \bibnamefont
  {Amaro}}, \bibinfo {author} {\bibfnamefont {Markus}\ \bibnamefont
  {M\"{u}ller}}, \ and\ \bibinfo {author} {\bibfnamefont {Amit~Kumar}\
  \bibnamefont {Pal}},\ }\bibfield  {title} {\enquote {\bibinfo {title}
  {Estimating localizable entanglement from witnesses},}\ }\href {\doibase
  10.1088/1367-2630/aac485} {\bibfield  {journal} {\bibinfo  {journal} {New J.
  Phys.}\ }\textbf {\bibinfo {volume} {20}},\ \bibinfo {pages} {063017}
  (\bibinfo {year} {2018})}\BibitemShut {NoStop}%
\bibitem [{\citenamefont {Amaro}\ \emph {et~al.}(2020)\citenamefont {Amaro},
  \citenamefont {M\"{u}ller},\ and\ \citenamefont {Pal}}]{amaro2020a}%
  \BibitemOpen
  \bibfield  {author} {\bibinfo {author} {\bibfnamefont {David}\ \bibnamefont
  {Amaro}}, \bibinfo {author} {\bibfnamefont {Markus}\ \bibnamefont
  {M\"{u}ller}}, \ and\ \bibinfo {author} {\bibfnamefont {Amit~Kumar}\
  \bibnamefont {Pal}},\ }\bibfield  {title} {\enquote {\bibinfo {title}
  {Scalable characterization of localizable entanglement in noisy topological
  quantum codes},}\ }\href {\doibase 10.1088/1367-2630/ab84b3} {\bibfield
  {journal} {\bibinfo  {journal} {New Journal of Physics}\ }\textbf {\bibinfo
  {volume} {22}},\ \bibinfo {pages} {053038} (\bibinfo {year}
  {2020})}\BibitemShut {NoStop}%
\bibitem [{\citenamefont {Jin}\ and\ \citenamefont {Korepin}(2004)}]{jin2004}%
  \BibitemOpen
  \bibfield  {author} {\bibinfo {author} {\bibfnamefont {B.-Q.}\ \bibnamefont
  {Jin}}\ and\ \bibinfo {author} {\bibfnamefont {V.~E.}\ \bibnamefont
  {Korepin}},\ }\bibfield  {title} {\enquote {\bibinfo {title} {Localizable
  entanglement in antiferromagnetic spin chains},}\ }\href {\doibase
  10.1103/PhysRevA.69.062314} {\bibfield  {journal} {\bibinfo  {journal} {Phys.
  Rev. A}\ }\textbf {\bibinfo {volume} {69}},\ \bibinfo {pages} {062314}
  (\bibinfo {year} {2004})}\BibitemShut {NoStop}%
\bibitem [{\citenamefont {Skr\o{}vseth}\ and\ \citenamefont
  {Bartlett}(2009)}]{skrovseth2009}%
  \BibitemOpen
  \bibfield  {author} {\bibinfo {author} {\bibfnamefont {Stein~Olav}\
  \bibnamefont {Skr\o{}vseth}}\ and\ \bibinfo {author} {\bibfnamefont
  {Stephen~D.}\ \bibnamefont {Bartlett}},\ }\bibfield  {title} {\enquote
  {\bibinfo {title} {Phase transitions and localizable entanglement in
  cluster-state spin chains with ising couplings and local fields},}\ }\href
  {\doibase 10.1103/PhysRevA.80.022316} {\bibfield  {journal} {\bibinfo
  {journal} {Phys. Rev. A}\ }\textbf {\bibinfo {volume} {80}},\ \bibinfo
  {pages} {022316} (\bibinfo {year} {2009})}\BibitemShut {NoStop}%
\bibitem [{\citenamefont {Smacchia}\ \emph {et~al.}(2011)\citenamefont
  {Smacchia}, \citenamefont {Amico}, \citenamefont {Facchi}, \citenamefont
  {Fazio}, \citenamefont {Florio}, \citenamefont {Pascazio},\ and\
  \citenamefont {Vedral}}]{smacchia2011}%
  \BibitemOpen
  \bibfield  {author} {\bibinfo {author} {\bibfnamefont {Pietro}\ \bibnamefont
  {Smacchia}}, \bibinfo {author} {\bibfnamefont {Luigi}\ \bibnamefont {Amico}},
  \bibinfo {author} {\bibfnamefont {Paolo}\ \bibnamefont {Facchi}}, \bibinfo
  {author} {\bibfnamefont {Rosario}\ \bibnamefont {Fazio}}, \bibinfo {author}
  {\bibfnamefont {Giuseppe}\ \bibnamefont {Florio}}, \bibinfo {author}
  {\bibfnamefont {Saverio}\ \bibnamefont {Pascazio}}, \ and\ \bibinfo {author}
  {\bibfnamefont {Vlatko}\ \bibnamefont {Vedral}},\ }\bibfield  {title}
  {\enquote {\bibinfo {title} {Statistical mechanics of the cluster ising
  model},}\ }\href {\doibase 10.1103/PhysRevA.84.022304} {\bibfield  {journal}
  {\bibinfo  {journal} {Phys. Rev. A}\ }\textbf {\bibinfo {volume} {84}},\
  \bibinfo {pages} {022304} (\bibinfo {year} {2011})}\BibitemShut {NoStop}%
\bibitem [{\citenamefont {Montes}\ and\ \citenamefont
  {Hamma}(2012)}]{montes2012}%
  \BibitemOpen
  \bibfield  {author} {\bibinfo {author} {\bibfnamefont {Sebasti\'an}\
  \bibnamefont {Montes}}\ and\ \bibinfo {author} {\bibfnamefont {Alioscia}\
  \bibnamefont {Hamma}},\ }\bibfield  {title} {\enquote {\bibinfo {title}
  {Phase diagram and quench dynamics of the cluster-$xy$ spin chain},}\ }\href
  {\doibase 10.1103/PhysRevE.86.021101} {\bibfield  {journal} {\bibinfo
  {journal} {Phys. Rev. E}\ }\textbf {\bibinfo {volume} {86}},\ \bibinfo
  {pages} {021101} (\bibinfo {year} {2012})}\BibitemShut {NoStop}%
\bibitem [{\citenamefont {Ac{\'i}n}\ \emph {et~al.}(2007)\citenamefont
  {Ac{\'i}n}, \citenamefont {Cirac},\ and\ \citenamefont
  {Lewenstein}}]{acin2007}%
  \BibitemOpen
  \bibfield  {author} {\bibinfo {author} {\bibfnamefont {Antonio}\ \bibnamefont
  {Ac{\'i}n}}, \bibinfo {author} {\bibfnamefont {J.~Ignacio}\ \bibnamefont
  {Cirac}}, \ and\ \bibinfo {author} {\bibfnamefont {Maciej}\ \bibnamefont
  {Lewenstein}},\ }\bibfield  {title} {\enquote {\bibinfo {title} {Entanglement
  percolation in quantum networks},}\ }\href {https://doi.org/10.1038/nphys549}
  {\bibfield  {journal} {\bibinfo  {journal} {Nat. Phys.}\ }\textbf {\bibinfo
  {volume} {3}},\ \bibinfo {pages} {256} (\bibinfo {year} {2007})}\BibitemShut
  {NoStop}%
\bibitem [{\citenamefont {Popescu}\ and\ \citenamefont
  {Rohrlich}(1997)}]{popescu1997}%
  \BibitemOpen
  \bibfield  {author} {\bibinfo {author} {\bibfnamefont {Sandu}\ \bibnamefont
  {Popescu}}\ and\ \bibinfo {author} {\bibfnamefont {Daniel}\ \bibnamefont
  {Rohrlich}},\ }\bibfield  {title} {\enquote {\bibinfo {title} {Thermodynamics
  and the measure of entanglement},}\ }\href {\doibase
  10.1103/PhysRevA.56.R3319} {\bibfield  {journal} {\bibinfo  {journal} {Phys.
  Rev. A}\ }\textbf {\bibinfo {volume} {56}},\ \bibinfo {pages} {R3319--R3321}
  (\bibinfo {year} {1997})}\BibitemShut {NoStop}%
\bibitem [{\citenamefont {Vedral}\ and\ \citenamefont
  {Plenio}(1998)}]{vedral1998}%
  \BibitemOpen
  \bibfield  {author} {\bibinfo {author} {\bibfnamefont {V.}~\bibnamefont
  {Vedral}}\ and\ \bibinfo {author} {\bibfnamefont {M.~B.}\ \bibnamefont
  {Plenio}},\ }\bibfield  {title} {\enquote {\bibinfo {title} {Entanglement
  measures and purification procedures},}\ }\href {\doibase
  10.1103/PhysRevA.57.1619} {\bibfield  {journal} {\bibinfo  {journal} {Phys.
  Rev. A}\ }\textbf {\bibinfo {volume} {57}},\ \bibinfo {pages} {1619--1633}
  (\bibinfo {year} {1998})}\BibitemShut {NoStop}%
\bibitem [{\citenamefont {Vidal}(2000)}]{vidal2000}%
  \BibitemOpen
  \bibfield  {author} {\bibinfo {author} {\bibfnamefont {Guifr\'{e}}\
  \bibnamefont {Vidal}},\ }\bibfield  {title} {\enquote {\bibinfo {title}
  {Entanglement monotones},}\ }\href {\doibase 10.1080/09500340008244048}
  {\bibfield  {journal} {\bibinfo  {journal} {Journal of Modern Optics}\
  }\textbf {\bibinfo {volume} {47}},\ \bibinfo {pages} {355--376} (\bibinfo
  {year} {2000})}\BibitemShut {NoStop}%
\bibitem [{\citenamefont {Smolin}\ \emph {et~al.}(2005)\citenamefont {Smolin},
  \citenamefont {Verstraete},\ and\ \citenamefont {Winter}}]{smolin2005}%
  \BibitemOpen
  \bibfield  {author} {\bibinfo {author} {\bibfnamefont {John~A.}\ \bibnamefont
  {Smolin}}, \bibinfo {author} {\bibfnamefont {Frank}\ \bibnamefont
  {Verstraete}}, \ and\ \bibinfo {author} {\bibfnamefont {Andreas}\
  \bibnamefont {Winter}},\ }\bibfield  {title} {\enquote {\bibinfo {title}
  {Entanglement of assistance and multipartite state distillation},}\ }\href
  {\doibase 10.1103/PhysRevA.72.052317} {\bibfield  {journal} {\bibinfo
  {journal} {Phys. Rev. A}\ }\textbf {\bibinfo {volume} {72}},\ \bibinfo
  {pages} {052317} (\bibinfo {year} {2005})}\BibitemShut {NoStop}%
\bibitem [{\citenamefont {Yang}\ and\ \citenamefont {Eisert}(2009)}]{yang2009}%
  \BibitemOpen
  \bibfield  {author} {\bibinfo {author} {\bibfnamefont {Dong}\ \bibnamefont
  {Yang}}\ and\ \bibinfo {author} {\bibfnamefont {Jens}\ \bibnamefont
  {Eisert}},\ }\bibfield  {title} {\enquote {\bibinfo {title} {Entanglement
  combing},}\ }\href {\doibase 10.1103/PhysRevLett.103.220501} {\bibfield
  {journal} {\bibinfo  {journal} {Phys. Rev. Lett.}\ }\textbf {\bibinfo
  {volume} {103}},\ \bibinfo {pages} {220501} (\bibinfo {year}
  {2009})}\BibitemShut {NoStop}%
\bibitem [{\citenamefont {Pollock}\ \emph {et~al.}(2021)\citenamefont
  {Pollock}, \citenamefont {Wang},\ and\ \citenamefont
  {Chitambar}}]{pollock2021}%
  \BibitemOpen
  \bibfield  {author} {\bibinfo {author} {\bibfnamefont {Kl\'ee}\ \bibnamefont
  {Pollock}}, \bibinfo {author} {\bibfnamefont {Ge}~\bibnamefont {Wang}}, \
  and\ \bibinfo {author} {\bibfnamefont {Eric}\ \bibnamefont {Chitambar}},\
  }\bibfield  {title} {\enquote {\bibinfo {title} {Entanglement of assistance
  in three-qubit systems},}\ }\href {\doibase 10.1103/PhysRevA.103.032428}
  {\bibfield  {journal} {\bibinfo  {journal} {Phys. Rev. A}\ }\textbf {\bibinfo
  {volume} {103}},\ \bibinfo {pages} {032428} (\bibinfo {year}
  {2021})}\BibitemShut {NoStop}%
\bibitem [{\citenamefont {D\"ur}\ \emph
  {et~al.}(2000{\natexlab{a}})\citenamefont {D\"ur}, \citenamefont {Vidal},\
  and\ \citenamefont {Cirac}}]{dur2000}%
  \BibitemOpen
  \bibfield  {author} {\bibinfo {author} {\bibfnamefont {W.}~\bibnamefont
  {D\"ur}}, \bibinfo {author} {\bibfnamefont {G.}~\bibnamefont {Vidal}}, \ and\
  \bibinfo {author} {\bibfnamefont {J.~I.}\ \bibnamefont {Cirac}},\ }\bibfield
  {title} {\enquote {\bibinfo {title} {Three qubits can be entangled in two
  inequivalent ways},}\ }\href {\doibase 10.1103/PhysRevA.62.062314} {\bibfield
   {journal} {\bibinfo  {journal} {Phys. Rev. A}\ }\textbf {\bibinfo {volume}
  {62}},\ \bibinfo {pages} {062314} (\bibinfo {year}
  {2000}{\natexlab{a}})}\BibitemShut {NoStop}%
\bibitem [{\citenamefont {Sen(De)}\ \emph {et~al.}(2003)\citenamefont
  {Sen(De)}, \citenamefont {Sen}, \citenamefont {Wie\ifmmode~\acute{s}\else
  \'{s}\fi{}niak}, \citenamefont {Kaszlikowski},\ and\ \citenamefont
  {\ifmmode~\dot{Z}\else \.{Z}\fi{}ukowski}}]{sende2003}%
  \BibitemOpen
  \bibfield  {author} {\bibinfo {author} {\bibfnamefont {Aditi}\ \bibnamefont
  {Sen(De)}}, \bibinfo {author} {\bibfnamefont {Ujjwal}\ \bibnamefont {Sen}},
  \bibinfo {author} {\bibfnamefont {Marcin}\ \bibnamefont
  {Wie\ifmmode~\acute{s}\else \'{s}\fi{}niak}}, \bibinfo {author}
  {\bibfnamefont {Dagomir}\ \bibnamefont {Kaszlikowski}}, \ and\ \bibinfo
  {author} {\bibfnamefont {Marek}\ \bibnamefont {\ifmmode~\dot{Z}\else
  \.{Z}\fi{}ukowski}},\ }\bibfield  {title} {\enquote {\bibinfo {title}
  {Multiqubit w states lead to stronger nonclassicality than
  greenberger-horne-zeilinger states},}\ }\href {\doibase
  10.1103/PhysRevA.68.062306} {\bibfield  {journal} {\bibinfo  {journal} {Phys.
  Rev. A}\ }\textbf {\bibinfo {volume} {68}},\ \bibinfo {pages} {062306}
  (\bibinfo {year} {2003})}\BibitemShut {NoStop}%
\bibitem [{\citenamefont {Dicke}(1954)}]{dicke1954}%
  \BibitemOpen
  \bibfield  {author} {\bibinfo {author} {\bibfnamefont {R.~H.}\ \bibnamefont
  {Dicke}},\ }\bibfield  {title} {\enquote {\bibinfo {title} {Coherence in
  spontaneous radiation processes},}\ }\href {\doibase 10.1103/PhysRev.93.99}
  {\bibfield  {journal} {\bibinfo  {journal} {Phys. Rev.}\ }\textbf {\bibinfo
  {volume} {93}},\ \bibinfo {pages} {99--110} (\bibinfo {year}
  {1954})}\BibitemShut {NoStop}%
\bibitem [{\citenamefont {Bergmann}\ and\ \citenamefont
  {Gühne}(2013)}]{bergmann2013}%
  \BibitemOpen
  \bibfield  {author} {\bibinfo {author} {\bibfnamefont {Marcel}\ \bibnamefont
  {Bergmann}}\ and\ \bibinfo {author} {\bibfnamefont {Otfried}\ \bibnamefont
  {Gühne}},\ }\bibfield  {title} {\enquote {\bibinfo {title} {Entanglement
  criteria for dicke states},}\ }\href {\doibase
  10.1088/1751-8113/46/38/385304} {\bibfield  {journal} {\bibinfo  {journal}
  {Journal of Physics A: Mathematical and Theoretical}\ }\textbf {\bibinfo
  {volume} {46}},\ \bibinfo {pages} {385304} (\bibinfo {year}
  {2013})}\BibitemShut {NoStop}%
\bibitem [{\citenamefont {L\"ucke}\ \emph {et~al.}(2014)\citenamefont
  {L\"ucke}, \citenamefont {Peise}, \citenamefont {Vitagliano}, \citenamefont
  {Arlt}, \citenamefont {Santos}, \citenamefont {T\'oth},\ and\ \citenamefont
  {Klempt}}]{lucke2014}%
  \BibitemOpen
  \bibfield  {author} {\bibinfo {author} {\bibfnamefont {Bernd}\ \bibnamefont
  {L\"ucke}}, \bibinfo {author} {\bibfnamefont {Jan}\ \bibnamefont {Peise}},
  \bibinfo {author} {\bibfnamefont {Giuseppe}\ \bibnamefont {Vitagliano}},
  \bibinfo {author} {\bibfnamefont {Jan}\ \bibnamefont {Arlt}}, \bibinfo
  {author} {\bibfnamefont {Luis}\ \bibnamefont {Santos}}, \bibinfo {author}
  {\bibfnamefont {G\'eza}\ \bibnamefont {T\'oth}}, \ and\ \bibinfo {author}
  {\bibfnamefont {Carsten}\ \bibnamefont {Klempt}},\ }\bibfield  {title}
  {\enquote {\bibinfo {title} {Detecting multiparticle entanglement of dicke
  states},}\ }\href {\doibase 10.1103/PhysRevLett.112.155304} {\bibfield
  {journal} {\bibinfo  {journal} {Phys. Rev. Lett.}\ }\textbf {\bibinfo
  {volume} {112}},\ \bibinfo {pages} {155304} (\bibinfo {year}
  {2014})}\BibitemShut {NoStop}%
\bibitem [{\citenamefont {Kumar}\ \emph {et~al.}(2017)\citenamefont {Kumar},
  \citenamefont {Dhar}, \citenamefont {Prabhu}, \citenamefont {Sen(De)},\ and\
  \citenamefont {Sen}}]{kumar2017}%
  \BibitemOpen
  \bibfield  {author} {\bibinfo {author} {\bibfnamefont {Asutosh}\ \bibnamefont
  {Kumar}}, \bibinfo {author} {\bibfnamefont {Himadri~Shekhar}\ \bibnamefont
  {Dhar}}, \bibinfo {author} {\bibfnamefont {R.}~\bibnamefont {Prabhu}},
  \bibinfo {author} {\bibfnamefont {Aditi}\ \bibnamefont {Sen(De)}}, \ and\
  \bibinfo {author} {\bibfnamefont {Ujjwal}\ \bibnamefont {Sen}},\ }\bibfield
  {title} {\enquote {\bibinfo {title} {Forbidden regimes in the distribution of
  bipartite quantum correlations due to multiparty entanglement},}\ }\href
  {\doibase https://doi.org/10.1016/j.physleta.2017.03.026} {\bibfield
  {journal} {\bibinfo  {journal} {Physics Letters A}\ }\textbf {\bibinfo
  {volume} {381}},\ \bibinfo {pages} {1701--1709} (\bibinfo {year}
  {2017})}\BibitemShut {NoStop}%
\bibitem [{\citenamefont {Nielsen}\ and\ \citenamefont
  {Chuang}(2010)}]{nielsen2010}%
  \BibitemOpen
  \bibfield  {author} {\bibinfo {author} {\bibfnamefont {M.~A.}\ \bibnamefont
  {Nielsen}}\ and\ \bibinfo {author} {\bibfnamefont {I.~L.}\ \bibnamefont
  {Chuang}},\ }\href@noop {} {\emph {\bibinfo {title} {Quantum Computation and
  Quantum Information}}}\ (\bibinfo  {publisher} {Cambridge University Press},\
  \bibinfo {year} {2010})\BibitemShut {NoStop}%
\bibitem [{\citenamefont {Holevo}\ and\ \citenamefont
  {Giovannetti}(2012)}]{holevo2012}%
  \BibitemOpen
  \bibfield  {author} {\bibinfo {author} {\bibfnamefont {A.~S.}\ \bibnamefont
  {Holevo}}\ and\ \bibinfo {author} {\bibfnamefont {V.}~\bibnamefont
  {Giovannetti}},\ }\bibfield  {title} {\enquote {\bibinfo {title} {Quantum
  channels and their entropic characteristics},}\ }\href {\doibase
  10.1088/0034-4885/75/4/046001} {\bibfield  {journal} {\bibinfo  {journal}
  {Rep. Prog. Phys.}\ }\textbf {\bibinfo {volume} {75}},\ \bibinfo {pages}
  {046001} (\bibinfo {year} {2012})}\BibitemShut {NoStop}%
\bibitem [{\citenamefont {Yu}\ and\ \citenamefont {Eberly}(2009)}]{yu2009}%
  \BibitemOpen
  \bibfield  {author} {\bibinfo {author} {\bibfnamefont {Ting}\ \bibnamefont
  {Yu}}\ and\ \bibinfo {author} {\bibfnamefont {J.~H.}\ \bibnamefont
  {Eberly}},\ }\bibfield  {title} {\enquote {\bibinfo {title} {Sudden death of
  entanglement},}\ }\href {\doibase 10.1126/science.1167343} {\bibfield
  {journal} {\bibinfo  {journal} {Science}\ }\textbf {\bibinfo {volume}
  {323}},\ \bibinfo {pages} {598--601} (\bibinfo {year} {2009})},\ \Eprint
  {http://arxiv.org/abs/https://science.sciencemag.org/content/323/5914/598.full.pdf}
  {https://science.sciencemag.org/content/323/5914/598.full.pdf} \BibitemShut
  {NoStop}%
\bibitem [{\citenamefont {Daffer}\ \emph {et~al.}(2004)\citenamefont {Daffer},
  \citenamefont {W\'odkiewicz}, \citenamefont {Cresser},\ and\ \citenamefont
  {McIver}}]{Daffer2004}%
  \BibitemOpen
  \bibfield  {author} {\bibinfo {author} {\bibfnamefont {Sonja}\ \bibnamefont
  {Daffer}}, \bibinfo {author} {\bibfnamefont {Krzysztof}\ \bibnamefont
  {W\'odkiewicz}}, \bibinfo {author} {\bibfnamefont {James~D.}\ \bibnamefont
  {Cresser}}, \ and\ \bibinfo {author} {\bibfnamefont {John~K.}\ \bibnamefont
  {McIver}},\ }\bibfield  {title} {\enquote {\bibinfo {title} {Depolarizing
  channel as a completely positive map with memory},}\ }\href {\doibase
  10.1103/PhysRevA.70.010304} {\bibfield  {journal} {\bibinfo  {journal} {Phys.
  Rev. A}\ }\textbf {\bibinfo {volume} {70}},\ \bibinfo {pages} {010304}
  (\bibinfo {year} {2004})}\BibitemShut {NoStop}%
\bibitem [{\citenamefont {Shrikant}\ \emph {et~al.}(2018)\citenamefont
  {Shrikant}, \citenamefont {Srikanth},\ and\ \citenamefont
  {Banerjee}}]{Shrikant2018}%
  \BibitemOpen
  \bibfield  {author} {\bibinfo {author} {\bibfnamefont {U.}~\bibnamefont
  {Shrikant}}, \bibinfo {author} {\bibfnamefont {R.}~\bibnamefont {Srikanth}},
  \ and\ \bibinfo {author} {\bibfnamefont {Subhashish}\ \bibnamefont
  {Banerjee}},\ }\bibfield  {title} {\enquote {\bibinfo {title} {Non-markovian
  dephasing and depolarizing channels},}\ }\href {\doibase
  10.1103/PhysRevA.98.032328} {\bibfield  {journal} {\bibinfo  {journal} {Phys.
  Rev. A}\ }\textbf {\bibinfo {volume} {98}},\ \bibinfo {pages} {032328}
  (\bibinfo {year} {2018})}\BibitemShut {NoStop}%
\bibitem [{\citenamefont {Gupta}\ \emph {et~al.}(2020)\citenamefont {Gupta},
  \citenamefont {Gupta}, \citenamefont {Mal}, \citenamefont {Sen(De)},\ and\
  \citenamefont {Sen}}]{Gupta2020}%
  \BibitemOpen
  \bibfield  {author} {\bibinfo {author} {\bibfnamefont {Ribhu}\ \bibnamefont
  {Gupta}}, \bibinfo {author} {\bibfnamefont {Shashank}\ \bibnamefont {Gupta}},
  \bibinfo {author} {\bibfnamefont {Shiladitya}\ \bibnamefont {Mal}}, \bibinfo
  {author} {\bibfnamefont {Aditi}\ \bibnamefont {Sen(De)}}, \ and\ \bibinfo
  {author} {\bibfnamefont {Ujjwal}\ \bibnamefont {Sen}},\ }\bibfield  {title}
  {\enquote {\bibinfo {title} {Constructive feedback of non-markovianity on
  resources in random quantum states},}\ }\href
  {https://arxiv.org/abs/2005.04009} {\bibfield  {journal} {\bibinfo  {journal}
  {arXiv:2005.04009}\ } (\bibinfo {year} {2020})}\BibitemShut {NoStop}%
\bibitem [{\citenamefont {De~Dominicis}\ and\ \citenamefont
  {Giardina}(2006)}]{de_dominicis_giardina_2006}%
  \BibitemOpen
  \bibfield  {author} {\bibinfo {author} {\bibfnamefont {Cirano}\ \bibnamefont
  {De~Dominicis}}\ and\ \bibinfo {author} {\bibfnamefont {Irene}\ \bibnamefont
  {Giardina}},\ }\href {\doibase 10.1017/CBO9780511534836} {\emph {\bibinfo
  {title} {Random Fields and Spin Glasses: A Field Theory Approach}}}\
  (\bibinfo  {publisher} {Cambridge University Press},\ \bibinfo {year}
  {2006})\BibitemShut {NoStop}%
\bibitem [{\citenamefont {Lieb}\ \emph {et~al.}(1961)\citenamefont {Lieb},
  \citenamefont {Schultz},\ and\ \citenamefont {Mattis}}]{Lieb1961}%
  \BibitemOpen
  \bibfield  {author} {\bibinfo {author} {\bibfnamefont {E.}~\bibnamefont
  {Lieb}}, \bibinfo {author} {\bibfnamefont {T.}~\bibnamefont {Schultz}}, \
  and\ \bibinfo {author} {\bibfnamefont {D.}~\bibnamefont {Mattis}},\
  }\bibfield  {title} {\enquote {\bibinfo {title} {Two soluble models of an
  antiferromagnetic chain},}\ }\href {\doibase
  https://doi.org/10.1016/0003-4916(61)90115-4} {\bibfield  {journal} {\bibinfo
   {journal} {Annals of Physics}\ }\textbf {\bibinfo {volume} {16}},\ \bibinfo
  {pages} {407--466} (\bibinfo {year} {1961})}\BibitemShut {NoStop}%
\bibitem [{\citenamefont {Barouch}\ \emph {et~al.}(1970)\citenamefont
  {Barouch}, \citenamefont {McCoy},\ and\ \citenamefont
  {Dresden}}]{Barouch1970}%
  \BibitemOpen
  \bibfield  {author} {\bibinfo {author} {\bibfnamefont {E.}~\bibnamefont
  {Barouch}}, \bibinfo {author} {\bibfnamefont {B.~M.}\ \bibnamefont {McCoy}},
  \ and\ \bibinfo {author} {\bibfnamefont {M.}~\bibnamefont {Dresden}},\
  }\bibfield  {title} {\enquote {\bibinfo {title} {Statistical mechanics of the
  \uppercase{XY} model. \uppercase{I}},}\ }\href {\doibase
  10.1103/PhysRevA.2.1075} {\bibfield  {journal} {\bibinfo  {journal} {Phys.
  Rev. A}\ }\textbf {\bibinfo {volume} {2}},\ \bibinfo {pages} {1075--1092}
  (\bibinfo {year} {1970})}\BibitemShut {NoStop}%
\bibitem [{\citenamefont {Barouch}\ and\ \citenamefont
  {McCoy}(1971{\natexlab{a}})}]{Barouch1971}%
  \BibitemOpen
  \bibfield  {author} {\bibinfo {author} {\bibfnamefont {E.}~\bibnamefont
  {Barouch}}\ and\ \bibinfo {author} {\bibfnamefont {B.~M.}\ \bibnamefont
  {McCoy}},\ }\bibfield  {title} {\enquote {\bibinfo {title} {Statistical
  mechanics of the \uppercase{XY} model. \uppercase{II}. spin-correlation
  functions},}\ }\href {\doibase 10.1103/PhysRevA.3.786} {\bibfield  {journal}
  {\bibinfo  {journal} {Phys. Rev. A}\ }\textbf {\bibinfo {volume} {3}},\
  \bibinfo {pages} {786--804} (\bibinfo {year}
  {1971}{\natexlab{a}})}\BibitemShut {NoStop}%
\bibitem [{\citenamefont {Barouch}\ and\ \citenamefont
  {McCoy}(1971{\natexlab{b}})}]{Barouch1971a}%
  \BibitemOpen
  \bibfield  {author} {\bibinfo {author} {\bibfnamefont {E.}~\bibnamefont
  {Barouch}}\ and\ \bibinfo {author} {\bibfnamefont {B.~M.}\ \bibnamefont
  {McCoy}},\ }\bibfield  {title} {\enquote {\bibinfo {title} {Statistical
  mechanics of the \uppercase{XY} model. \uppercase{III}},}\ }\href {\doibase
  10.1103/PhysRevA.3.2137} {\bibfield  {journal} {\bibinfo  {journal} {Phys.
  Rev. A}\ }\textbf {\bibinfo {volume} {3}},\ \bibinfo {pages} {2137--2140}
  (\bibinfo {year} {1971}{\natexlab{b}})}\BibitemShut {NoStop}%
\bibitem [{\citenamefont {Chakrabarti}\ \emph {et~al.}(1996)\citenamefont
  {Chakrabarti}, \citenamefont {Dutta},\ and\ \citenamefont {Sen}}]{dutta1996}%
  \BibitemOpen
  \bibfield  {author} {\bibinfo {author} {\bibfnamefont {Bikas~K}\ \bibnamefont
  {Chakrabarti}}, \bibinfo {author} {\bibfnamefont {Amit}\ \bibnamefont
  {Dutta}}, \ and\ \bibinfo {author} {\bibfnamefont {Parongama}\ \bibnamefont
  {Sen}},\ }\href {\doibase https://www.springer.com/gp/book/9783540498650}
  {\emph {\bibinfo {title} {Quantum Ising Phases and Transitions in Transverse
  Ising Models}}},\ Vol.~\bibinfo {volume} {41}\ (\bibinfo  {publisher}
  {Springer, Heidelberg},\ \bibinfo {year} {1996})\BibitemShut {NoStop}%
\bibitem [{\citenamefont {Sachdev}(2011)}]{sachdev_2011}%
  \BibitemOpen
  \bibfield  {author} {\bibinfo {author} {\bibfnamefont {Subir}\ \bibnamefont
  {Sachdev}},\ }\href {\doibase 10.1017/CBO9780511973765} {\emph {\bibinfo
  {title} {Quantum Phase Transitions}}},\ \bibinfo {edition} {2nd}\ ed.\
  (\bibinfo  {publisher} {Cambridge University Press},\ \bibinfo {year}
  {2011})\BibitemShut {NoStop}%
\bibitem [{\citenamefont {Yang}\ and\ \citenamefont
  {Yang}(1966{\natexlab{a}})}]{Yang1966}%
  \BibitemOpen
  \bibfield  {author} {\bibinfo {author} {\bibfnamefont {C.~N.}\ \bibnamefont
  {Yang}}\ and\ \bibinfo {author} {\bibfnamefont {C.~P.}\ \bibnamefont
  {Yang}},\ }\bibfield  {title} {\enquote {\bibinfo {title} {One-dimensional
  chain of anisotropic spin-spin interactions. i. proof of bethe's hypothesis
  for ground state in a finite system},}\ }\href {\doibase
  10.1103/PhysRev.150.321} {\bibfield  {journal} {\bibinfo  {journal} {Phys.
  Rev.}\ }\textbf {\bibinfo {volume} {150}},\ \bibinfo {pages} {321--327}
  (\bibinfo {year} {1966}{\natexlab{a}})}\BibitemShut {NoStop}%
\bibitem [{\citenamefont {Yang}\ and\ \citenamefont
  {Yang}(1966{\natexlab{b}})}]{Yang1966a}%
  \BibitemOpen
  \bibfield  {author} {\bibinfo {author} {\bibfnamefont {C.~N.}\ \bibnamefont
  {Yang}}\ and\ \bibinfo {author} {\bibfnamefont {C.~P.}\ \bibnamefont
  {Yang}},\ }\bibfield  {title} {\enquote {\bibinfo {title} {One-dimensional
  chain of anisotropic spin-spin interactions. ii. properties of the
  ground-state energy per lattice site for an infinite system},}\ }\href
  {\doibase 10.1103/PhysRev.150.327} {\bibfield  {journal} {\bibinfo  {journal}
  {Phys. Rev.}\ }\textbf {\bibinfo {volume} {150}},\ \bibinfo {pages}
  {327--339} (\bibinfo {year} {1966}{\natexlab{b}})}\BibitemShut {NoStop}%
\bibitem [{\citenamefont {Yang}\ and\ \citenamefont
  {Yang}(1966{\natexlab{c}})}]{Yang1966b}%
  \BibitemOpen
  \bibfield  {author} {\bibinfo {author} {\bibfnamefont {C.~N.}\ \bibnamefont
  {Yang}}\ and\ \bibinfo {author} {\bibfnamefont {C.~P.}\ \bibnamefont
  {Yang}},\ }\bibfield  {title} {\enquote {\bibinfo {title} {One-dimensional
  chain of anisotropic spin-spin interactions. iii. applications},}\ }\href
  {\doibase 10.1103/PhysRev.151.258} {\bibfield  {journal} {\bibinfo  {journal}
  {Phys. Rev.}\ }\textbf {\bibinfo {volume} {151}},\ \bibinfo {pages}
  {258--264} (\bibinfo {year} {1966}{\natexlab{c}})}\BibitemShut {NoStop}%
\bibitem [{\citenamefont {Langari}(1998)}]{Langari1998}%
  \BibitemOpen
  \bibfield  {author} {\bibinfo {author} {\bibfnamefont {A.}~\bibnamefont
  {Langari}},\ }\bibfield  {title} {\enquote {\bibinfo {title} {Phase diagram
  of the antiferromagnetic xxz model in the presence of an external magnetic
  field},}\ }\href {\doibase 10.1103/PhysRevB.58.14467} {\bibfield  {journal}
  {\bibinfo  {journal} {Phys. Rev. B}\ }\textbf {\bibinfo {volume} {58}},\
  \bibinfo {pages} {14467--14475} (\bibinfo {year} {1998})}\BibitemShut
  {NoStop}%
\bibitem [{\citenamefont {Mikeska}\ and\ \citenamefont
  {Kolezhuk}(2004)}]{Mikeska2004}%
  \BibitemOpen
  \bibfield  {author} {\bibinfo {author} {\bibfnamefont {Hans-J{\"u}rgen}\
  \bibnamefont {Mikeska}}\ and\ \bibinfo {author} {\bibfnamefont {Alexei~K.}\
  \bibnamefont {Kolezhuk}},\ }\enquote {\bibinfo {title} {One-dimensional
  magnetism},}\ in\ \href {\doibase 10.1007/BFb0119591} {\emph {\bibinfo
  {booktitle} {Quantum Magnetism}}},\ \bibinfo {editor} {edited by\ \bibinfo
  {editor} {\bibfnamefont {Ulrich}\ \bibnamefont {Schollw{\"o}ck}}, \bibinfo
  {editor} {\bibfnamefont {Johannes}\ \bibnamefont {Richter}}, \bibinfo
  {editor} {\bibfnamefont {Damian J.~J.}\ \bibnamefont {Farnell}}, \ and\
  \bibinfo {editor} {\bibfnamefont {Raymod~F.}\ \bibnamefont {Bishop}}}\
  (\bibinfo  {publisher} {Springer Berlin Heidelberg},\ \bibinfo {address}
  {Berlin, Heidelberg},\ \bibinfo {year} {2004})\ pp.\ \bibinfo {pages}
  {1--83}\BibitemShut {NoStop}%
\bibitem [{\citenamefont {Giamarchi}(2004)}]{Giamarchi2004}%
  \BibitemOpen
  \bibfield  {author} {\bibinfo {author} {\bibfnamefont {T.}~\bibnamefont
  {Giamarchi}},\ }\href {\doibase 10.1093/acprof:oso/9780198525004.001.0001}
  {\emph {\bibinfo {title} {{Quantum physics in one dimension}}}},\
  International series of monographs on physics\ (\bibinfo  {publisher}
  {Clarendon Press},\ \bibinfo {address} {Oxford},\ \bibinfo {year}
  {2004})\BibitemShut {NoStop}%
\bibitem [{\citenamefont {Peres}(1996)}]{peres1996}%
  \BibitemOpen
  \bibfield  {author} {\bibinfo {author} {\bibfnamefont {Asher}\ \bibnamefont
  {Peres}},\ }\bibfield  {title} {\enquote {\bibinfo {title} {Separability
  criterion for density matrices},}\ }\href {\doibase
  10.1103/PhysRevLett.77.1413} {\bibfield  {journal} {\bibinfo  {journal}
  {Phys. Rev. Lett.}\ }\textbf {\bibinfo {volume} {77}},\ \bibinfo {pages}
  {1413--1415} (\bibinfo {year} {1996})}\BibitemShut {NoStop}%
\bibitem [{\citenamefont {Horodecki}\ \emph {et~al.}(1996)\citenamefont
  {Horodecki}, \citenamefont {Horodecki},\ and\ \citenamefont
  {Horodecki}}]{horodecki1996}%
  \BibitemOpen
  \bibfield  {author} {\bibinfo {author} {\bibfnamefont {Michal}\ \bibnamefont
  {Horodecki}}, \bibinfo {author} {\bibfnamefont {Pawel}\ \bibnamefont
  {Horodecki}}, \ and\ \bibinfo {author} {\bibfnamefont {Ryszard}\ \bibnamefont
  {Horodecki}},\ }\bibfield  {title} {\enquote {\bibinfo {title} {Separability
  of mixed states: necessary and sufficient conditions},}\ }\href {\doibase
  https://doi.org/10.1016/S0375-9601(96)00706-2} {\bibfield  {journal}
  {\bibinfo  {journal} {Phys. Lett. A}\ }\textbf {\bibinfo {volume} {223}},\
  \bibinfo {pages} {1 -- 8} (\bibinfo {year} {1996})}\BibitemShut {NoStop}%
\bibitem [{\citenamefont {Vidal}\ and\ \citenamefont
  {Werner}(2002)}]{vidal2002}%
  \BibitemOpen
  \bibfield  {author} {\bibinfo {author} {\bibfnamefont {G.}~\bibnamefont
  {Vidal}}\ and\ \bibinfo {author} {\bibfnamefont {R.~F.}\ \bibnamefont
  {Werner}},\ }\bibfield  {title} {\enquote {\bibinfo {title} {Computable
  measure of entanglement},}\ }\href {\doibase 10.1103/PhysRevA.65.032314}
  {\bibfield  {journal} {\bibinfo  {journal} {Phys. Rev. A}\ }\textbf {\bibinfo
  {volume} {65}},\ \bibinfo {pages} {032314} (\bibinfo {year}
  {2002})}\BibitemShut {NoStop}%
\bibitem [{\citenamefont {\ifmmode~\dot{Z}\else \.{Z}\fi{}yczkowski}\ \emph
  {et~al.}(1998)\citenamefont {\ifmmode~\dot{Z}\else \.{Z}\fi{}yczkowski},
  \citenamefont {Horodecki}, \citenamefont {Sanpera},\ and\ \citenamefont
  {Lewenstein}}]{zyczkowski1998}%
  \BibitemOpen
  \bibfield  {author} {\bibinfo {author} {\bibfnamefont {Karol}\ \bibnamefont
  {\ifmmode~\dot{Z}\else \.{Z}\fi{}yczkowski}}, \bibinfo {author}
  {\bibfnamefont {Pawe\l{}}\ \bibnamefont {Horodecki}}, \bibinfo {author}
  {\bibfnamefont {Anna}\ \bibnamefont {Sanpera}}, \ and\ \bibinfo {author}
  {\bibfnamefont {Maciej}\ \bibnamefont {Lewenstein}},\ }\bibfield  {title}
  {\enquote {\bibinfo {title} {Volume of the set of separable states},}\ }\href
  {\doibase 10.1103/PhysRevA.58.883} {\bibfield  {journal} {\bibinfo  {journal}
  {Phys. Rev. A}\ }\textbf {\bibinfo {volume} {58}},\ \bibinfo {pages}
  {883--892} (\bibinfo {year} {1998})}\BibitemShut {NoStop}%
\bibitem [{\citenamefont {Lee}\ \emph {et~al.}(2000)\citenamefont {Lee},
  \citenamefont {Kim}, \citenamefont {Park},\ and\ \citenamefont
  {Lee}}]{lee2000}%
  \BibitemOpen
  \bibfield  {author} {\bibinfo {author} {\bibfnamefont {Jinhyoung}\
  \bibnamefont {Lee}}, \bibinfo {author} {\bibfnamefont {M.~S.}\ \bibnamefont
  {Kim}}, \bibinfo {author} {\bibfnamefont {Y.~J.}\ \bibnamefont {Park}}, \
  and\ \bibinfo {author} {\bibfnamefont {S.}~\bibnamefont {Lee}},\ }\bibfield
  {title} {\enquote {\bibinfo {title} {Partial teleportation of entanglement in
  a noisy environment},}\ }\href {\doibase 10.1080/09500340008235138}
  {\bibfield  {journal} {\bibinfo  {journal} {Journal of Modern Optics}\
  }\textbf {\bibinfo {volume} {47}},\ \bibinfo {pages} {2151--2164} (\bibinfo
  {year} {2000})}\BibitemShut {NoStop}%
\bibitem [{\citenamefont {Horodecki}\ \emph {et~al.}(1997)\citenamefont
  {Horodecki}, \citenamefont {Horodecki},\ and\ \citenamefont
  {Horodecki}}]{Horodecki1997}%
  \BibitemOpen
  \bibfield  {author} {\bibinfo {author} {\bibfnamefont {Micha\l{}}\
  \bibnamefont {Horodecki}}, \bibinfo {author} {\bibfnamefont {Pawe\l{}}\
  \bibnamefont {Horodecki}}, \ and\ \bibinfo {author} {\bibfnamefont {Ryszard}\
  \bibnamefont {Horodecki}},\ }\bibfield  {title} {\enquote {\bibinfo {title}
  {Inseparable two spin- $\frac{1}{2}$ density matrices can be distilled to a
  singlet form},}\ }\href {\doibase 10.1103/PhysRevLett.78.574} {\bibfield
  {journal} {\bibinfo  {journal} {Phys. Rev. Lett.}\ }\textbf {\bibinfo
  {volume} {78}},\ \bibinfo {pages} {574--577} (\bibinfo {year}
  {1997})}\BibitemShut {NoStop}%
\bibitem [{\citenamefont {D\"ur}\ \emph
  {et~al.}(2000{\natexlab{b}})\citenamefont {D\"ur}, \citenamefont {Cirac},
  \citenamefont {Lewenstein},\ and\ \citenamefont {Bru\ss{}}}]{Dur2000a}%
  \BibitemOpen
  \bibfield  {author} {\bibinfo {author} {\bibfnamefont {W.}~\bibnamefont
  {D\"ur}}, \bibinfo {author} {\bibfnamefont {J.~I.}\ \bibnamefont {Cirac}},
  \bibinfo {author} {\bibfnamefont {M.}~\bibnamefont {Lewenstein}}, \ and\
  \bibinfo {author} {\bibfnamefont {D.}~\bibnamefont {Bru\ss{}}},\ }\bibfield
  {title} {\enquote {\bibinfo {title} {Distillability and partial transposition
  in bipartite systems},}\ }\href {\doibase 10.1103/PhysRevA.61.062313}
  {\bibfield  {journal} {\bibinfo  {journal} {Phys. Rev. A}\ }\textbf {\bibinfo
  {volume} {61}},\ \bibinfo {pages} {062313} (\bibinfo {year}
  {2000}{\natexlab{b}})}\BibitemShut {NoStop}%
\bibitem [{\citenamefont {D\"ur}\ and\ \citenamefont {Cirac}(2000)}]{Dur2000b}%
  \BibitemOpen
  \bibfield  {author} {\bibinfo {author} {\bibfnamefont {W.}~\bibnamefont
  {D\"ur}}\ and\ \bibinfo {author} {\bibfnamefont {J.~I.}\ \bibnamefont
  {Cirac}},\ }\bibfield  {title} {\enquote {\bibinfo {title} {Activating bound
  entanglement in multiparticle systems},}\ }\href {\doibase
  10.1103/PhysRevA.62.022302} {\bibfield  {journal} {\bibinfo  {journal} {Phys.
  Rev. A}\ }\textbf {\bibinfo {volume} {62}},\ \bibinfo {pages} {022302}
  (\bibinfo {year} {2000})}\BibitemShut {NoStop}%
\bibitem [{\citenamefont {Plenio}(2005)}]{plenio2005}%
  \BibitemOpen
  \bibfield  {author} {\bibinfo {author} {\bibfnamefont {M.~B.}\ \bibnamefont
  {Plenio}},\ }\bibfield  {title} {\enquote {\bibinfo {title} {Logarithmic
  negativity: A full entanglement monotone that is not convex},}\ }\href
  {\doibase 10.1103/PhysRevLett.95.090503} {\bibfield  {journal} {\bibinfo
  {journal} {Phys. Rev. Lett.}\ }\textbf {\bibinfo {volume} {95}},\ \bibinfo
  {pages} {090503} (\bibinfo {year} {2005})}\BibitemShut {NoStop}%
\bibitem [{\citenamefont {Bennett}\ \emph
  {et~al.}(1996{\natexlab{a}})\citenamefont {Bennett}, \citenamefont
  {DiVincenzo}, \citenamefont {Smolin},\ and\ \citenamefont
  {Wootters}}]{bennett1996}%
  \BibitemOpen
  \bibfield  {author} {\bibinfo {author} {\bibfnamefont {Charles~H.}\
  \bibnamefont {Bennett}}, \bibinfo {author} {\bibfnamefont {David~P.}\
  \bibnamefont {DiVincenzo}}, \bibinfo {author} {\bibfnamefont {John~A.}\
  \bibnamefont {Smolin}}, \ and\ \bibinfo {author} {\bibfnamefont {William~K.}\
  \bibnamefont {Wootters}},\ }\bibfield  {title} {\enquote {\bibinfo {title}
  {Mixed-state entanglement and quantum error correction},}\ }\href {\doibase
  10.1103/PhysRevA.54.3824} {\bibfield  {journal} {\bibinfo  {journal} {Phys.
  Rev. A}\ }\textbf {\bibinfo {volume} {54}},\ \bibinfo {pages} {3824--3851}
  (\bibinfo {year} {1996}{\natexlab{a}})}\BibitemShut {NoStop}%
\bibitem [{\citenamefont {Bennett}\ \emph
  {et~al.}(1996{\natexlab{b}})\citenamefont {Bennett}, \citenamefont
  {Bernstein}, \citenamefont {Popescu},\ and\ \citenamefont
  {Schumacher}}]{bennett1996a}%
  \BibitemOpen
  \bibfield  {author} {\bibinfo {author} {\bibfnamefont {Charles~H.}\
  \bibnamefont {Bennett}}, \bibinfo {author} {\bibfnamefont {Herbert~J.}\
  \bibnamefont {Bernstein}}, \bibinfo {author} {\bibfnamefont {Sandu}\
  \bibnamefont {Popescu}}, \ and\ \bibinfo {author} {\bibfnamefont {Benjamin}\
  \bibnamefont {Schumacher}},\ }\bibfield  {title} {\enquote {\bibinfo {title}
  {Concentrating partial entanglement by local operations},}\ }\href {\doibase
  10.1103/PhysRevA.53.2046} {\bibfield  {journal} {\bibinfo  {journal} {Phys.
  Rev. A}\ }\textbf {\bibinfo {volume} {53}},\ \bibinfo {pages} {2046--2052}
  (\bibinfo {year} {1996}{\natexlab{b}})}\BibitemShut {NoStop}%
\bibitem [{\citenamefont {Bengtsson}\ and\ \citenamefont
  {Zyczkowski}(2006)}]{bengtsson_zyczkowski_2006}%
  \BibitemOpen
  \bibfield  {author} {\bibinfo {author} {\bibfnamefont {Ingemar}\ \bibnamefont
  {Bengtsson}}\ and\ \bibinfo {author} {\bibfnamefont {Karol}\ \bibnamefont
  {Zyczkowski}},\ }\href {\doibase 10.1017/CBO9780511535048} {\emph {\bibinfo
  {title} {Geometry of Quantum States: An Introduction to Quantum
  Entanglement}}}\ (\bibinfo  {publisher} {Cambridge University Press},\
  \bibinfo {year} {2006})\BibitemShut {NoStop}%
\bibitem [{\citenamefont {Banerjee}\ \emph {et~al.}(2020)\citenamefont
  {Banerjee}, \citenamefont {Pal},\ and\ \citenamefont
  {Sen(De)}}]{banerjee2020}%
  \BibitemOpen
  \bibfield  {author} {\bibinfo {author} {\bibfnamefont {Ratul}\ \bibnamefont
  {Banerjee}}, \bibinfo {author} {\bibfnamefont {Amit~Kumar}\ \bibnamefont
  {Pal}}, \ and\ \bibinfo {author} {\bibfnamefont {Aditi}\ \bibnamefont
  {Sen(De)}},\ }\bibfield  {title} {\enquote {\bibinfo {title} {Uniform
  decoherence effect on localizable entanglement in random multiqubit pure
  states},}\ }\href {\doibase 10.1103/PhysRevA.101.042339} {\bibfield
  {journal} {\bibinfo  {journal} {Phys. Rev. A}\ }\textbf {\bibinfo {volume}
  {101}},\ \bibinfo {pages} {042339} (\bibinfo {year} {2020})}\BibitemShut
  {NoStop}%
\bibitem [{\citenamefont {Chen}\ and\ \citenamefont {Jiang}(2020)}]{chen2020}%
  \BibitemOpen
  \bibfield  {author} {\bibinfo {author} {\bibfnamefont {Xiao-yu}\ \bibnamefont
  {Chen}}\ and\ \bibinfo {author} {\bibfnamefont {Li-zhen}\ \bibnamefont
  {Jiang}},\ }\bibfield  {title} {\enquote {\bibinfo {title} {Noise tolerance
  of dicke states},}\ }\href {\doibase 10.1103/PhysRevA.101.012308} {\bibfield
  {journal} {\bibinfo  {journal} {Phys. Rev. A}\ }\textbf {\bibinfo {volume}
  {101}},\ \bibinfo {pages} {012308} (\bibinfo {year} {2020})}\BibitemShut
  {NoStop}%
\bibitem [{\citenamefont {Sanpera}\ \emph {et~al.}(1998)\citenamefont
  {Sanpera}, \citenamefont {Tarrach},\ and\ \citenamefont
  {Vidal}}]{sanpera1998}%
  \BibitemOpen
  \bibfield  {author} {\bibinfo {author} {\bibfnamefont {Anna}\ \bibnamefont
  {Sanpera}}, \bibinfo {author} {\bibfnamefont {Rolf}\ \bibnamefont {Tarrach}},
  \ and\ \bibinfo {author} {\bibfnamefont {Guifr\'e}\ \bibnamefont {Vidal}},\
  }\bibfield  {title} {\enquote {\bibinfo {title} {Local description of quantum
  inseparability},}\ }\href {\doibase 10.1103/PhysRevA.58.826} {\bibfield
  {journal} {\bibinfo  {journal} {Phys. Rev. A}\ }\textbf {\bibinfo {volume}
  {58}},\ \bibinfo {pages} {826--830} (\bibinfo {year} {1998})}\BibitemShut
  {NoStop}%
\bibitem [{\citenamefont {Rana}(2013)}]{rana2013}%
  \BibitemOpen
  \bibfield  {author} {\bibinfo {author} {\bibfnamefont {Swapan}\ \bibnamefont
  {Rana}},\ }\bibfield  {title} {\enquote {\bibinfo {title} {Negative
  eigenvalues of partial transposition of arbitrary bipartite states},}\ }\href
  {\doibase 10.1103/PhysRevA.87.054301} {\bibfield  {journal} {\bibinfo
  {journal} {Phys. Rev. A}\ }\textbf {\bibinfo {volume} {87}},\ \bibinfo
  {pages} {054301} (\bibinfo {year} {2013})}\BibitemShut {NoStop}%
\bibitem [{\citenamefont {Moreno}\ and\ \citenamefont
  {Parisio}(2018)}]{Moreno2018}%
  \BibitemOpen
  \bibfield  {author} {\bibinfo {author} {\bibfnamefont {M.~G.~M.}\
  \bibnamefont {Moreno}}\ and\ \bibinfo {author} {\bibfnamefont {Fernando}\
  \bibnamefont {Parisio}},\ }\href {\doibase 10.48550/ARXIV.1801.00762}
  {\enquote {\bibinfo {title} {All bipartitions of arbitrary dicke states},}\ }
  (\bibinfo {year} {2018})\BibitemShut {NoStop}%
\bibitem [{\citenamefont {Banerjee}\ \emph {et~al.}(2022)\citenamefont
  {Banerjee}, \citenamefont {Pal},\ and\ \citenamefont
  {Sen(De)}}]{Banerjee2022}%
  \BibitemOpen
  \bibfield  {author} {\bibinfo {author} {\bibfnamefont {Ratul}\ \bibnamefont
  {Banerjee}}, \bibinfo {author} {\bibfnamefont {Amit~Kumar}\ \bibnamefont
  {Pal}}, \ and\ \bibinfo {author} {\bibfnamefont {Aditi}\ \bibnamefont
  {Sen(De)}},\ }\bibfield  {title} {\enquote {\bibinfo {title} {Hierarchies of
  localizable entanglement due to spatial distribution of local noise},}\
  }\href {\doibase 10.1103/PhysRevResearch.4.023035} {\bibfield  {journal}
  {\bibinfo  {journal} {Phys. Rev. Res.}\ }\textbf {\bibinfo {volume} {4}},\
  \bibinfo {pages} {023035} (\bibinfo {year} {2022})}\BibitemShut {NoStop}%
\bibitem [{\citenamefont {Pfeuty}(1970)}]{Pfeuty1970}%
  \BibitemOpen
  \bibfield  {author} {\bibinfo {author} {\bibfnamefont {P.}~\bibnamefont
  {Pfeuty}},\ }\bibfield  {title} {\enquote {\bibinfo {title} {The
  one-dimensional \uppercase{I}sing model with a transverse field},}\ }\href
  {\doibase https://doi.org/10.1016/0003-4916(70)90270-8} {\bibfield  {journal}
  {\bibinfo  {journal} {Annals of Physics}\ }\textbf {\bibinfo {volume} {57}},\
  \bibinfo {pages} {79--90} (\bibinfo {year} {1970})}\BibitemShut {NoStop}%
\bibitem [{\citenamefont {Osterloh}\ \emph {et~al.}(2002)\citenamefont
  {Osterloh}, \citenamefont {Amico}, \citenamefont {Falci},\ and\ \citenamefont
  {Fazio}}]{osterloh2002}%
  \BibitemOpen
  \bibfield  {author} {\bibinfo {author} {\bibfnamefont {A.}~\bibnamefont
  {Osterloh}}, \bibinfo {author} {\bibfnamefont {Luigi}\ \bibnamefont {Amico}},
  \bibinfo {author} {\bibfnamefont {G.}~\bibnamefont {Falci}}, \ and\ \bibinfo
  {author} {\bibfnamefont {Rosario}\ \bibnamefont {Fazio}},\ }\bibfield
  {title} {\enquote {\bibinfo {title} {Scaling of entanglement close to a
  quantum phase transition},}\ }\href {\doibase 10.1038/416608a} {\bibfield
  {journal} {\bibinfo  {journal} {Nature}\ }\textbf {\bibinfo {volume} {416}},\
  \bibinfo {pages} {608--610} (\bibinfo {year} {2002})}\BibitemShut {NoStop}%
\bibitem [{\citenamefont {Osborne}\ and\ \citenamefont
  {Nielsen}(2002)}]{osborne2002}%
  \BibitemOpen
  \bibfield  {author} {\bibinfo {author} {\bibfnamefont {Tobias~J.}\
  \bibnamefont {Osborne}}\ and\ \bibinfo {author} {\bibfnamefont {Michael~A.}\
  \bibnamefont {Nielsen}},\ }\bibfield  {title} {\enquote {\bibinfo {title}
  {Entanglement in a simple quantum phase transition},}\ }\href {\doibase
  10.1103/PhysRevA.66.032110} {\bibfield  {journal} {\bibinfo  {journal} {Phys.
  Rev. A}\ }\textbf {\bibinfo {volume} {66}},\ \bibinfo {pages} {032110}
  (\bibinfo {year} {2002})}\BibitemShut {NoStop}%
\bibitem [{\citenamefont {Schindler}\ \emph {et~al.}(2013)\citenamefont
  {Schindler}, \citenamefont {Nigg}, \citenamefont {Monz}, \citenamefont
  {Barreiro}, \citenamefont {Martinez}, \citenamefont {Wang}, \citenamefont
  {Quint}, \citenamefont {Brandl}, \citenamefont {Nebendahl}, \citenamefont
  {Roos}, \citenamefont {Chwalla}, \citenamefont {Hennrich},\ and\
  \citenamefont {Blatt}}]{schindler2013}%
  \BibitemOpen
  \bibfield  {author} {\bibinfo {author} {\bibfnamefont {Philipp}\ \bibnamefont
  {Schindler}}, \bibinfo {author} {\bibfnamefont {Daniel}\ \bibnamefont
  {Nigg}}, \bibinfo {author} {\bibfnamefont {Thomas}\ \bibnamefont {Monz}},
  \bibinfo {author} {\bibfnamefont {Julio~T}\ \bibnamefont {Barreiro}},
  \bibinfo {author} {\bibfnamefont {Esteban}\ \bibnamefont {Martinez}},
  \bibinfo {author} {\bibfnamefont {Shannon~X}\ \bibnamefont {Wang}}, \bibinfo
  {author} {\bibfnamefont {Stephan}\ \bibnamefont {Quint}}, \bibinfo {author}
  {\bibfnamefont {Matthias~F}\ \bibnamefont {Brandl}}, \bibinfo {author}
  {\bibfnamefont {Volckmar}\ \bibnamefont {Nebendahl}}, \bibinfo {author}
  {\bibfnamefont {Christian~F}\ \bibnamefont {Roos}}, \bibinfo {author}
  {\bibfnamefont {Michael}\ \bibnamefont {Chwalla}}, \bibinfo {author}
  {\bibfnamefont {Markus}\ \bibnamefont {Hennrich}}, \ and\ \bibinfo {author}
  {\bibfnamefont {Rainer}\ \bibnamefont {Blatt}},\ }\bibfield  {title}
  {\enquote {\bibinfo {title} {A quantum information processor with trapped
  ions},}\ }\href {\doibase 10.1088/1367-2630/15/12/123012} {\bibfield
  {journal} {\bibinfo  {journal} {New Journal of Physics}\ }\textbf {\bibinfo
  {volume} {15}},\ \bibinfo {pages} {123012} (\bibinfo {year}
  {2013})}\BibitemShut {NoStop}%
\bibitem [{\citenamefont {Bermudez}\ \emph {et~al.}(2017)\citenamefont
  {Bermudez}, \citenamefont {Xu}, \citenamefont {Nigmatullin}, \citenamefont
  {O'Gorman}, \citenamefont {Negnevitsky}, \citenamefont {Schindler},
  \citenamefont {Monz}, \citenamefont {Poschinger}, \citenamefont {Hempel},
  \citenamefont {Home}, \citenamefont {Schmidt-Kaler}, \citenamefont {Biercuk},
  \citenamefont {Blatt}, \citenamefont {Benjamin},\ and\ \citenamefont
  {M\"uller}}]{bermudez2017}%
  \BibitemOpen
  \bibfield  {author} {\bibinfo {author} {\bibfnamefont {A.}~\bibnamefont
  {Bermudez}}, \bibinfo {author} {\bibfnamefont {X.}~\bibnamefont {Xu}},
  \bibinfo {author} {\bibfnamefont {R.}~\bibnamefont {Nigmatullin}}, \bibinfo
  {author} {\bibfnamefont {J.}~\bibnamefont {O'Gorman}}, \bibinfo {author}
  {\bibfnamefont {V.}~\bibnamefont {Negnevitsky}}, \bibinfo {author}
  {\bibfnamefont {P.}~\bibnamefont {Schindler}}, \bibinfo {author}
  {\bibfnamefont {T.}~\bibnamefont {Monz}}, \bibinfo {author} {\bibfnamefont
  {U.~G.}\ \bibnamefont {Poschinger}}, \bibinfo {author} {\bibfnamefont
  {C.}~\bibnamefont {Hempel}}, \bibinfo {author} {\bibfnamefont
  {J.}~\bibnamefont {Home}}, \bibinfo {author} {\bibfnamefont {F.}~\bibnamefont
  {Schmidt-Kaler}}, \bibinfo {author} {\bibfnamefont {M.}~\bibnamefont
  {Biercuk}}, \bibinfo {author} {\bibfnamefont {R.}~\bibnamefont {Blatt}},
  \bibinfo {author} {\bibfnamefont {S.}~\bibnamefont {Benjamin}}, \ and\
  \bibinfo {author} {\bibfnamefont {M.}~\bibnamefont {M\"uller}},\ }\bibfield
  {title} {\enquote {\bibinfo {title} {Assessing the progress of trapped-ion
  processors towards fault-tolerant quantum computation},}\ }\href {\doibase
  10.1103/PhysRevX.7.041061} {\bibfield  {journal} {\bibinfo  {journal} {Phys.
  Rev. X}\ }\textbf {\bibinfo {volume} {7}},\ \bibinfo {pages} {041061}
  (\bibinfo {year} {2017})}\BibitemShut {NoStop}%
\bibitem [{\citenamefont {Breuer}\ and\ \citenamefont
  {Petruccione}(2002)}]{breuer2002}%
  \BibitemOpen
  \bibfield  {author} {\bibinfo {author} {\bibfnamefont {H.-P.}\ \bibnamefont
  {Breuer}}\ and\ \bibinfo {author} {\bibfnamefont {F.}~\bibnamefont
  {Petruccione}},\ }\href@noop {} {\emph {\bibinfo {title} {The Theory of Open
  Quantum Systems}}}\ (\bibinfo  {publisher} {Oxford University Press,
  Oxford},\ \bibinfo {year} {2002})\BibitemShut {NoStop}%
\end{thebibliography}%

\end{document}